\documentclass[12pt,a4paper]{article}
\usepackage[margin=1in]{geometry}
\usepackage[T1]{fontenc}
\usepackage{lmodern}
\usepackage{amsmath,amssymb,amsthm,bm,mathtools}
\usepackage{graphicx,booktabs,array,longtable}
\usepackage{csquotes}
\usepackage[english]{babel}
\usepackage[style=apa6,natbib=true,backend=biber,
 doi=true,isbn=false,url=false,eprint=false]{biblatex}
\DeclareLanguageMapping{english}{english-apa6}
\AtEveryBibitem{\clearfield{number}\clearfield{issn}}
\usepackage{setspace,microtype,enumitem}
\usepackage{caption,subcaption,placeins,needspace}
\usepackage{etoolbox}
\usepackage[hidelinks]{hyperref}
\hypersetup{pdftitle={Optimal Radial Weighting for Robust Mutual Fund Selection},
 pdfauthor={Ping Zhao, Qianya Sun, and Long Feng},pdflang={en-US},
 pdfsubject={Radially weighted alpha estimation and false discovery control},
 pdfkeywords={Mutual fund selection, false discovery rate, radial weighting}}
\allowdisplaybreaks[2]

\AtBeginEnvironment{longtable}{}
\setlist[enumerate]{itemsep=2pt,topsep=4pt}
\setlist[itemize]{itemsep=2pt,topsep=4pt}
\numberwithin{equation}{section}
\newtheorem{assumption}{Assumption}
\newtheorem{theorem}{Theorem}
\newtheorem*{theorem*}{Theorem}

\newtheorem{corollary}{Corollary}
\newtheorem*{corollary*}{Corollary}
\newtheorem{lemma}{Lemma}[section]
\newtheorem*{lemma*}{Lemma}
\BeforeBeginEnvironment{assumption}{\Needspace{3\baselineskip}}
\BeforeBeginEnvironment{theorem}{\Needspace{3\baselineskip}}
\BeforeBeginEnvironment{corollary}{\Needspace{3\baselineskip}}
\BeforeBeginEnvironment{lemma}{\Needspace{3\baselineskip}}
\BeforeBeginEnvironment{proof}{\Needspace{3\baselineskip}}
\theoremstyle{remark}

\newcommand{\E}{\mathbb E}
\newcommand{\Pp}{\mathbb P}
\newcommand{\Var}{\operatorname{Var}}
\newcommand{\Cov}{\operatorname{Cov}}
\newcommand{\diag}{\operatorname{diag}}
\newcommand{\tr}{\operatorname{tr}}
\newcommand{\rank}{\operatorname{rank}}
\newcommand{\FDR}{\operatorname{FDR}}
\newcommand{\FDP}{\operatorname{FDP}}
\newcommand{\TDP}{\operatorname{TDP}}

\newcommand{\op}{\mathrm{op}}
\newcommand{\ind}{\mathbb I}
\newcommand{\one}{\bm 1}
\newcommand{\R}{\mathbb R}
\newcommand{\barPhi}{\overline\Phi}

\newcommand{\LS}{\mathrm{LS}}
\newcommand{\SSm}{\mathrm{SS}}
\newcommand{\IN}{\mathrm{IN}}
\newcommand{\INlam}{\mathrm{IN},\lambda}
\makeatletter
\renewcommand{\@maketitle}{%
  \begin{center}
    {\Large \@title\par}
    \vspace{1em}
    {\normalsize\begin{tabular}[t]{c}\@author\end{tabular}\par}
  \end{center}}
\makeatother

\title{\Large Optimal Radial Weighting for Robust Mutual Fund Selection}
\author{Ping Zhao$^{1}$\qquad Qianya Sun$^{2}$\qquad Long Feng$^{2}$\\[3pt]
  \small $^{1}$School of Mathematical Sciences, Tiangong University\\[2pt]
  \small $^{2}$School of Statistics and Data Science, Nankai University}
\date{}

\begin{document}
\maketitle
\begin{abstract}

Identifying skilled mutual funds requires methods that distinguish positive
alphas from sampling variation when many funds are evaluated simultaneously.
Heavy-tailed returns and dependence across funds complicate this task. We
propose a robust selection procedure based on radially weighted alpha
estimation in linear factor pricing models. The method uses both the
directions and lengths of standardized residuals, estimates marginal scales
from pairwise residual differences, and accounts for the estimation of
factor loadings. Within the proposed class, inverse-norm weighting minimizes
the asymptotic variance of the alpha estimators and maximizes the local
asymptotic power of individual tests at a common significance level. For
radial weights fixed conditional on the factor design, we establish
asymptotic false discovery rate control under elliptically symmetric errors
and show that the proportion of sufficiently strong positive-alpha funds
selected tends to one. The theory accommodates comparable cross-sectional
and time-series dimensions and nonvanishing alphas. We also develop a
smoothed implementation and assess its finite-sample properties. Simulations
show higher true discovery proportions than least-squares and spatial-sign
methods under heavy-tailed distributions and improved numerical stability
from smoothing. An application to U.S. equity mutual funds examines the
subsequent performance of portfolios formed from the selected funds.

\end{abstract}

\section{Introduction}

Evaluating mutual fund performance is an important problem for investors
and fund managers. A common approach measures performance by the alpha in
a linear factor pricing model, which represents the part of expected
return not explained by the chosen pricing factors
\citep{Jensen1968,FamaFrench1993}. \citet{Carhart1997} shows that common
return factors and fund expenses explain much of the persistence in fund
performance, highlighting the importance of risk adjustment when assessing
managerial skill. A positive population alpha indicates performance above
the benchmark, whereas a positive estimate may reflect sampling variation.
Distinguishing skill from luck is therefore essential when selecting
funds from a large cross-section.

A substantial literature studies this distinction from the investor's
perspective. \citet{Baks2001} develop a Bayesian approach to performance
evaluation that incorporates prior beliefs about managerial ability into
portfolio decisions. \citet{PastorStambaugh2002} combine information from
fund returns and passive benchmarks with beliefs about asset pricing and
managerial skill when selecting equity mutual funds. Bootstrap methods
provide another approach to assessing whether exceptional estimated
performance can be explained by luck. \citet{Kosowski2006} and
\citet{FamaFrench2010} use such methods to examine the cross-section of
fund performance. \citet{HarveyLiu2022} revisit these studies and show
that the treatment of fund histories and resampling can materially
affect their conclusions. These contributions establish the importance
of accounting for estimation uncertainty when evaluating apparent
outperformance.

The distinction becomes more difficult when many funds are evaluated
simultaneously. Even if each individual test has a small Type I error
probability, testing a large number of funds can produce a substantial
number of false discoveries. Related concerns arise in searches across technical
trading rules, calendar effects, and pricing factors
\citep{Sullivan1999,Sullivan2001,HarveyLiuZhu2016}.
\citet{White2000} and \citet{Hansen2005} develop bootstrap-based
tests of whether any model in a collection outperforms a benchmark.
\citet{RomanoWolf2005} develop a stepwise procedure that accounts for
dependence among test statistics and asymptotically controls the
familywise error rate. These results motivate evaluating an individual fund's apparent
outperformance in relation to the full set of funds examined.

The false discovery rate (FDR), introduced by
\citet{BenjaminiHochberg1995}, provides a useful criterion for fund
selection. In this setting, it is the expected proportion of selected
funds whose population alphas are nonpositive.
\citet{BenjaminiYekutieli2001} study FDR control under dependence, while
\citet{Storey2002} and \citet{StoreyTaylorSiegmund2004} develop approaches
to estimating FDR and using these estimates to construct testing
procedures. In finance, \citet{Barras2010} use false-discovery methods
to account for luck when estimating the proportions of funds with
different levels of performance. \citet{BajgrowiczScaillet2012} apply
this perspective to technical trading rules. \citet{HarveyLiu2020}
develop a double-bootstrap approach that considers both false and
missed discoveries. Together, these studies motivate selection rules
that limit false discoveries while retaining power to identify funds
with positive alphas.

Cross-sectional dependence is a central feature of this selection
problem because funds can have similar exposures and holdings.
\citet{FanHanGu2012} develop principal factor approximation for
estimating the false discovery proportion under general covariance
dependence. For mutual fund selection, \citet{LanDu2019} construct
factor-adjusted p-values after estimating latent common factors and
establish consistent false discovery proportion estimation under their
conditions. \citet{GiglioLiaoXiu2021} develop large-scale alpha testing
that accounts for omitted factors and missing observations, with a
wild-bootstrap implementation. \citet{WangHanTong2023} take an
empirical Bayes approach, using a dependent mixture model for
standardized alpha estimates. Their optimal multiple-testing rule
balances false discoveries and missed discoveries through conditional
probabilities of funds being unskilled. These approaches demonstrate
the value of using information across funds when making individual
selection decisions.

Heavy-tailed returns raise further questions about the accuracy and
power of large-scale testing procedures. \citet{LiuShao2014} study how
tail behavior and bootstrap calibration affect FDR control in
large-scale t-tests. Robust estimation offers a way to reduce the
influence of unusually large observations. \citet{ZhouBoseFanLiu2018}
develop robust M-estimation theory with applications to
dependence-adjusted multiple testing, and \citet{FanKeSunZhou2019}
combine robust estimation with factor adjustment in FarmTest.
These contributions show that accounting for dependence and obtaining
reliable marginal statistics are both important for testing with
heavy-tailed data.

Another approach uses symmetrized data aggregation.
\citet{DuGuoSunZou2023} combine sample splitting, screening, and
information pooling to construct multiple-testing statistics with
useful symmetry properties. \citet{Feng2024} develop a one-sided
refinement for robust mutual fund selection in linear factor models.
\citet{Wang2025} propose the spatial-sign based SS-BH procedure,
which combines robust location and scale estimation with one-sided
testing and FDR control. Building on this approach, we study the efficiency of residual-based
alpha estimation in a model with observed pricing factors and
cross-sectional dependence.

Conventional least-squares alpha statistics are sensitive to extreme
returns and can lose efficiency under heavy tails. SS-BH reduces this
sensitivity by replacing residual vectors with their directions.
The efficiency of this construction depends on how the location and
scale equations use the projected returns.
After factor projection, the alpha signal varies across dates through
the transformed intercept. The original spatial-median equation fits
a common location without using this intercept as a score weight.
It also assigns the same weight to every spatial direction, leaving
potentially useful residual-length information unused.

The original diagonal scale estimate is fitted jointly with that
location. Consequently, the remaining projected alpha signal and
location estimation error enter its scale residuals. Each squared
sign also contains a denominator involving every fund, making the
marginal-scale approximation sensitive to cross-sectional dependence
when residual norms do not concentrate well. These issues motivate
three changes: estimate scales from pairwise quantiles of unrestricted
regression residuals, use the transformed intercept in the alpha
equation, and incorporate residual norms through radial weights.
Section~\ref{sec:motivation} relates these features to the estimating
equations and their finite-sample behavior.

Previous work on inverse-norm tests provides a useful starting point.
\citet{Feng2021} introduce the inverse norm sign test (INST) for
high-dimensional one-sample location problems and establish its
optimality within a class of weighted spatial-sign tests.
\citet{Huang2023} extend the approach to two-sample location testing,
and \citet{Yan2025} develop inverse-norm weighted max-type and max-sum
tests. These results show that using information about observation
lengths, in addition to directions, can improve the power of location
tests. These
findings motivate using radial information for fund selection.

This paper develops a robust procedure for selecting mutual funds with
positive alphas in a linear factor pricing model. We estimate marginal
scales from quantiles of pairwise differences of unrestricted regression
residuals and hold these scales fixed during alpha estimation. We then
weight residual directions according to their lengths and incorporate
the projected intercept in the estimating equation. The resulting
studentized alpha estimates provide one-sided p-values for the
Benjamini--Hochberg procedure. This construction separates marginal-scale
estimation from alpha estimation and uses radial information to improve
the precision of individual performance measures.

The first contribution is an efficiency characterization of this class
of alpha estimators. We show that inverse-norm weighting minimizes their
asymptotic variance and therefore maximizes the local asymptotic power
of individual alpha tests at a common significance level. An exact
expression quantifies the efficiency loss from alternative weights.
Comparisons with least-squares and spatial-sign procedures identify the
gains from radial weighting and from accounting for the projected
intercept, clarifying the role of residual magnitudes in fund selection.

The second contribution is a theory for simultaneous fund selection
that accounts for estimating factor loadings, marginal scales, and
standard errors from the same observations. For radial weights fixed
conditional on the factor design, we establish uniform Gaussian
approximations and asymptotic FDR control under elliptically symmetric
errors and suitable dependence and signal conditions. The conditions
permit nonvanishing alphas and comparable cross-sectional and time-series
dimensions. The fixed-order radial moment conditions accommodate
standardized multivariate $t$ distributions with degrees of freedom
greater than two. We also show that the proportion of sufficiently
strong positive-alpha funds selected tends to one.

We implement the inverse-norm procedure, denoted IN-BH, using a smoothed
estimating equation and study its finite-sample performance. The
simulations show higher true discovery proportions than least-squares
and spatial-sign procedures under heavy-tailed distributions, together with
improved numerical stability
from smoothing. An application to U.S. equity mutual funds illustrates
rolling selection and subsequent portfolio performance, with higher
terminal wealth and smaller drawdowns in the reported comparisons.

Section~\ref{sec:method} introduces the model, the benchmark procedures,
and the proposed weighted estimator.
Section~\ref{sec:theory} gives its theoretical properties.
Section~\ref{sec:optimal} derives the optimal weight and the benchmark
efficiency comparisons. Section~\ref{sec:simulation} reports the numerical
experiments. Section~\ref{sec:empirical} applies the procedures to U.S.
equity mutual funds, and Section~\ref{sec:conclusion} concludes. The
appendices contain the proofs.

\noindent\textit{Notation.}
All vectors are column vectors. We use $\|\cdot\|$ for the Euclidean
vector norm and $\|\cdot\|_{\op}$ for the matrix operator norm. For
$\mathbf A=(A_{ij})$, write
$\|\mathbf A\|_\infty=\max_i\sum_j|A_{ij}|$.
The vector $\one_m$ contains $m$ ones, $\mathbf I_m$ is the
$m\times m$ identity matrix, and $\bm e_j$ is the $j$th coordinate
vector of the appropriate dimension. The unit sphere in $\R^m$
is denoted by $\mathbb S^{m-1}$.
For a vector, $\diag$ forms a diagonal matrix; for a square matrix,
it retains only its diagonal; for a list of square matrices, it
forms a block-diagonal matrix.
We write $\Phi$ and $\phi$ for the standard normal distribution
function and density, respectively, and $\barPhi=1-\Phi$.

\section{Radially weighted fund selection}\label{sec:method}

\subsection{Model and selection target}

Let $Y_{ti}$ be the excess return of fund $i$ at date $t$. We consider
\begin{equation}\label{eq:model}
 Y_{ti}=\alpha_i+\bm\beta_i^\top\bm f_t+\varepsilon_{ti},
 \qquad t=1,\ldots,T,\quad i=1,\ldots,N,
\end{equation}
where $\bm f_t\in\R^p$ contains observable pricing factors and $p$ is fixed.
Write $\bm\alpha=(\alpha_1,\ldots,\alpha_N)^\top$ and
$\bm\varepsilon_t=(\varepsilon_{t1},\ldots,\varepsilon_{tN})^\top$.
The intercept $\alpha_i$ measures performance relative to these factors.
The selection problem is
\begin{equation}\label{eq:hypotheses}
 H_{0i}:\alpha_i\leq0
 \quad\text{against}\quad
 H_{1i}:\alpha_i>0,\qquad i=1,\ldots,N.
\end{equation}
Write $\mathcal H_0=\{i:\alpha_i\leq0\}$,
$\mathcal H_1=\{i:\alpha_i>0\}$, and $N_0=|\mathcal H_0|$.
For a selected set $\widehat{\mathcal H}$, define
\begin{equation}\label{eq:fdr}
 \FDP=\frac{|\widehat{\mathcal H}\cap\mathcal H_0|}
                 {|\widehat{\mathcal H}|\vee1},
 \qquad
 \FDR=\E(\FDP),\qquad
 \TDP=\frac{|\widehat{\mathcal H}\cap\mathcal H_1|}
                 {|\mathcal H_1|\vee1}.
\end{equation}
A selected negative-alpha fund is included in the false-discovery count.

Let $\mathbf Y=(Y_{ti})\in\R^{T\times N}$, with $i$th column
$\bm Y_i=(Y_{1i},\ldots,Y_{Ti})^\top$, and let
$\mathbf F=(\bm f_1,\ldots,\bm f_T)^\top\in\R^{T\times p}$ contain the
factor observations, without an intercept column. Define
\begin{equation}\label{eq:projection}
 \begin{gathered}
 \mathbf P_F=\mathbf F(\mathbf F^\top\mathbf F)^{-1}\mathbf F^\top,
 \qquad \mathbf M_F=\mathbf I_T-\mathbf P_F,
 \qquad \mathbf Z=\mathbf M_F\mathbf Y,\\
 \bm h=\mathbf M_F\one_T,\qquad
 \omega_T=\frac{\bm h^\top\bm h}{T},\qquad \kappa_T=1-\omega_T.
 \end{gathered}
\end{equation}
Write $\mathbf Z=(Z_{ti})$ and let $\bm Z_t$ denote the column-vector
representation of its $t$th row. We require $\omega_T>0$. Projection removes the observable factor component
and transforms the alpha contribution into $\bm h\bm\alpha^\top$.
The projected return vector at date $t$ therefore has signal $h_t\bm\alpha$.
The identities
\begin{equation}\label{eq:design-orthogonality}
 \mathbf F^\top\bm h=\bm0,\qquad
 \frac1T\sum_t h_t=\frac1T\sum_t h_t^2=\omega_T
\end{equation}
allow the estimating equation to represent this signal and remove the
average first-order effect of estimating factor loadings.

\subsection{LS-BH and SS-BH}\label{sec:benchmarks}

The least-squares alpha estimator used in LS-BH
\citep[Section~3.4]{GiglioLiaoXiu2021} is
\begin{equation}\label{eq:LSmean}
 \widehat{\bm\alpha}_{\LS}
 =\frac{\mathbf Y^\top\bm h}{\bm h^\top\bm h}
 =\frac{\mathbf Y^\top\mathbf M_F\one_T}{T\omega_T}.
\end{equation}
Let $\mathbf X=(\one_T,\mathbf F)$ and define the unrestricted
residual projection by
\[
 \mathbf M_X=\mathbf I_T-\mathbf X(\mathbf X^\top\mathbf X)^{-1}
                                      \mathbf X^\top,
 \qquad \mathbf M_{(\one_T,\mathbf F)}=\mathbf M_X.
\]
The conventional LS-BH statistic is
\begin{equation}\label{eq:LSstat}
 S_{\LS,i}=\frac{\sqrt{T\omega_T}\widehat\alpha_{\LS,i}}
                         {\widehat\sigma_i},\qquad
 \widehat\sigma_i^2=\frac1T\bm Y_i^\top\mathbf M_X\bm Y_i.
\end{equation}
It uses the one-sided p-value $p_{\LS,i}=\barPhi(S_{\LS,i})$.
The benchmark variance uses divisor $T$.
The linear estimator and squared-residual variance are sensitive to
extreme observations, so heavy-tailed returns can reduce the efficiency
of LS-BH and make its finite-sample normal calibration less reliable.

Spatial signs reduce the influence of large residual vectors and
provide a robust alternative. For $\bm x\ne\bm0$, write
$\bm U(\bm x)=\bm x/\|\bm x\|$.
The SS-BH procedure of \citet{Wang2025} estimates a location
$\widehat{\bm\theta}_{\SSm}$ from the projected observations
$\bm Z_t$. Its location and scale equations are
\begin{align}
 \frac1T\sum_{t=1}^T\widehat{\bm U}_{\SSm,t}&=\bm0,
                 \label{eq:ss-location}\\
 \frac1T\sum_{t=1}^T
   \diag(\widehat{\bm U}_{\SSm,t}\widehat{\bm U}_{\SSm,t}^{\top})
     &=N^{-1}\mathbf I_N,\label{eq:scale}
\end{align}
where
\[
 \widehat{\bm U}_{\SSm,t}
 =\bm U\{\widehat{\mathbf D}_{\SSm}^{-1/2}
                  (\bm Z_t-\widehat{\bm\theta}_{\SSm})\},\qquad
 \widehat r_{\SSm,t}
 =\|\widehat{\mathbf D}_{\SSm}^{-1/2}
                  (\bm Z_t-\widehat{\bm\theta}_{\SSm})\|.
\]
Here $\widehat{\mathbf D}_{\SSm}
=\diag(\widehat d_{\SSm,1}^{\,2},\ldots,
\widehat d_{\SSm,N}^{\,2})$ and $\prod_i\widehat d_{\SSm,i}=1$.
At iteration $m$, let $r_{\SSm,t}^{(m)}$ and
$U_{\SSm,ti}^{(m)}$ be the radius and spatial-sign coordinate computed
from the current location $\bm\theta_{\SSm}^{(m)}$ and diagonal scale
$\mathbf D_{\SSm}^{(m)}=\diag\{(d_{\SSm,i}^{(m)})^2:i\leq N\}$.
The original simultaneous updates are
\begin{align*}
 \bm\theta_{\SSm}^{(m+1)}
   &=\frac{\sum_t(r_{\SSm,t}^{(m)})^{-1}\bm Z_t}
           {\sum_t(r_{\SSm,t}^{(m)})^{-1}},
      \\
 d_{\SSm,i}^{(m+1),\mathrm{raw}}
   &=d_{\SSm,i}^{(m)}
          \left\{\frac NT\sum_t(U_{\SSm,ti}^{(m)})^2\right\}^{1/2},
          \\
 d_{\SSm,i}^{(m+1)}
   &=\frac{d_{\SSm,i}^{(m+1),\mathrm{raw}}}
          {\{\prod_jd_{\SSm,j}^{(m+1),\mathrm{raw}}\}^{1/N}}.
\end{align*}
Let $\widehat m_{j,\SSm}=T^{-1}\sum_t\widehat r_{\SSm,t}^{\,j}$.
The original studentization can be written as
\begin{align}
 \widehat Q_{\SSm}
 &=\widehat m_{-1,\SSm}^{-2}
       -2\kappa_T\frac{\widehat m_{1,\SSm}}{\widehat m_{-1,\SSm}}
       +\kappa_T\widehat m_{2,\SSm},\nonumber\\
 S_{\SSm,i}
 &=\frac{\sqrt{TN}\,\widehat\theta_{\SSm,i}}
           {\widehat d_{\SSm,i}\sqrt{\widehat Q_{\SSm}}},\qquad
 \widehat\alpha_{\SSm,i}=\omega_T^{-1}\widehat\theta_{\SSm,i}.
      \label{eq:ss-statistic}
\end{align}
The expression
\[
 \widehat Q_{\SSm}
 =\frac1T\sum_t
       (\widehat m_{-1,\SSm}^{-1}-\kappa_T\widehat r_{\SSm,t})^2
       +\kappa_T\omega_T\widehat m_{2,\SSm}
\]
provides an equivalent nonnegative calculation.
The SS-BH p-values are $p_{\SSm,i}=\barPhi(S_{\SSm,i})$.
Both benchmarks apply the rule of \citet{BenjaminiHochberg1995} at a
common nominal level $q\in(0,1)$. For any supplied p-values $p_1,\ldots,p_N$, order
them as $p_{(1)}\leq\cdots\leq p_{(N)}$ and set
\begin{equation}\label{eq:bh}
 \widehat k=\max\bigl(\{k:p_{(k)}\leq qk/N\}\cup\{0\}\bigr).
\end{equation}
Select the funds with $p_i\leq q\widehat k/N$ when $\widehat k>0$,
and select none otherwise.

\subsection{Residual information and estimation efficiency}\label{sec:motivation}

The efficiency of spatial-sign alpha estimation depends on the use of
the projected intercept, residual norms, and marginal scales. The
location equation in SS-BH assigns equal weights to residual directions,
while its scale equation depends jointly on the location fit and
cross-sectional dependence. These features motivate the weighted
construction below.

The projected signal is $h_t\bm\alpha$, whereas
\eqref{eq:ss-location} fits a common location to all dates and uses no
outer factor $h_t$ in the score. Rescaling the fitted location by
$\omega_T^{-1}$ does not recover an estimating equation that fits
$h_t\bm\alpha$ at each date. Using $h_t$ both in the signal and as
a score weight exploits $\mathbf F^\top\bm h=\bm0$ and improves
the first-order efficiency comparison. Section~\ref{sec:optimal}
quantifies the nonnegative variance loss from the original fit.

In addition, every spatial sign in \eqref{eq:ss-location} has unit
length. The original score therefore does not use differences in
residual norms to weight the directions. Such differences contain
information about the precision of a date's return vector, especially
under common heavy-tailed scale variation. A radial weight can use
this information while continuing to limit the contribution of large
residual vectors.

Marginal-scale estimation is also affected by the projected alpha signal. Write
$\widetilde{\bm\varepsilon}_{F,t}$ for the column-vector version of
row $t$ of $\mathbf M_F\mathbf E$, where
$\mathbf E=(\varepsilon_{ti})$. The residual used in the joint SS fit is
\[
 \bm Z_t-\widehat{\bm\theta}_{\SSm}
 =\widetilde{\bm\varepsilon}_{F,t}
  +(h_t-\omega_T)\bm\alpha
  -(\widehat{\bm\theta}_{\SSm}-\omega_T\bm\alpha).
\]
Thus even centering at $\omega_T\bm\alpha$ leaves the time-varying
signal $(h_t-\omega_T)\bm\alpha$ in the scale residuals. Location
estimation error also enters the squared signs. This dependence on
the alpha signal differs from a common translation, under which the
spatial-median scale fit is invariant.

Moreover, the $i$th coordinate of \eqref{eq:scale} is
\[
 \frac1T\sum_t
 \frac{(Z_{ti}-\widehat\theta_{\SSm,i})^2/
                    \widehat d_{\SSm,i}^{\,2}}
      {\sum_j(Z_{tj}-\widehat\theta_{\SSm,j})^2/
                    \widehat d_{\SSm,j}^{\,2}}=\frac1N.
\]
Its denominator involves every fund. Relating these diagonal sign
moments to marginal scales uses concentration of the joint residual
norm under the dependence conditions. Stronger cross-sectional
dependence can weaken this approximation and make the scale fit more
sensitive in finite samples. The residual decomposition above describes
additional sources of finite-sample variation in the scale estimator.

We therefore separate scale estimation from alpha estimation. A
pairwise-quantile residual (PQR) scale pilot uses unrestricted
regression residuals, which remove alpha and factor exposures exactly.
We then hold these marginal scales fixed, fit the projected signal
$h_t\bm\alpha$, and weight each spatial sign by both $h_t$ and a
function of its residual norm. These changes yield the weighted
spatial-sign method developed next.

\subsection{Pairwise-quantile scales and weighted alpha estimation}\label{sec:weighted-estimation}

Pairwise differences provide location-invariant robust scale estimates
\citep{RousseeuwCroux1993}. We estimate relative marginal scales from
the unrestricted regression residuals
\begin{equation}\label{eq:pqr-pilot}
 \widetilde{\mathbf E}=\mathbf M_X\mathbf Y.
\end{equation}
Because $\mathbf M_X\one_T=\bm0$ and $\mathbf M_X\mathbf F=\mathbf0$,
the pilot residuals equal $\mathbf M_X\mathbf E$, where
$\mathbf E=(\varepsilon_{ti})$. Thus the scale pilot removes both
alpha and observable factor exposures. Alpha estimation uses
$\mathbf Z$ in \eqref{eq:projection}, which retains the projected signal.

For the fixed probability $\tau_D=1/2$, define
\begin{align}
 \widehat H_i(x)&=\binom T2^{-1}
    \sum_{1\le s<t\le T}
       \ind\{|\widetilde E_{ti}-\widetilde E_{si}|\le x\},
       \qquad x\ge0,\nonumber\\
 \widehat q_i&=\inf\{x\ge0:\widehat H_i(x)\ge\tau_D\}.
 \label{eq:pqr-quantile}
\end{align}
The empirical quantile is the order statistic of rank
$\lceil\tau_D\binom T2\rceil$. The trace-normalized scale matrix is
\begin{equation}\label{eq:pqr-scale}
 \widehat{\mathbf D}_Q
  =\frac{\diag(\widehat q_1^2,\ldots,\widehat q_N^2)}
          {N^{-1}\sum_{j=1}^N\widehat q_j^2}
  =\diag(\widehat d_{Q,1}^{\,2},\ldots,\widehat d_{Q,N}^{\,2}).
\end{equation}
The PQR scale estimate is computed once and held fixed
throughout the weighted alpha fit. Lemma~\ref{lem:pqr-scale} in
Appendix~\ref{app:linearization-main} establishes its uniform consistency
under the model conditions in Section~\ref{sec:theory}.
Under the elliptical model, standardized marginal errors have a common
distribution, so their population pairwise quantiles differ only by
the marginal scale factors. Each marginal scale is therefore identified from its coordinatewise
residual distribution. Cross-sectional dependence enters the subsequent
simultaneous-selection analysis through the conditions in
Section~\ref{sec:theory}.

A common scale convention is needed to specify an arbitrary radial weight.
The geometrically normalized representative is
\begin{equation}\label{eq:pqr-geometric}
 \rho_Q=(\det\widehat{\mathbf D}_Q)^{1/(2N)},\qquad
 \widehat{\mathbf D}_Q^{\,g}=\rho_Q^{-2}\widehat{\mathbf D}_Q,
 \qquad
 \widehat d_{Q,i}^{\,g}
   =\frac{\widehat q_i}{(\prod_j\widehat q_j)^{1/N}}.
\end{equation}
If $w^g$ is specified under this convention, its trace-normalized
representative is $w^Q(r)=w^g(\rho_Qr)$. The resulting alpha estimate and
studentized statistic agree exactly. For inverse-norm weighting,
changing between these conventions only rescales the weight by an
immaterial common factor. The implementation uses \eqref{eq:pqr-scale}.

Given a nonnegative radial weight $w$ under the chosen convention,
define $\widehat{\bm\alpha}_w$ by
\begin{equation}\label{eq:location}
 \frac1T\sum_{t=1}^T h_t w(\widehat r_{w,t})
       \widehat{\bm U}_{w,t}=\bm0,
\end{equation}
where $\widehat{\bm e}_{w,t}=
\widehat{\mathbf D}_Q^{-1/2}(\bm Z_t-h_t\widehat{\bm\alpha}_w)$,
$\widehat r_{w,t}=\|\widehat{\bm e}_{w,t}\|$, and
$\widehat{\bm U}_{w,t}=\bm U(\widehat{\bm e}_{w,t})$.
The scale is common across radial weights, while the fitted
alpha vectors and their associated radii depend on $w$. The two appearances
of $h_t$ serve different roles: $h_t\widehat{\bm\alpha}_w$ represents the
projected signal, and the outer factor $h_t$ uses the orthogonality in
\eqref{eq:design-orthogonality}. The entries of $\bm h$ may have either
sign; the radial weight $w$ is nonnegative.
For $w(r)\equiv1$, this is a design-weighted spatial-sign equation,
distinct from the original SS-BH fit in \eqref{eq:ss-location}.
For $w(r)=r$, its solution is exactly the OLS intercept in
\eqref{eq:LSmean}. The choice $w(r)=r^{-1}$ additionally downweights
directions associated with large norms and defines the theoretical
IN-BH benchmark. Section~\ref{sec:smoothing} specifies the smoothed
implementation used in the numerical study.

\subsection{Studentization and selection}

Let $\mathbf D=\diag(d_1^2,\ldots,d_N^2)$, with $d_i>0$ and
$\prod_i d_i=1$, denote the population diagonal shape matrix in the
elliptical model of Section~\ref{sec:theory}.
The population radial quantities are
\begin{equation}\label{eq:AB}
 A_w=\E\{w(r)/r\},\qquad B_w=\E\{w(r)^2\},\qquad
 J(w)=\frac{B_w}{A_w^2},
\end{equation}
where $\bm\varepsilon$ has the common error distribution and
$r=\|\mathbf D^{-1/2}\bm\varepsilon\|$. These quantities use the
geometric scale representative and its weight $w^g$; the superscript is
suppressed in \eqref{eq:AB}. Expectations involving a radial weight
integrate over the error distribution, holding the factor design and
that weight fixed. Section~\ref{sec:theory} derives the
high-dimensional approximation under which the first-order variance of
$\widehat\alpha_{w,i}$ is $d_i^2J(w)/(NT\omega_T)$.
The factor design enters through $\omega_T$, while $J(w)$ determines
the comparison of radial weights.

The fitted studentizer retains the estimated projection term in the
conditional linearization. At the solution, define the $T$-vectors
$\widehat{\bm a}_w$ and $\widehat{\bm\pi}_w$ by
\begin{equation}\label{eq:projected-score-coefficients}
 \widehat a_{w,t}=h_t\frac{w(\widehat r_{w,t})}{\widehat r_{w,t}},
 \qquad \widehat{\bm\pi}_w=\mathbf P_F\widehat{\bm a}_w,
 \qquad
 \widehat c_{w,t}=h_tw(\widehat r_{w,t})
                    -\widehat r_{w,t}\widehat\pi_{w,t}.
\end{equation}
Set
\begin{align*}
 \widehat A_{h,w}&=\frac1T\sum_{t=1}^T
             h_t^2\frac{w(\widehat r_{w,t})}{\widehat r_{w,t}},\\
 \widehat B_{c,w}&=\frac1T\sum_{t=1}^T\widehat c_{w,t}^{\,2}.
\end{align*}
The coefficients in \eqref{eq:projected-score-coefficients} account for
the use of estimated factor loadings in the residuals. Their population
first-order variance remains $\omega_T B_w$ because
$\mathbf F^\top\bm h=\bm0$. The sample calculation retains the
corresponding fluctuation in the weighted factor average.
The standardized alpha statistic is
\begin{equation}\label{eq:statistic}
 S_{w,i}
   =\frac{\sqrt{TN}\widehat A_{h,w}\widehat\alpha_{w,i}}
               {\widehat d_{Q,i}\sqrt{\widehat B_{c,w}}}.
\end{equation}
We apply a residual degrees-of-freedom calibration to its fitted variance.
For the unrestricted design $\mathbf X=(\one_T,\mathbf F)$, let
\begin{equation*}
 n_T=T-\rank(\mathbf X)=T-p-1>0,\qquad
 c_{\mathrm{df},T}=\frac{n_T}{T}.
\end{equation*}
The estimated alpha variance and its calibrated version are
\begin{equation}\label{eq:df-variance}
 \widehat{\operatorname{se}}_{w,i}^{\,2}
  =\frac{\widehat d_{Q,i}^{\,2}\widehat B_{c,w}}
           {NT\widehat A_{h,w}^{\,2}},\qquad
 \widehat{\operatorname{se}}_{w,i}^{\,2,\mathrm{df}}
  =c_{\mathrm{df},T}^{-1}
       \widehat{\operatorname{se}}_{w,i}^{\,2}.
\end{equation}
The statistic used for selection and its upper-tail probability are
\begin{equation}\label{eq:statistic-df}
 S_{w,i}^{\mathrm{df}}
   =\sqrt{c_{\mathrm{df},T}}\,S_{w,i},\qquad
 p_{w,i}=\barPhi(S_{w,i}^{\mathrm{df}}),
\end{equation}
The multiplier uses the residual degrees of
freedom of the unrestricted regression underlying the PQR pilot.
Appendix~\ref{app:df-calibration} gives the corresponding residual
second-moment identity, and Theorem~\ref{thm:main-statistic} establishes
the calibrated statistic's centered Gaussian representation. Multiplication
of $w$ by a positive constant, or the normalization transformation after
\eqref{eq:pqr-geometric}, leaves both statistics unchanged.

The construction also preserves shifts in the alpha vector. If
$\mathbf Y^{a}=\mathbf Y+\one_T\bm a^\top$, then the unrestricted
pilot and PQR scales are unchanged, while
$\mathbf Z^{a}=\mathbf Z+\bm h\bm a^\top$. The OLS initializer and
the weighted updates consequently satisfy, on a successful fit,
\begin{equation}\label{eq:alpha-equivariance}
 \widehat{\bm\alpha}_w^{\,a}
       =\widehat{\bm\alpha}_w+\bm a,\qquad
 \widehat{\operatorname{se}}_{w,i}^{\,a}
       =\widehat{\operatorname{se}}_{w,i}.
\end{equation}
The fitted residuals, stopping criteria, and success event are invariant
under this transformation. This property permits inference on the
centered estimation error throughout the alpha parameter space.

Apply the BH rule in \eqref{eq:bh} to $p_{w,1},\ldots,p_{w,N}$,
and denote the resulting rejection index by $\widehat k_w$. This procedure
applies BH to p-values obtained from radially weighted alpha estimates.
The nominal
level $q$ has the false-discovery interpretation in \eqref{eq:fdr}.
For $w(r)=r$, \eqref{eq:location} gives the OLS alpha estimate in
\eqref{eq:LSmean}.

For inverse-norm weighting, $w_{\IN}(r)=r^{-1}$, let
$\widehat C_h=T^{-1}\sum_t h_t^2/\widehat r_{\IN,t}^{\,2}$.
Since $\bm h^\top\widehat{\bm\pi}_{\IN}=0$,
\begin{align}
 \widehat A_{h,\IN}&=\widehat C_h,\qquad
 \widehat B_{c,\IN}
   =\widehat C_h+\frac1T\sum_t
          \widehat r_{\IN,t}^{\,2}\widehat\pi_{\IN,t}^{\,2},
          \label{eq:in-variance}\\
 S_{\IN,i}&=\frac{\sqrt{NT}\,\widehat C_h\widehat\alpha_{\IN,i}}
                  {\widehat d_{Q,i}\sqrt{\widehat B_{c,\IN}}}.
                  \label{eq:in-statistic}
\end{align}
Its calibrated statistic is
\begin{equation}\label{eq:in-statistic-df}
 S_{\IN,i}^{\mathrm{df}}
   =\frac{\sqrt{Nn_T}\,\widehat C_h\widehat\alpha_{\IN,i}}
                  {\widehat d_{Q,i}\sqrt{\widehat B_{c,\IN}}},
 \qquad p_{\IN,i}=\barPhi(S_{\IN,i}^{\mathrm{df}}).
\end{equation}
Thus the fitted variance is positive whenever $\widehat C_h>0$.
The second term in \eqref{eq:in-variance} has smaller asymptotic order
and leaves the first-order radial efficiency criterion unchanged.
BH applied to these calibrated p-values defines the unregularized
IN-BH procedure. Its smoothed implementation is specified below. Section~\ref{sec:optimal}
establishes inverse-norm optimality within the class in
\eqref{eq:location} and compares the procedure with the original
spatial-sign and least-squares benchmarks.

\subsection{Implemention}\label{sec:smoothing}

Compute the PQR scale matrix once from the unrestricted residuals and
hold it fixed. Starting from the OLS alpha
$\bm\alpha_w^{(0)}=\mathbf Y^\top\bm h/(\bm h^\top\bm h)$,
form the residual radii
\[
 r_{w,t}^{(m)}
 =\|\widehat{\mathbf D}_Q^{-1/2}
                 (\bm Z_t-h_t\bm\alpha_w^{(m)})\|
\]
and update
\begin{equation}\label{eq:updates}
 \bm\alpha_w^{(m+1)}
   =\frac{\sum_t h_t\{w(r_{w,t}^{(m)})/r_{w,t}^{(m)}\}\bm Z_t}
           {\sum_t h_t^2w(r_{w,t}^{(m)})/r_{w,t}^{(m)}}.
\end{equation}
For inverse-norm weighting, the ratio $w(r)/r$ is $r^{-2}$.

For the numerical implementation, let $\widetilde{\bm e}_t$ be the
transpose of row $t$ of the unrestricted residual matrix in
\eqref{eq:pqr-pilot}, and set
\begin{equation}\label{eq:smooth-pilot}
 r_t^{\mathrm{pil}}
 =\|\widehat{\mathbf D}_Q^{-1/2}\widetilde{\bm e}_t\|,
 \qquad
 \lambda=c\,\operatorname{med}_{t\leq T}r_t^{\mathrm{pil}},
 \qquad c=0.5.
\end{equation}
The PQR scale and $\lambda$ remain fixed during alpha estimation.
The smoothed weight and estimating equation are
\begin{equation}\label{eq:smooth-location}
 w_\lambda(r)=\frac{r}{r^2+\lambda^2},\qquad
 \frac1T\sum_{t=1}^T
 \frac{h_t\widehat{\bm e}_{\INlam,t}}
      {\widehat r_{\INlam,t}^{\,2}+\lambda^2}=\bm0,
\end{equation}
where $\widehat{\bm e}_{\INlam,t}
=\widehat{\mathbf D}_Q^{-1/2}
(\bm Z_t-h_t\widehat{\bm\alpha}_{\INlam})$ and
$\widehat r_{\INlam,t}=\|\widehat{\bm e}_{\INlam,t}\|$.
For $\lambda>0$, the score is continuous at a zero residual.
Starting from the OLS alpha, update
\begin{equation}\label{eq:smooth-update}
 \bm\alpha^{(m+1)}=
 \frac{\sum_t h_t\bm Z_t/\{(r_t^{(m)})^2+\lambda^2\}}
      {\sum_t h_t^2/\{(r_t^{(m)})^2+\lambda^2\}},
\end{equation}
where $r_t^{(m)}$ is the residual radius at $\bm\alpha^{(m)}$.
The general projection-adjusted studentizer in
\eqref{eq:projected-score-coefficients}--\eqref{eq:statistic-df} becomes
\begin{equation}\label{eq:smooth-studentizer}
 \begin{gathered}
 q_t=\widehat r_{\INlam,t}^{\,2}+\lambda^2,\qquad
 \widehat A_{h,\INlam}=\frac1T\sum_t\frac{h_t^2}{q_t},\qquad
 \widehat{\bm\pi}_{\INlam}=\mathbf P_F(h_t/q_t)_{t=1}^T,\\
 \widehat B_{c,\INlam}=\frac1T\sum_t
 \widehat r_{\INlam,t}^{\,2}
       \left(\frac{h_t}{q_t}-\widehat\pi_{\INlam,t}\right)^2,\qquad
 S_{\INlam,i}^{\mathrm{df}}=
 \frac{\sqrt{Nn_T}\,\widehat A_{h,\INlam}\widehat\alpha_{\INlam,i}}
      {\widehat d_{Q,i}\sqrt{\widehat B_{c,\INlam}}}.
 \end{gathered}
\end{equation}
BH applied to $\barPhi(S_{\INlam,i}^{\mathrm{df}})$ defines the smoothed
IN-BH procedure.
The theory below covers radial profiles fixed conditional on the factor
design; the pilot-dependent choice of $\lambda$ requires a separate analysis.

\section{Theoretical properties}\label{sec:theory}

We study the local solution of the weighted alpha equation for radial
profiles fixed conditional on the factor design. Throughout the
asymptotic analysis, $N,T\to\infty$. The analysis accounts
for estimating factor loadings, PQR scales, alpha, and variance from the
same observations. A uniform centered Gaussian representation yields
false discovery control and marginal local power. Detailed proofs appear
in Appendices~\ref{app:linearization-main}--\ref{app:efficiency}.

\subsection{Model conditions}

In model~\eqref{eq:model}, the errors have the elliptical representation
\begin{equation}\label{eq:radial-model}
 \bm\varepsilon_t=\sqrt N\tau_t\mathbf D^{1/2}
                          \mathbf R_N^{1/2}\bm U_t^\circ,
 \qquad \bm U_t^\circ\sim\operatorname{Unif}(\mathbb S^{N-1}),
\end{equation}
where $\tau_t>0$ is independent of $\bm U_t^\circ$, and the pairs are
independent and identically distributed over time. The entire factor
process is independent of the error sequence. The radial law may depend
on $N$; its index is suppressed in sample quantities. Using the geometric
normalization of $\mathbf D$, put
$\bm v_t=\sqrt N\mathbf R_N^{1/2}\bm U_t^\circ$.
The scatter matrix is
$\mathbf\Sigma_N=\mathbf D^{1/2}\mathbf R_N\mathbf D^{1/2}$, and
$\Cov(\bm\varepsilon_t)=\E(\tau_t^2)\mathbf\Sigma_N$.
The correlation matrix $\mathbf R_N$ is positive semidefinite, with
$\diag(\mathbf R_N)=\mathbf I_N$.
Write $\bm\tau=(\tau_1,\ldots,\tau_T)^\top$,
$L=\log(NT)$, $\eta_{NT}=\sqrt{L/T}$,
$\mathbf Q_T=T^{-1}\mathbf F^\top\mathbf F$,
$\bar{\bm f}=T^{-1}\sum_t\bm f_t$, and
$b_N=\|\mathbf R_N\|_{\op}$.
For constants $0<c<C<\infty$, define the design event
\[
 \mathcal F_{NT}=\left\{
 \rank(\mathbf F)=p,\quad \omega_T\geq c,\quad
 \max_{t\leq T}(\mathbf P_F)_{tt}\leq CL/T
 \right\}.
\]

\begin{assumption}[Elliptical errors and factor design]
\label{ass:radial-primitive}
The representation \eqref{eq:radial-model}, with its stated independence
conditions, holds for an arbitrary deterministic alpha sequence.
The factor dimension $p$ is fixed and
$\Pp(\mathcal F_{NT})\to1$. The marginal scales satisfy $d_i>0$ and
$\prod_i d_i=1$. The families $\{\tau_N^2\}$ and
$\{\tau_N^{-4}\}$ are uniformly integrable:
\begin{equation}\label{eq:radial-uniform-integrability}
 \lim_{K\to\infty}\sup_N
 \left[
 \E\{\tau_N^2\ind(\tau_N^2>K)\}
 +\E\{\tau_N^{-4}\ind(\tau_N^{-4}>K)\}
 \right]=0.
\end{equation}
\end{assumption}

The fixed factor dimension, factor--error independence, and
nondegenerate projected intercept are also assumed in
\citet[Section~2]{Wang2025}, whose error model contains the elliptical
representation in \eqref{eq:radial-model}.
The leverage bound limits the influence of a single observation on the
factor projection, while $\omega_T\geq c$ preserves information about
alpha after projecting out the factors. These conditions apply to
deterministic designs and are invariant to nonsingular changes of factor
coordinates. For a radial law that does not vary with $N$,
\eqref{eq:radial-uniform-integrability} reduces to
$\E\tau^2<\infty$ and $\E\tau^{-4}<\infty$.
Inverse moments of normalized observation radii through order four
are likewise imposed in \citet[Section~2]{Wang2025} and
\citet[Condition~C2]{Cheng2026}. The uniform-integrability requirement
in \eqref{eq:radial-uniform-integrability} provides tail control when
the radial law varies with $N$.
The positive moment controls factor projection and radial variances;
the inverse moment controls sensitivity and small residual radii.
The PQR pilot removes alpha exactly, and
\eqref{eq:alpha-equivariance} permits fixed or increasing nonzero alphas.
Appendix~\ref{app:elliptical-examples} verifies these conditions for
standardized multivariate $t$ distributions with any fixed degrees of
freedom greater than two and for other elliptical distributions.

Specify the radial weight by a profile $k$ on the geometric scale:
\begin{equation}\label{eq:normalized-profile}
 w_N^{\,g}(r)=k(r/\sqrt N),\qquad
 w_N^{\,Q}(r)=w_N^{\,g}(\rho_Qr),\qquad g(s)=k(s)/s,
\end{equation}
where $\rho_Q$ is defined in \eqref{eq:pqr-geometric}.
The profile $k$ may depend on $N$ and the factor design. All expectations
involving a radial weight or profile integrate over the error distribution
conditional on $\mathbf F$, holding that weight or profile fixed.
Write $a=\E g(\tau)$ under this convention.

\begin{assumption}[Weight regularity and growth]
\label{ass:weight-primitive}
The function $g\geq0$ is twice continuously differentiable on
$(0,\infty)$, $a\geq c>0$, and, for $j=1,2$,
\[
 |g(s)|\leq C(1+s^{-2}),\qquad
 |s^jg^{(j)}(s)|\leq C(s^{-1}+s^{-2}).
\]
The constants are uniform and
\begin{equation}\label{eq:strong-growth}
 \frac{L^{11}}T+\frac{b_NL^9}{N}\longrightarrow0.
\end{equation}
\end{assumption}

Conditions on radial weights, their derivatives, and the associated
moments are discussed in \citet[Remark~1]{Yan2025}.
The bounds in Assumption~\ref{ass:weight-primitive} control changes
in the weighted score induced by estimating alpha, factor loadings,
and marginal scales. Related logarithmic growth restrictions appear
in \citet[Theorem~1]{Cheng2026} and \citet[Section~2]{Wang2025}.
The powers in \eqref{eq:strong-growth} provide sufficient rates for
the uniform estimation and studentization remainders in the present
factor model.
When $b_N$ is bounded, these conditions allow
$N=\lceil T^b\rceil$ for any fixed $b>0$, including proportional
cross-sectional and time dimensions. The constant, linear, and
inverse-norm weights have normalized profiles $k_0(s)=1$,
$k_{\mathrm{lin}}(s)=s$, and $k_{\IN}(s)=s^{-1}$, respectively.
Their geometric weights are proportional to $1$, $r$, and $r^{-1}$;
the common proportionality constants cancel from the estimating equation
and studentized statistic. The sensitivity involves $h_t^2$, although
the individual $h_t$ may have either sign.

For a predetermined $0\leq\ell\leq M<\infty$, the smoothed profile
$k_\ell(s)=s/(s^2+\ell^2)$ has
$g_\ell(s)=(s^2+\ell^2)^{-1}$. It satisfies the derivative bounds in
Assumption~\ref{ass:weight-primitive}, and
\[
 \E g_\ell(\tau)
 \geq\{\E\tau^2+\ell^2\}^{-1}
 \geq(C_2+M^2)^{-1}>0,
 \qquad C_2=\sup_N\E\tau_N^2<\infty.
\]

\subsection{Distributional approximation and local power}

With $a=\E\{k(\tau)/\tau\}$ as defined above, let
\begin{align}
 J_{\tau,w}&=\frac{\E k(\tau)^2}{a^2},\qquad
 \mathcal V_{w,T}=\frac{J_{\tau,w}}{\omega_T},
       \label{eq:population-V-profile}\\
 \nu_{w,i}&=\frac{\sqrt T\alpha_i}{d_i\sqrt{\mathcal V_{w,T}}}
       =\frac{\sqrt{T\omega_T}\alpha_i}{d_i\sqrt{J_{\tau,w}}}.
       \label{eq:main-noncentrality}
\end{align}
Here $\mathcal V_{w,T}$ is the variance factor for the alpha estimator.
The identity $\mathbf F^\top\bm h=\bm0$ removes the first-order
loading-estimation term from the weighted score. Thus the factor design
enters through $\omega_T$, while the radial efficiency criterion is
$J_{\tau,w}$. For the true geometric radius,
\[
 \frac{\E w(r)^2}{N[\E\{w(r)/r\}]^2}
   =J_{\tau,w}+O\{(b_N/N)^{1/2}\}.
\]

Put
\begin{equation}\label{eq:explicit-rates}
 \begin{gathered}
 e_{NT}=\frac{L^5}{T}+L^4\sqrt{\frac{b_N}{NT}},\qquad
 d_{NT}=\frac L{\sqrt T}+\sqrt{\frac{b_NL}{N}},\qquad
 r_{NT}=\sqrt T\,e_{NT}+\sqrt L\,d_{NT}.
 \end{gathered}
\end{equation}
Lemma~\ref{lem:local-root} establishes a local solution of
\eqref{eq:location} with standardized estimation error
$O_p(\eta_{NT})$. The following results concern the equivariant
selection of this solution specified in its proof and the resulting
fitted standard errors. Let the standard errors denote the positive
square roots in \eqref{eq:df-variance}, and define
\begin{equation}\label{eq:centered-pivots}
 Z_{w,i}^{\circ}
   =\frac{\widehat\alpha_{w,i}-\alpha_i}
          {\widehat{\operatorname{se}}_{w,i}},\qquad
 Z_{w,i}^{\circ,\mathrm{df}}
   =\frac{\widehat\alpha_{w,i}-\alpha_i}
          {\widehat{\operatorname{se}}_{w,i}^{\,\mathrm{df}}}.
\end{equation}
If the local solution or its studentizer is undefined, assign zero to
the centered pivots and one to all p-values.

\begin{theorem}[Uniform centered Gaussian representation]\label{thm:main-statistic}
Under Assumptions~\ref{ass:radial-primitive}--\ref{ass:weight-primitive},
on an extension of the probability space there is a vector $\bm G_w$ with
$\bm G_w\mid(\bm\tau,\mathbf F)\sim N(\bm0,\mathbf R_N)$ such that
\begin{equation}\label{eq:main-gaussian-representation}
 \max_{i\leq N}
 \left\{|Z_{w,i}^{\circ}-G_{w,i}|,
        |Z_{w,i}^{\circ,\mathrm{df}}-G_{w,i}|\right\}
       =O_p(r_{NT}),\qquad
 r_{NT}\sqrt{\log N}\longrightarrow0.
\end{equation}
The local solution has positive fitted radii and standard errors with
probability tending to one, and
\begin{equation}\label{eq:standard-error-consistency}
 \max_{i\leq N}
 \left|
  \frac{\sqrt T\,\widehat{\operatorname{se}}_{w,i}^{\,\mathrm{df}}}
       {d_i\sqrt{\mathcal V_{w,T}}}-1
 \right|=o_p(1).
\end{equation}
These statements hold for arbitrary deterministic alpha sequences.
\end{theorem}

Equation~\eqref{eq:alpha-equivariance} makes the centered error and its
studentizer invariant to alpha. For the local solution,
\begin{equation}\label{eq:random-drift-representation}
 S_{w,i}^{\mathrm{df}}
  =G_{w,i}+\widehat\nu_{w,i}+\xi_{w,i},\qquad
 \widehat\nu_{w,i}
  =\frac{\alpha_i}{\widehat{\operatorname{se}}_{w,i}^{\,\mathrm{df}}},
 \qquad
 \max_i|\xi_{w,i}|=O_p(r_{NT}).
\end{equation}
The random drift retains the fitted standard error. For a nonpositive
alpha it is nonpositive, which supplies the null comparison for FDR
control. The degrees-of-freedom multiplier changes the centered pivot by
$O_p\{(p+1)\sqrt L/T\}$ and preserves the first-order variance factor.

\begin{corollary}[Marginal local power]\label{cor:radial-local-power}
Under the conditions of Theorem~\ref{thm:main-statistic}, suppose
$\mathcal V_{w,T}\to\mathcal V_w\in(0,\infty)$ in probability and
$\sqrt T\alpha_i/d_i\to b_i^\circ$ for a coordinate sequence. Then
\[
 S_{w,i}\Rightarrow N(b_i^\circ/\sqrt{\mathcal V_w},1),\qquad
 S_{w,i}^{\mathrm{df}}\Rightarrow
                      N(b_i^\circ/\sqrt{\mathcal V_w},1).
\]
\end{corollary}

At a common marginal cutoff $c$, the limiting power is
$\barPhi(c-b_i^\circ/\sqrt{\mathcal V_w})$. For a positive local alpha,
minimizing $J_{\tau,w}$ therefore maximizes this power within the
$h_t$-weighted class. Section~\ref{sec:optimal} derives the inverse-norm
optimum. The Gaussian representation is proved from the elliptical
directions and the estimating equations, including PQR estimation and
studentization from the same sample.

\subsection{False discovery control}
\label{subsec:weighted-fdr}

For the hypotheses in \eqref{eq:hypotheses}, the uniform representation
in Theorem~\ref{thm:main-statistic} provides the approximation needed for
simultaneous selection. The remaining conditions concern cross-sectional
dependence and the number and magnitude of positive population alphas.
Fix $q\in(0,1)$ and let
\begin{equation}\label{eq:fdr-floor}
 z(u)=\Phi^{-1}(1-u),\qquad
 u_{0,N}=\frac{qk_N}{N},\qquad z_{0,N}=z(u_{0,N}),
\end{equation}
where $1\leq k_N\leq N/2$ is an integer. Write
$s_N=\max_i|\{j\ne i:R_{N,ij}\ne0\}|$ for the largest number of
nonzero off-diagonal entries in a row of $\mathbf R_N$.
For a weighted procedure covered by Theorem~\ref{thm:main-statistic},
let $\mathcal V_{w,T}$ be its alpha variance factor. For deterministic
sets $\mathcal S_N\subseteq\mathcal H_1$ with $|\mathcal S_N|=2k_N$,
the population signal requirement is
\begin{equation}\label{eq:fdr-signal-barrier}
 \min_{i\in\mathcal S_N}\frac{\alpha_i}{d_i}
 \geq\sqrt{\frac{\mathcal V_{w,T}}T}
                 2\sqrt{\log N}.
\end{equation}

\begin{assumption}[Cross-sectional dependence and detectable alphas]
\label{ass:fdr-signals}
The sequence $k_N$ tends to infinity, and the correlation matrices
satisfy either
\[
 \frac{b_N}{k_N}\longrightarrow0,
\]
or, for a fixed $0<\varrho<1$,
\[
 \max_{i<j}|R_{N,ij}|\leq\varrho,\qquad
 \frac{s_N}{N^{1-\beta_\varrho}k_N^{\beta_\varrho}}
       \longrightarrow0,\qquad
 \beta_\varrho=\frac{2\varrho}{1+\varrho}.
\]
There are deterministic sets $\mathcal S_N$ as above for which
\eqref{eq:fdr-signal-barrier} holds with probability tending to one.
\end{assumption}

For the proposed $h_t$-weighted procedure,
$\mathcal V_{w,T}=J_{\tau,w}/\omega_T$, so that
\eqref{eq:fdr-signal-barrier} is equivalent to
\[
 \min_{i\in\mathcal S_N}\frac{\alpha_i}{d_i}
 \geq2\sqrt{\frac{J_{\tau,w}\log N}{T\omega_T}}.
\]
Spectral restrictions ensuring weak Gaussian residual dependence
are used by \citet[Section~3.1]{FanHanGu2012}. Restrictions on the
number of correlated coordinates and on their absolute correlations
appear in \citet[Section~4]{LiuShao2014} and
\citet[Section~2]{Wang2025}.
The spectral condition relates the strength of residual dependence to
the number of detectable funds and permits a diverging spectral norm.
A bounded spectral norm, including that of a fixed-parameter AR(1)
correlation matrix, suffices for any $k_N\to\infty$. The sparse
alternative allows the number of correlated pairs to increase with
$k_N$. Both alternatives permit correlations of either sign.
No positive lower bound is imposed on the proportion of true nulls.

The population noncentralities in \eqref{eq:main-noncentrality} satisfy
$\min_{i\in\mathcal S_N}\nu_{w,i}\geq2\sqrt{\log N}$, following the
detectability condition in \citet[Section~2]{Wang2025}. Their signal lower bound
has order $\sqrt{\log N/T}$; fixed nonzero and larger positive alphas
are also admissible. The standardized margin ensures a BH threshold
of at least $qk_N/N$ without a separate restriction on dependence
within the signal set.

\begin{theorem}[False discovery control]\label{thm:fdr-main}
Under Assumptions~\ref{ass:radial-primitive}--\ref{ass:fdr-signals},
the BH rule in \eqref{eq:bh}, applied to the p-values in
\eqref{eq:statistic-df}, satisfies, for every fixed $\epsilon>0$,
\begin{equation}\label{eq:fdr-asymptotic}
 \FDR\leq q\frac{N_0}{N}+o(1),\qquad
 \Pp\!\left\{\FDP>q\frac{N_0}{N}+\epsilon\right\}\longrightarrow0.
\end{equation}
\end{theorem}

Negative-alpha funds belong to the null set and are included in the
false-discovery count, with no restriction on the magnitude of their
alphas. The theorem concerns the local weighted estimator and its
positive studentizer. On the complementary event where either is
undefined, the selection is empty.

For a nonempty deterministic set
$\mathcal D_N\subseteq\mathcal H_1$, write
$\TDP(\mathcal D_N)=
|\widehat{\mathcal H}\cap\mathcal D_N|/|\mathcal D_N|$.
Recovery under a population signal condition also requires the fitted
standard error to approximate its population counterpart sufficiently
accurately near the selection threshold. Write $\Psi(K)$ for the
uniform radial tail expectation in \eqref{eq:radial-tail-envelope}.

\begin{corollary}[Recovery of sufficiently strong funds]\label{cor:strong-fund-recovery}
Under Assumptions~\ref{ass:radial-primitive}--\ref{ass:fdr-signals},
suppose that there is a deterministic sequence $K_{NT}$ such that
\begin{equation}\label{eq:recovery-tail-rate}
 K_{NT}\longrightarrow\infty,\qquad
 \frac{K_{NT}L^2}{T}\longrightarrow0,\qquad
 \sqrt L\,\Psi(K_{NT})\longrightarrow0.
\end{equation}
If
\begin{equation}\label{eq:fdr-target-barrier}
 \min_{i\in\mathcal D_N}\nu_{w,i}-z_{0,N}
                  \xrightarrow{p}\infty,
\end{equation}
then $\E\{1-\TDP(\mathcal D_N)\}\to0$ and
$\TDP(\mathcal D_N)\to1$ in probability.
\end{corollary}

Taking $\mathcal D_N=\mathcal H_1$ yields power consistency when
all positive alphas satisfy \eqref{eq:fdr-target-barrier}.
Appendix~\ref{app:weighted-fdr-proofs} gives the quantitative
probability bounds and the complete proofs.

\section{Optimal radial weighting and efficiency}\label{sec:optimal}

The first-order variance separates the contribution of radial weighting
from that of the factor design. This separation yields an optimal
weight and permits direct comparisons with the benchmarks in
Section~\ref{sec:benchmarks}.
For $r=\|\mathbf D^{-1/2}\bm\varepsilon\|$, write
$m_j=\E(r^j)$ whenever the moment exists. The radial efficiency criterion is
\begin{equation*}
 J(w)=\frac{\E\{w(r)^2\}}{A_w^2},\qquad
 A_w=\E\{w(r)/r\}.
\end{equation*}
Under the conditions of Theorem~\ref{thm:main-statistic},
$J(w)/N$ is asymptotically equivalent to $J_{\tau,w}$, and the
first-order variance of $\widehat\alpha_{w,i}$ is
$d_i^2J_{\tau,w}/(T\omega_T)$. The same factor-design multiplier
$\omega_T^{-1}$ applies to every admissible radial weight in the
$h_t$-weighted estimating equation. The calibration factor
$c_{\mathrm{df},T}$ is also common to these weights and converges to
one, so the first-order efficiency criterion remains $J_{\tau,w}$.

\subsection{Inverse-norm weighting and the optimum}

For the unregularized benchmark $w_{\IN}(r)=r^{-1}$, the estimating equation becomes
\begin{equation*}
 \sum_{t=1}^T
 \frac{h_t(\bm Z_t-h_t\widehat{\bm\alpha}_{\IN})}
      {\widehat r_{\IN,t}^{\,2}}=\bm0,\qquad
 \widehat r_{\IN,t}
 =\|\widehat{\mathbf D}_Q^{-1/2}
               (\bm Z_t-h_t\widehat{\bm\alpha}_{\IN})\|.
\end{equation*}
The scale $\widehat{\mathbf D}_Q$ remains fixed during the location fit.
For this weight the empirical sensitivity and projection-adjusted
variance terms satisfy
\[
 \widehat A_{h,\IN}=\widehat C_h,\qquad
 \widehat B_{c,\IN}
   =\widehat C_h+\frac1T\sum_t
          \widehat r_{\IN,t}^{\,2}\widehat\pi_{\IN,t}^{\,2}
   \geq\widehat C_h.
\]
Here $\widehat{\bm\pi}_{\IN}$ is the fitted projection defined in
\eqref{eq:projected-score-coefficients}. The identity follows from
$\mathbf F^\top\bm h=\bm0$.
The statistic in \eqref{eq:in-statistic-df} retains this variance term
and applies the residual degrees-of-freedom calibration. The resulting
selection procedure is IN-BH.

Assume $0<m_{-2}<\infty$ and consider measurable nonnegative weights
with $0<A_w<\infty$ and $\E w(r)^2<\infty$. Define
\begin{equation}\label{eq:wstar}
 w_\star(r)=\frac1{m_{-2}r},\qquad
 J_\star=\frac1{m_{-2}}.
\end{equation}
The normalization $A_{w_\star}=1$ is immaterial to the estimating
equation and its studentized statistic.

\Needspace{3\baselineskip}
\begin{theorem}[Optimal radial weight]\label{thm:optimal-radial-weight}
For every admissible weight,
\begin{equation*}
 J(w)-J_\star
 =\E\left[\left\{
       \frac{w(r)}{A_w}-\frac1{m_{-2}r}
       \right\}^{2}\right]\geq0.
\end{equation*}
Equality holds if and only if $w(r)$ is proportional to $r^{-1}$
almost surely under the radial distribution.
\end{theorem}

The exact identity quantifies the loss from any other normalized
radial weight. In the normalized radial representation, put
$\zeta=\E\tau^{-2}$. The same identity gives
\begin{equation}\label{eq:profile-efficiency-gap}
 J_{\tau,w}-\zeta^{-1}
 =\E\left[\left\{
      \frac{k(\tau)}{\E\{k(\tau)/\tau\}}
           -\frac1{\zeta\tau}\right\}^{2}\right].
\end{equation}
Inverse-norm weighting therefore minimizes the first-order alpha
variance and maximizes marginal local power at a common cutoff among
weights satisfying the conditions of
Corollary~\ref{cor:radial-local-power}. This conclusion agrees with the
inverse-norm principle for factor-pricing score tests in
\citet{Zhao2022}; the present construction supplies individual alpha
statistics and their simultaneous-selection theory. The comparison concerns marginal local power at a common cutoff;
BH rejection sets also depend on the joint collection of p-values.

The unregularized estimating equation can use $r^{-1}$ directly: the population moment
$m_{-2}$ in \eqref{eq:wstar} is a normalization constant and cancels
from both the equation and the statistic. The sample inverse moments
in \eqref{eq:in-statistic} estimate its variance, rather than an
additional tuning parameter.

\subsection{Efficiency relative to the benchmarks}

Write $a_{\SSm}=\E\tau^{-1}$, $\zeta=\E\tau^{-2}$, and
$\mu_2=\E\tau^2$. For the efficiency comparison, we use the
least-squares and spatial-sign limits in \citet[Section~2]{Wang2025},
under the corresponding regularity conditions. Together with
Theorem~\ref{thm:main-statistic}, these give the leading alpha
variance factors
\begin{align}
 \mathcal V_{\IN,T}&=\frac1{\omega_T\zeta},&
 \mathcal V_{\LS,T}&=\frac{\mu_2}{\omega_T},
       \label{eq:benchmark-variances}\\
 \mathcal V_{\SSm,T}
   &=\frac{a_{\SSm}^{-2}
         -2\kappa_T\E\tau/a_{\SSm}+\kappa_T\mu_2}{\omega_T^2}.
       \nonumber
\end{align}
The constant radial profile in the new $h_t$-weighted equation has
variance factor $a_{\SSm}^{-2}/\omega_T$. Relative to this constant-profile estimator, the original SS-BH
location fit has the additional variance
\[
 \mathcal V_{\SSm,T}-\frac{a_{\SSm}^{-2}}{\omega_T}
 =\frac{\kappa_T}{\omega_T^2}
       \E(a_{\SSm}^{-1}-\tau)^2\geq0.
\]
This isolates the gain from using the projected intercept in the
location equation. Inverse-norm weighting provides the further gain
from using residual lengths; the next result separates these two
contributions.

\begin{corollary}[Benchmark efficiency gaps]\label{cor:benchmark-efficiency-gaps}
For $0<\omega_T\leq1$,
\begin{align*}
 \mathcal V_{\LS,T}-\mathcal V_{\IN,T}
   &=\frac{\mu_2-\zeta^{-1}}{\omega_T}\geq0,
       \\
 \mathcal V_{\SSm,T}-\mathcal V_{\IN,T}
   &=\frac1{\omega_T}
      \E\left(a_{\SSm}^{-1}-\frac1{\zeta\tau}\right)^2
     +\frac{\kappa_T}{\omega_T^2}
      \E(a_{\SSm}^{-1}-\tau)^2\geq0.
\end{align*}
Each gap is zero if and only if $\tau$ is constant almost surely.
\end{corollary}

These are population first-order comparisons. A strict limiting
power gain requires that the corresponding variance gap persists
as $N,T$ increase. Under multivariate normal errors, radial concentration can
make the gaps converge to zero even though the finite-dimensional
radius is nonconstant. The simulations compare the original SS-BH
and LS-BH implementations with IN-BH at a common nominal FDR level.

\section{Monte Carlo experiments}\label{sec:simulation}

We compare IN-BH with the least-squares BH (LS-BH) benchmark
considered by \citet{GiglioLiaoXiu2021} and the spatial-sign BH
(SS-BH) procedure of \citet{Wang2025}. Following Scenario I of
the latter study, we generate data from
model~\eqref{eq:model} with $N=200$ funds, $p=3$ factors, and
$T=60$ or $120$. IN-BH is implemented as in
Section~\ref{sec:smoothing}.

In each replication, the coordinates of $\bm\mu_f$ are drawn
independently from $\operatorname{Unif}(0,1)$, and the factors are
generated independently across dates from
$N(\bm\mu_f,\mathbf\Sigma_f)$, where
$(\Sigma_f)_{jk}=0.5^{|j-k|}$. The loading coefficients
$\beta_{i1}$, $\beta_{i2}$, and $\beta_{i3}$ are drawn independently
from $\operatorname{Unif}(0.2,2)$, $\operatorname{Unif}(-1,1.5)$,
and $\operatorname{Unif}(-1.5,1.5)$, respectively. Twenty randomly
selected funds have $\alpha_i\sim\operatorname{Unif}(0.15,0.30)+\delta$,
and twenty other funds have
$\alpha_i\sim\operatorname{Unif}(-0.30,-0.20)$; the remaining
alphas are zero.

The errors are independent across dates and independent of the
factors and loadings. Let
$\mathbf R_N=(\rho^{|i-j|})_{i,j=1}^N$. We consider three
distributions:
\begin{enumerate}[label=(\roman*)]
\item Multivariate normal distribution:
      $\bm\varepsilon_t\sim N(\bm0,\mathbf R_N)$.
\item Multivariate $t$ distribution:
      $\sqrt{3}\,\bm\varepsilon_t\sim t_{N,3}(\bm0,\mathbf R_N)$,
      with three degrees of freedom and scale matrix $\mathbf R_N$.
\item Multivariate normal mixture distribution:
      $\bm\varepsilon_t\sim
      0.8N(\bm0,\mathbf R_N/2.6)+0.2N(\bm0,9\mathbf R_N/2.6)$.
\end{enumerate}
All three distributions have mean zero and covariance $\mathbf R_N$.

To examine the effect of signal strength, we fix $\rho=0.5$ and set
$\delta=0.06,0.12,\ldots,0.54$ for $T=60$ and
$\delta=0.04,0.08,\ldots,0.36$ for $T=120$.
All procedures use the nominal FDR level $q=0.10$.
We report empirical FDR and mean TDP over 2,000 replications.

\begin{figure}[!htbp]
\centering
\includegraphics[width=\textwidth]{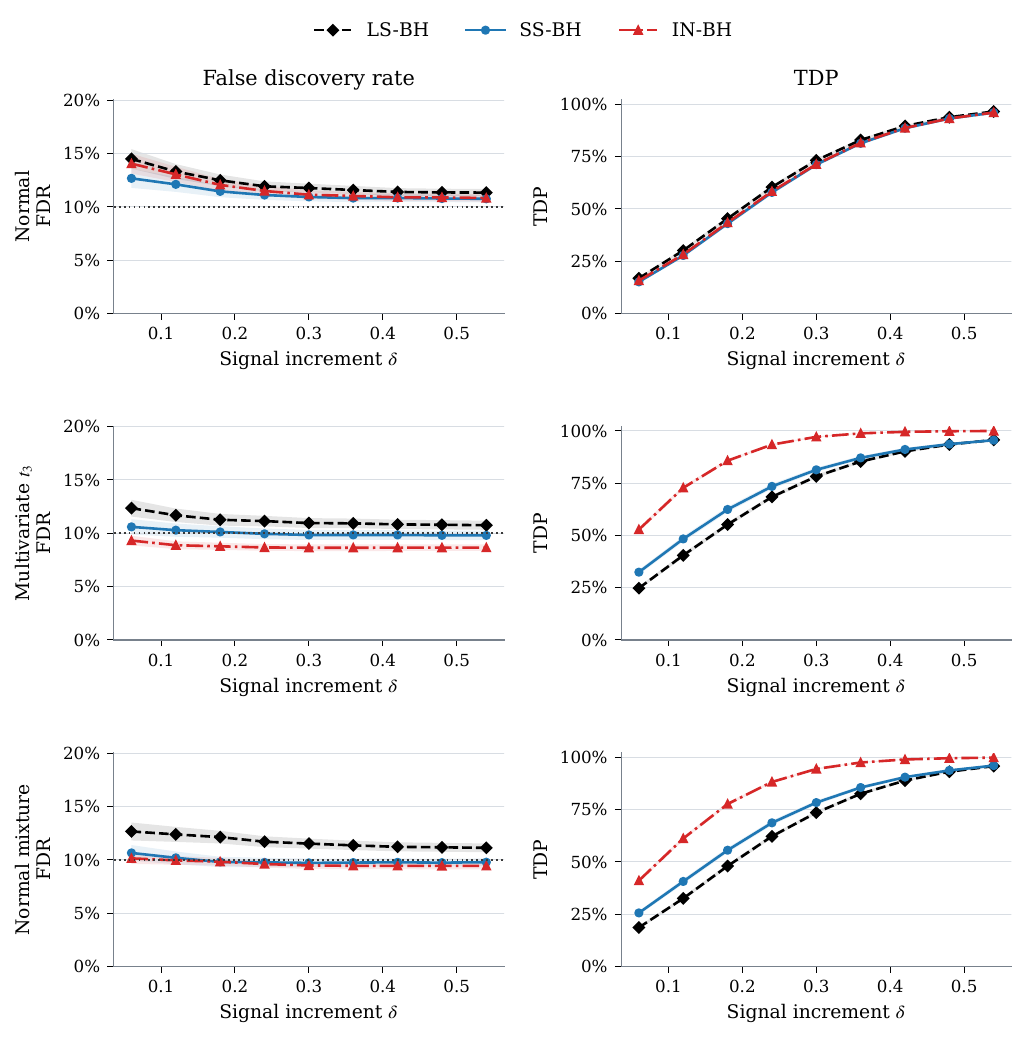}
\caption{Signal-strength experiment with $T=60$, $N=200$, and
$\rho=0.5$. Rows correspond to the multivariate normal, standardized
multivariate $t_3$, and multivariate normal mixture distributions;
columns report FDR and TDP. IN-BH, SS-BH, and LS-BH are red, blue,
and black, respectively. The dotted line marks the nominal FDR level,
and shaded bands show pointwise 95\% Monte Carlo intervals.}
\label{fig:signal60}
\end{figure}

\begin{figure}[!htbp]
\centering
\includegraphics[width=\textwidth]{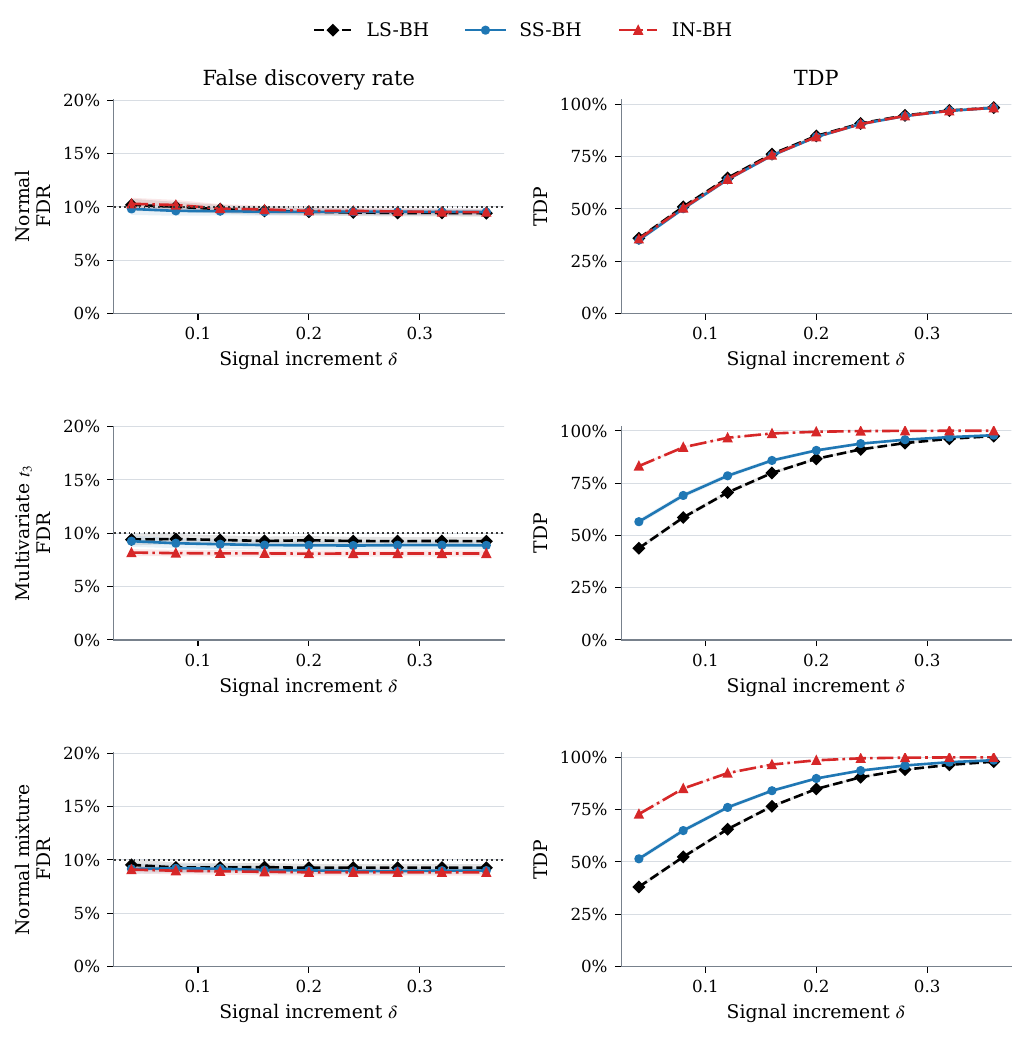}
\caption{Signal-strength experiment with $T=120$, $N=200$, and
$\rho=0.5$. Row ordering, line colors, and shaded bands are as in
Figure~\ref{fig:signal60}.}
\label{fig:signal120}
\end{figure}

Figures~\ref{fig:signal60} and \ref{fig:signal120} show the signal-strength
experiments. Under the multivariate $t$ and normal mixture distributions,
IN-BH has higher TDP than both comparison methods at all signal strengths
considered.
The improvement in TDP is greatest before the methods select nearly all
positive-alpha funds.

The higher TDP under heavy-tailed distributions is consistent with the use
of residual norms in the design-weighted spatial-sign estimating equation.
In the signal-strength experiments, IN-BH's empirical FDR remains below
the nominal level under the multivariate $t$ distribution. Under the normal
mixture distribution, it is close to or below the target, apart from a small
exceedance at the weakest signal strength with the smaller time-series
sample size. With the larger time-series sample size, empirical FDR remains
below the target under both heavy-tailed distributions.

Under the multivariate normal distribution, the methods have similar TDP.
LS-BH has a modestly higher TDP than IN-BH, whereas IN-BH and SS-BH
perform similarly. This pattern is
consistent with the smaller efficiency gains from radial weighting when
the residual norms concentrate under the multivariate normal distribution. With the smaller time-series sample size, IN-BH's empirical
FDR exceeds the nominal level when signals are weak.

\begin{figure}[!htbp]
\centering
\includegraphics[width=\textwidth]{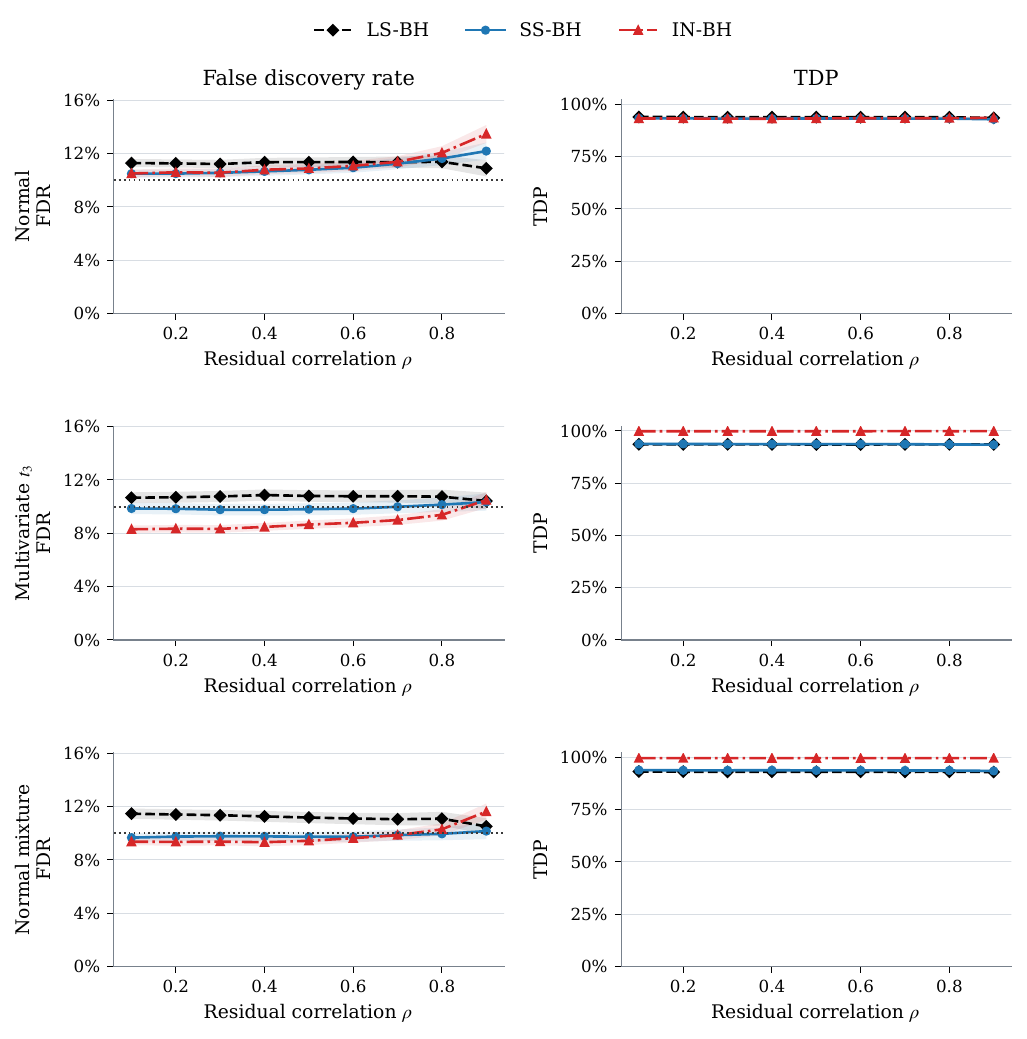}
\caption{Residual-correlation experiment with $N=200$, $T=60$,
and $\delta=0.48$. Row ordering, line colors, and shaded bands are
as in Figure~\ref{fig:signal60}. The dotted line marks the nominal
FDR level.}
\label{fig:correlation}
\end{figure}

We next examine the effect of residual correlation by fixing $N=200$,
$T=60$, and $\delta=0.48$ and varying $\rho=0.1,0.2,\ldots,0.9$.
The factor and loading distributions, the proportions of positive,
negative, and zero alphas, the three families of error distributions, the nominal
FDR level, and the number of replications remain as specified above.
Figure~\ref{fig:correlation} shows that the TDP of each procedure changes
little across the correlation levels considered. Under heavy-tailed
distributions, IN-BH retains its TDP advantage over LS-BH and SS-BH,
whereas the three procedures have similar TDP under the multivariate
normal distribution. IN-BH's empirical FDR generally increases as
residual correlation strengthens. Under heavy-tailed distributions,
it remains below the nominal level over much of the correlation range;
under the multivariate normal distribution, it is already above the
nominal level at weaker correlations. At the strongest correlation
considered, IN-BH's empirical FDR exceeds the nominal level under all
three distributions.

\section{Real data application}\label{sec:empirical}

We use the mutual-fund panel accompanying \citet{GaoPengYan2024},
originally constructed from the CRSP Survivor-Bias-Free Mutual Fund
Database. The source study considers actively managed U.S. equity funds,
excludes passive index funds, and requires initial total net assets
above \$10 million and more than 80\% of holdings in equities.
The unbalanced panel contains 726 fund series over 368 months, from
February 1987 to September 2017. Fund eligibility is assessed within
each historical estimation window.

\begin{figure}[p]
\centering
\includegraphics[width=\textwidth,height=0.92\textheight,keepaspectratio]{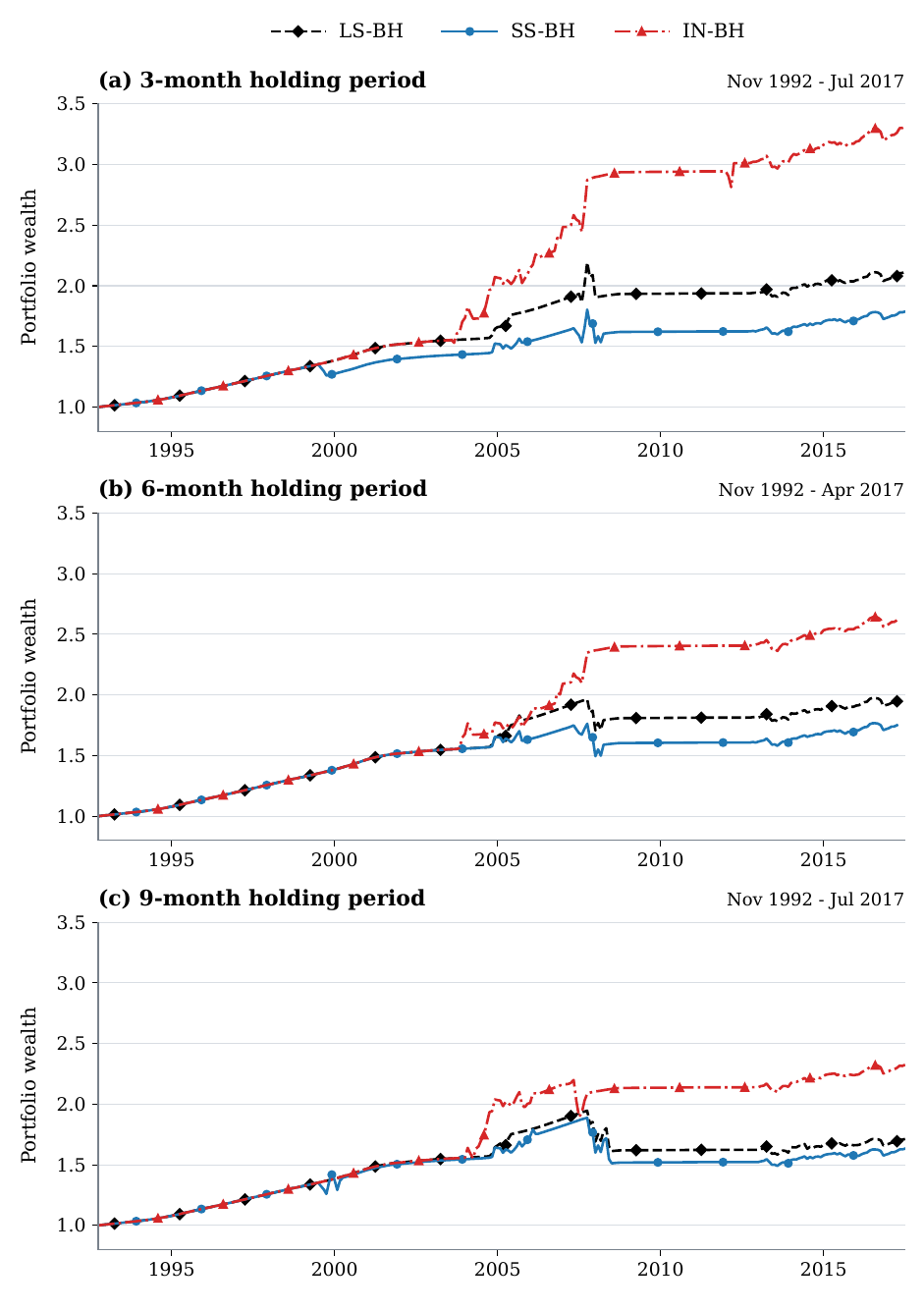}
\caption{Subsequent portfolio wealth after rolling 69-month estimation.
Panels use holding periods of three, six, and nine months.}
\label{fig:empirical69}
\end{figure}

Let $R_{ti}$ denote the monthly fund return, reported net of fees and
expenses in the source study, and let $R_{f,t}$ denote the aligned
monthly risk-free return. We form $Y_{ti}=R_{ti}-R_{f,t}$ before
fitting the selection model. Returns and factors are expressed in
decimal units. Following the source data convention, zero entries in
the fund-return file are treated as missing observations. The analysis
uses the factor observations archived with the fund panel.

The pricing benchmark is the Fama--French--Carhart four-factor model
\citep{FamaFrench1993,Carhart1997}. Within each estimation window $b$,
we fit
\[
 Y_{ti}=\alpha_i^{(b)}
       +\beta_{i,M}^{(b)}\mathrm{MKT}_t
       +\beta_{i,S}^{(b)}\mathrm{SMB}_t
       +\beta_{i,H}^{(b)}\mathrm{HML}_t
       +\beta_{i,U}^{(b)}\mathrm{MOM}_t
       +\varepsilon_{ti}^{(b)}.
\]
Here $\mathrm{MKT}_t$ is the market return in excess of the risk-free
rate, and the remaining factors represent size, value, and momentum.
Intercepts and factor exposures are fund specific and are re-estimated
at each portfolio formation date. The selection target is a positive
alpha relative to the four-factor benchmark.

The estimation window contains $T=69$ months, and the portfolio is
reconstituted every $H\in\{3,6,9\}$ months. Index February 1987 by one
and write $s_b=1+(b-1)H$. The estimation and subsequent holding windows
are
\[
 \mathcal I_b=\{s_b,\ldots,s_b+68\},\qquad
 \mathcal O_b=\{s_b+69,\ldots,s_b+68+H\}.
\]
A fund is eligible if its returns are observed throughout
$\mathcal I_b$. LS-BH, SS-BH, and IN-BH are implemented as in
Section~\ref{sec:method}, with nominal FDR level $q=0.05$.
The first estimation window runs from February 1987 to October 1992,
and investment starts in November 1992. We retain complete holding
periods, ending in April 2017 for the six-month design and July 2017
for the other two designs.

At a formation date, the available wealth is allocated equally among
the selected funds. These positions are held until the next formation
date, so portfolio weights drift with realized returns within a holding
period. If no fund is selected, the portfolio earns the monthly
risk-free return. Let $\mathcal S_{b,m}$ be the selected set for method
$m$ and let $\widetilde R_{ti}$ denote the return after the missing-value
convention described below. For a nonempty set, wealth at month $\ell$
of the holding period is
\[
 W_{b,m}(\ell)=\frac{W_{b,m}(0)}{|\mathcal S_{b,m}|}
       \sum_{i\in\mathcal S_{b,m}}
       \prod_{j=1}^{\ell}(1+\widetilde R_{s_b+68+j,i}),
       \qquad \ell=1,\ldots,H.
\]
The terminal wealth becomes the next period's initial wealth.
Portfolio performance compounds raw fund returns, with no deduction
for transaction costs.

If a selected fund has an unobserved holding-period return, its position
is converted to cash at the last observed month-end value and earns
the risk-free return until the next formation date. We also consider
a conservative alternative that assigns zero value to the affected
position from its first missing month until the next formation date.
The other positions and selection rules are unchanged.

Figure~\ref{fig:empirical69} plots wealth from an initial investment of
one for each holding-period design. IN-BH produces higher terminal
wealth than both benchmarks in all three designs. With three-month
reconstitution, its compound annual growth rate is 4.96\%, compared
with 3.07\% for LS-BH and 2.37\% for SS-BH. With six-month
reconstitution, the corresponding rates are 4.00\%, 2.76\%, and
2.31\%; with nine-month reconstitution, they are 3.47\%, 2.20\%,
and 2.00\%. The return rankings and maximum drawdowns are unchanged
when all wealth paths are evaluated through April 2017.

The wealth advantage is accompanied by smaller historical drawdowns.
For the three-, six-, and nine-month designs, the largest peak-to-trough
loss for IN-BH is 5.44\%, 5.35\%, and 13.84\%, respectively. The
corresponding losses are 13.06\%, 13.06\%, and 18.57\% for LS-BH,
and 15.24\%, 15.24\%, and 21.00\% for SS-BH. Higher return and
smaller drawdown therefore occur together in these comparisons.
More frequent reconstitution is associated with higher growth for
IN-BH over the holding periods considered.

IN-BH forms a nonempty portfolio at least as often as either benchmark
in each holding-period design. It also retains the highest compound
annual growth rate under the conservative alternative for unobserved
holding-period returns.

\FloatBarrier
\section{Conclusion}\label{sec:conclusion}

This paper develops radially weighted alpha estimators for robust mutual
fund selection. For fixed radial profiles, we establish asymptotic FDR
control and identify inverse-norm weighting as optimal for local power.
Simulations and an application to U.S. equity mutual funds illustrate
the gains from using both residual directions and magnitudes.

Two extensions merit further study. First, the proposed procedure could
be combined with latent-factor adjustment to accommodate common
dependence remaining after projection on the observed factors
\citep{LanDu2019,Wang2025}. In applications with a rich set of observed
pricing factors, additional adjustment may be unnecessary if little
common variation remains in the residuals. Second, extending radial
weighting to varying-coefficient models would allow fund alphas and
factor loadings to evolve over time \citep{MaLanSuTsai2020}.
This extension would require incorporating the estimation of the
coefficient functions into the alpha statistics and establishing FDR
control for the resulting fund-selection procedure.

\printbibliography[title={References}]
\par\medskip
\noindent\textbf{Data availability statement.}
The mutual fund data are available from CRSP under a commercial license
and are subject to redistribution restrictions.
The R code used to reproduce the simulation and empirical results is
publicly available at \url{https://github.com/flnankai/radialFundSelect}.

\appendix
\section*{Appendices}
\Needspace{3\baselineskip}
\noindent
Section~\ref{app:linearization-main} establishes the distributional and
local power results for the weighted alpha estimator, including PQR scale
consistency and studentization, and verifies the radial conditions.
Section~\ref{app:weighted-fdr-proofs} provides Gaussian tail-count bounds
and proves false discovery control and recovery of sufficiently strong
positive-alpha funds. Section~\ref{app:efficiency} proves the optimality of
inverse-norm weighting and derives the first-order efficiency gaps relative
to the least-squares and spatial-sign benchmarks.

\section{Proofs for the weighted alpha estimator}
\label{app:linearization-main}

The calculations are conditional on the factor design. On
$\mathcal F_{NT}$, replace $\mathbf F$ by
$\mathbf F^*=\mathbf F\mathbf Q_T^{-1/2}$. Then
\[
 T^{-1}(\mathbf F^*)^\top\mathbf F^*=\mathbf I_p,\qquad
 \mathbf P_{F^*}=\mathbf P_F,\qquad
 \|\bm f_t^*\|^2=T(\mathbf P_F)_{tt}\leq CL.
\]
For the fixed profile, this transformation preserves the estimators and
studentizers. Suppressing the superscript, we use
\[
 \mathbf Q_T=\mathbf I_p,\qquad
 \max_t\|\bm f_t\|\leq C\sqrt L,\qquad
 T^{-1}\sum_t\|\bm f_t\|^2=p.
\]

\subsection{Spherical directions and radial bounds}
\label{app:elliptical-geometry}

By \eqref{eq:radial-model},
\[
 \varpi_t=\|\bm v_t\|/\sqrt N,\qquad
 \varpi_t^2=(\bm U_t^\circ)^\top\mathbf R_N\bm U_t^\circ,\qquad
 r_t/\sqrt N=\tau_t\varpi_t,\qquad \tau_t\perp\bm U_t^\circ.
\]
The spherical coordinates satisfy
\begin{equation}\label{eq:spherical-moments}
 \begin{gathered}
 \E e^{\lambda\bm a^\top\bm v_t}
       \leq e^{\lambda^2\bm a^\top\mathbf R_N\bm a/2},\qquad
 \E\bm v_t\bm v_t^\top=\mathbf R_N,\\
 \E(v_{ti}^2v_{tj}^2)=\frac{N}{N+2}(1+2R_{N,ij}^2),
 \qquad \bm a\in\R^N,\quad\lambda\in\R.
 \end{gathered}
\end{equation}
We use separate events for the directions and the radial variables:
\begin{equation}\label{eq:local-envelope-event}
 \mathcal A_{NT}^{\mathrm{sph}}(K)
 =\left\{\max_{t,i}|v_{ti}|\leq K\sqrt L,\quad
          \max_t|\varpi_t-1|\leq K\sqrt{b_NL/N}\right\},
\end{equation}
and
\begin{equation}\label{eq:radial-truncation-event}
 \mathcal R_{\tau,T}=\{\max_t\tau_t^{-1}\leq T^{1/4}\}.
\end{equation}
The inverse-moment assumption gives the deterministic tail bound
\begin{equation}\label{eq:inverse-fourth-tail}
 \zeta_{\tau,T}
 =\sup_N\E\{\tau_N^{-4}\ind(\tau_N^{-1}>T^{1/4})\}
 \longrightarrow0.
\end{equation}

\begin{lemma}[Spherical bounds]\label{lem:spherical-geometry}
Under Assumptions~\ref{ass:radial-primitive}--\ref{ass:weight-primitive},
\eqref{eq:spherical-moments} holds. For every fixed $r>0$ and
integer $m\geq1$, uniformly for all sufficiently large $N$,
\begin{gather}
 \E(\varpi_N-1)^2\leq 2b_N/N,\qquad
 \E(\varpi_N^r+\varpi_N^{-r})\leq C_r,\nonumber\\
 \E|\varpi_N^2-1|^{2m}
 \leq C_m\left\{\frac{\tr((\mathbf R_N-\mathbf I_N)^2)}{N^2}\right\}^{m}
 \leq C_m(b_N/N)^m,
 \label{eq:spherical-centered-moments}
\end{gather}
where $\varpi_N$ has the law of $\varpi_t$. For each $A>0$, there is
$C_A<\infty$ such that
\[
 \Pp\{\mathcal A_{NT}^{\mathrm{sph}}(C_A)^c\}
       \leq C_A(NT)^{-A},\qquad
 \Pp(\mathcal R_{\tau,T}^c)\leq\zeta_{\tau,T}=o(1).
\]
\end{lemma}

\begin{proof}[Proof of Lemma~\ref{lem:spherical-geometry}]
Let $\bm Z^\circ\sim N(\bm0,\mathbf I_N)$,
$R_{Z^\circ}=\|\bm Z^\circ\|$, and
$\bm U^\circ=\bm Z^\circ/R_{Z^\circ}$. Polar integration gives
\[
 R_{Z^\circ}\sim\chi_N,\qquad
 \bm U^\circ\sim\operatorname{Unif}(\mathbb S^{N-1}),\qquad
 R_{Z^\circ}\perp\bm U^\circ.
\]
For $m\geq0$, symmetry and independence imply
\[
 \E(\sqrt N U_1^\circ)^{2m+1}=0,\qquad
 \E(\sqrt N U_1^\circ)^{2m}
 =\frac{N^m(2m-1)!!}{N(N+2)\cdots(N+2m-2)}
 \leq(2m-1)!!.
\]
Rotational invariance and the exponential series therefore yield
\[
 \E e^{\lambda\bm a^\top\bm v_t}
 \leq\sum_{m=0}^{\infty}
        \frac{\{\lambda^2\bm a^\top\mathbf R_N\bm a/2\}^m}{m!}
 =e^{\lambda^2\bm a^\top\mathbf R_N\bm a/2}.
\]
For $\bm a_i=\mathbf R_N^{1/2}\bm e_i$, $\|\bm a_i\|^2=1$ and
$\bm a_i^\top\bm a_j=R_{N,ij}$. Gaussian moments give
\[
 \E(v_{ti}v_{tj})=R_{N,ij},\qquad
 \E(v_{ti}^2v_{tj}^2)
 =\frac{N^2}{\E R_{Z^\circ}^4}
    \E\{(\bm a_i^\top\bm Z^\circ)^2
                    (\bm a_j^\top\bm Z^\circ)^2\}
 =\frac{N}{N+2}(1+2R_{N,ij}^2).
\]
All spherical monomials of odd total degree have expectation zero.

For symmetric $\mathbf B$ with eigenvalues $\beta_j$, put
$B_2=\sum_j\beta_j^2$ and $B_\infty=\max_j|\beta_j|$.
For $|u|B_\infty\leq1/4$,
\[
 \begin{aligned}
 \log\E\exp\{u((\bm Z^\circ)^\top\mathbf B\bm Z^\circ-\tr\mathbf B)\}
 &=\frac12\sum_j\{-\log(1-2u\beta_j)-2u\beta_j\}\\
 &\leq C u^2B_2.
 \end{aligned}
\]
Exponential Markov's inequality, with
$|u|=c\min\{\sqrt{x/B_2},B_\infty^{-1}\}$ when $B_2>0$, gives
\[
 \Pp\{|(\bm Z^\circ)^\top\mathbf B\bm Z^\circ-\tr\mathbf B|
       >C(\sqrt{B_2x}+B_\infty x)\}\leq2e^{-x},\qquad x\geq1.
\]
Apply this bound to $\mathbf R_N$ and $\mathbf I_N$.
Since $\tr\mathbf R_N=N$ and $\tr(\mathbf R_N^2)\leq Nb_N$,
for $1\leq x\leq c_0N$ the two events imply $R_{Z^\circ}^2\geq N/2$ and
\[
 |\varpi_N^2-1|
 =\frac{|(\bm Z^\circ)^\top\mathbf R_N\bm Z^\circ
                         -R_{Z^\circ}^2|}{R_{Z^\circ}^2}
 \leq C\{\sqrt{b_Nx/N}+b_Nx/N\}.
\]
Consequently,
\begin{equation}\label{eq:spherical-norm-tail}
 \Pp\{|\varpi_N^2-1|>C(\sqrt{b_Nx/N}+b_Nx/N)\}
 \leq4e^{-x},\qquad 1\leq x\leq c_0N.
\end{equation}
The fourth-moment identity gives
\[
 \E\varpi_N^2=1,\qquad
 \E(\varpi_N^2-1)^2
 =\frac{2}{N+2}\left\{\frac{\tr(\mathbf R_N^2)}N-1\right\}
 \leq\frac{2b_N}N.
\]
Since $|\varpi_N-1|\leq|\varpi_N^2-1|$, this also proves the
claimed second-moment bound.

For $\mathbf B=\mathbf R_N-\mathbf I_N$, $\tr\mathbf B=0$ and
$B_\infty^2\leq B_2$. Integrating the preceding tail bound yields
\[
 \E|(\bm Z^\circ)^\top\mathbf B\bm Z^\circ|^{2m}
 \leq C_m B_2^m.
\]
The quadratic form equals $R_{Z^\circ}^2(\varpi_N^2-1)$ with independent
factors. Since $\E R_{Z^\circ}^{4m}=N(N+2)\cdots(N+4m-2)\geq N^{2m}$,
\eqref{eq:spherical-centered-moments} follows, including $\mathbf B=\mathbf0$.

For any integer $h\geq1$, convexity gives
\[
 \E\varpi_N^{2h}
 =\E\left\{\sum_j\frac{\lambda_j}{N}N(U_j^\circ)^2\right\}^{h}
 \leq\sum_j\frac{\lambda_j}{N}
                 \E|\sqrt N U_j^\circ|^{2h}\leq C_h,
\]
where $\lambda_j$ are the eigenvalues of $\mathbf R_N$.
For negative moments, concavity of the logarithm implies
\[
 \log(1+2s\lambda_j/N)
 \geq(\lambda_j/b_N)\log(1+2sb_N/N),\qquad s\geq0.
\]
With $a_N=N/(2b_N)\to\infty$,
\[
 \E e^{-s(\bm Z^\circ)^\top\mathbf R_N\bm Z^\circ/N}
 =\prod_j(1+2s\lambda_j/N)^{-1/2}
 \leq(1+s/a_N)^{-a_N}.
\]
For $a_N\geq2h$, the Laplace integral for a negative power gives
\[
 \begin{aligned}
 \E\{(\bm Z^\circ)^\top\mathbf R_N\bm Z^\circ/N\}^{-h}
 &\leq\frac1{\Gamma(h)}
          \int_0^\infty s^{h-1}(1+s/a_N)^{-a_N}\,ds\\
 &=a_N^h\frac{\Gamma(a_N-h)}{\Gamma(a_N)}
   =\prod_{j=1}^{h}\frac{a_N}{a_N-j}\leq2^h.
 \end{aligned}
\]
Independence and $\E(R_{Z^\circ}^2/N)^{-h}\geq1$ give
$\E\varpi_N^{-2h}\leq2^h$. Moment monotonicity with an integer
$h\geq r/2$ proves the stated bounds.

Finally, \eqref{eq:spherical-moments} and a union bound imply
\[
 \Pp\{\max_{t,i}|v_{ti}|>K\sqrt L\}
 \leq2NT e^{-K^2L/2}.
\]
Since $b_NL/N\to0$ and $L/N\to0$,
\eqref{eq:spherical-norm-tail} with $x=(A+3)L$ gives the bound for
$\mathcal A_{NT}^{\mathrm{sph}}(C_A)$. By \eqref{eq:inverse-fourth-tail},
\[
 \Pp(\mathcal R_{\tau,T}^c)
 \leq T\Pp(\tau^{-1}>T^{1/4})
 \leq\E\{\tau^{-4}\ind(\tau^{-1}>T^{1/4})\}
 \leq\zeta_{\tau,T}\longrightarrow0.
\]
The argument allows singular $\mathbf R_N$.
\end{proof}

\subsection{Uniform consistency of the residual PQR scale}
\label{app:pqr}

Let $W_N\overset{d}=\sqrt N U_1^\circ$ be independent of $\tau$;
primes denote independent copies. Define
\[
 \begin{gathered}
 m_{D,N}=\frac1N\sum_i d_i^2,\qquad
 H_N(x)=\Pp\{|\tau W_N-\tau'W_N'|\leq x\},\\
 q_{D,N}=\inf\{x:H_N(x)\geq\tau_D\},\qquad 0<\tau_D<1.
 \end{gathered}
\]
The unrestricted projection satisfies
\[
 \mathbf M_X\one=\bm0,\qquad \mathbf M_X\mathbf F=\mathbf0,
 \qquad\widetilde{\mathbf E}=\mathbf M_X\mathbf Y=\mathbf M_X\mathbf E.
\]
Hence the PQR scales are invariant to arbitrary deterministic shifts in
$\bm\alpha$.

\begin{lemma}[Residual PQR consistency]\label{lem:pqr-scale}
Under the error and factor-design conditions of
Assumption~\ref{ass:radial-primitive}, if $L/T\to0$, then
$\min_i\widehat q_i>0$ with probability tending to one and
\begin{align}
 \max_i\left|\widehat q_i/d_i-q_{D,N}\right|
 &=O_p(\eta_{NT}),\label{eq:pqr-quantile-rate}\\
 \max_i\left\{
 \left|\log\frac{\widehat d_{Q,i}^{\,g}}{d_i}\right|
 +\left|\log\frac{\widehat d_{Q,i}}{d_i/\sqrt{m_{D,N}}}\right|
 \right\}&=O_p(\eta_{NT}).\label{eq:pqr-scale-rate}
\end{align}
\end{lemma}

\begin{proof}[Proof of Lemma~\ref{lem:pqr-scale}]
Condition on $\mathbf F\in\mathcal F_{NT}$. For each $i$,
\[
 e_{ti}=\varepsilon_{ti}/d_i,\qquad E_N=\tau W_N,\qquad
 \mathcal L(e_{1i},\ldots,e_{Ti}\mid\mathbf F)
       =\mathcal L(E_N)^{\otimes T}.
\]
For $N\geq8$, the density of $W_N$ is
\[
 \begin{aligned}
 f_N(x)&=c_N(1-x^2/N)_+^{(N-3)/2},
 &c_N&=\frac{\Gamma(N/2)}{\sqrt{\pi N}\Gamma((N-1)/2)},\\
 f_N'(x)&=-c_N\frac{N-3}{N}x(1-x^2/N)_+^{(N-5)/2}.
 \end{aligned}
\]
Using $\log(1-y)\leq-y$ for $0\leq y<1$ and
$\log(1-y)\geq-2y$ for $0\leq y\leq1/2$, for $N\geq8$,
\[
 \begin{gathered}
 2e^{-1}\leq\int_{-1}^1(1-x^2/N)^{(N-3)/2}\,dx
 \leq c_N^{-1}
 \leq\int_{\R}e^{-x^2/4}\,dx=2\sqrt\pi,\\
 (2\sqrt\pi)^{-1}\leq c_N\leq e/2,\qquad
 |f_N'(x)|\leq C|x|e^{-x^2/8},\\
 \sup_x\{|f_N(x)|+|f_N'(x)|\}\leq C.
 \end{gathered}
\]
For fixed $M<\infty$ and $N\geq8\vee2M^2$,
\[
 \begin{gathered}
 \frac{N-3}{2}\log(1-x^2/N)\geq-x^2\geq-M^2,
                 \qquad |x|\leq M,\\
 \inf_{|x|\leq M}f_N(x)\geq(2\sqrt\pi)^{-1}e^{-M^2}=:c_M>0.
 \end{gathered}
\]
Since $|\tau^{-2}f_N'(x/\tau)|\leq C\tau^{-2}$ and $\E\tau^{-2}<\infty$,
\[
 \begin{aligned}
 h_N(x)&=\E\{\tau^{-1}f_N(x/\tau)\},&
 h_N'(x)&=\E\{\tau^{-2}f_N'(x/\tau)\},\\
 \|h_N\|_\infty&\leq C\E\tau^{-1}\leq C,&
 \|h_N'\|_\infty&\leq C\E\tau^{-2}\leq C.
 \end{aligned}
\]
Put $C_2=1\vee\sup_N\E\tau^{-2}\vee\sup_N\E\tau^2$,
$a_D=(2\sqrt{C_2})^{-1}$, and $b_D=2\sqrt{C_2}$. Then
\[
 \Pp(\tau<a_D)+\Pp(\tau>b_D)
 \leq a_D^2\E\tau^{-2}+b_D^{-2}\E\tau^2\leq\tfrac12.
\]
For $|x|\leq M$,
\[
 h_N(x)\geq b_D^{-1}
     \inf_{|u|\leq M/a_D}f_N(u)\Pp(a_D\leq\tau\leq b_D)
 \geq c_{M/a_D}/(2b_D)>0.
\]
For $D_{0,N}=E_N-E_N'$, convolution gives
\begin{align*}
 f_{\Delta,N}(y)&=\int_{\R}h_N(x)h_N(x-y)\,dx=f_{\Delta,N}(-y),\\
 f_{\Delta,N}'(y)&=-\int_{\R}h_N(x)h_N'(x-y)\,dx,\\
 \|f_{\Delta,N}\|_\infty+\|f_{\Delta,N}'\|_\infty
 &\leq\|h_N\|_1(\|h_N\|_\infty+\|h_N'\|_\infty)\leq C.
\end{align*}
For fixed $M$,
\[
 \inf_{|y|\leq M}f_{\Delta,N}(y)
 \geq2\left(\inf_{|x|\leq1}h_N(x)\right)
        \left(\inf_{|z|\leq M+1}h_N(z)\right)>0.
\]
Choose $q_\ell=\tau_D/(4C)$ and
$q_u=\{4C_2/(1-\tau_D)\}^{1/2}$. Then
\[
 \begin{gathered}
 H_N(x)=\int_{-x}^x f_{\Delta,N}(u)\,du,\qquad
 H_N'(x)=f_{\Delta,N}(x)+f_{\Delta,N}(-x)=2f_{\Delta,N}(x),\quad x>0,\\
 H_N(q_\ell)\leq2Cq_\ell=\tau_D/2<\tau_D,\\
 1-H_N(q_u)\leq\E D_{0,N}^2/q_u^2
       =2\E\tau^2/q_u^2\leq(1-\tau_D)/2,\\
 q_\ell<q_{D,N}<q_u,\qquad H_N(q_{D,N})=\tau_D.
 \end{gathered}
\]
Taking $h_D=q_\ell/2$, for some fixed $c_D>0$,
\[
 \inf_{|x-q_{D,N}|\leq h_D}H_N'(x)
 \geq2\inf_{|x|\leq q_u+h_D}f_{\Delta,N}(x)\geq c_D.
\]
Put $\bm x_t^X=(1,\bm f_t^\top)^\top$,
$\mathbf Q_X=T^{-1}\mathbf X^\top\mathbf X$, and
$\bm u_X=\mathbf Q_T^{-1}\bar{\bm f}$. Then
\[
 \mathbf Q_X=
 \begin{pmatrix}1&\bar{\bm f}^{\top}\\\bar{\bm f}&\mathbf Q_T\end{pmatrix},
 \qquad \omega_T=1-\bar{\bm f}^{\top}\bm u_X,\qquad
 \mathbf Q_X\binom{1}{-\bm u_X}=\binom{\omega_T}{\bm0}.
\]
Block inversion gives
\begin{equation}\label{eq:augmented-gram-inverse}
 \mathbf Q_X^{-1}=
 \begin{pmatrix}
 \omega_T^{-1}&-\omega_T^{-1}\bar{\bm f}^{\top}\mathbf Q_T^{-1}\\
 -\omega_T^{-1}\mathbf Q_T^{-1}\bar{\bm f}&
 \mathbf Q_T^{-1}+\omega_T^{-1}\mathbf Q_T^{-1}\bar{\bm f}
                   \bar{\bm f}^{\top}\mathbf Q_T^{-1}
 \end{pmatrix}.
\end{equation}
The design bounds yield
\[
 \begin{gathered}
 \|\bar{\bm f}\|^2\leq T^{-1}\sum_t\|\bm f_t\|^2
   =\tr(\mathbf Q_T)\leq C,\qquad
 \|\mathbf Q_X\|_{\op}\leq\tr(\mathbf Q_X)=1+\tr(\mathbf Q_T)\leq C,\\
 \|\mathbf Q_X^{-1}\|_{\op}
 \leq\|\mathbf Q_T^{-1}\|_{\op}
       +\omega_T^{-1}(1+\|\bm u_X\|^2)\leq C.
 \end{gathered}
\]
The standardized OLS coefficient error is
\[
 \widehat{\bm c}_i=\mathbf Q_X^{-1}\frac1T
                    \sum_t\bm x_t^X\tau_tv_{ti},\qquad
 \mathbf H_T=\frac1T\sum_t\tau_t^2\bm x_t^X(\bm x_t^X)^\top.
\]
By \eqref{eq:spherical-moments},
\[
 \E\!\left(e^{\lambda\bm a^\top\widehat{\bm c}_i}
        \mid\bm\tau,\mathbf F\right)
 \leq\exp\left\{\frac{\lambda^2}{2T}
    \bm a^\top\mathbf Q_X^{-1}\mathbf H_T\mathbf Q_X^{-1}\bm a\right\}.
\]
For $B,K>0$,
\[
 \begin{aligned}
 \E(\tr\mathbf H_T\mid\mathbf F)
 &=\E\tau^2\,\tr\mathbf Q_X\leq C,\\
 \Pp\{\max_i\|\widehat{\bm c}_i\|>K\eta_{NT}\mid\mathbf F\}
 &\leq\Pp(\tr\mathbf H_T>B\mid\mathbf F)
       +2N(p+1)e^{-cK^2L/B}\\
 &\leq C/B+2N(p+1)e^{-cK^2L/B},\\
 \lim_{K\to\infty}\limsup_{N,T\to\infty}
 \Pp\{\max_i\|\widehat{\bm c}_i\|>K\eta_{NT}\mid\mathbf F\}
 &\leq\lim_{B\to\infty}C/B=0.
 \end{aligned}
\]
Thus, with $\widehat{\bm\vartheta}_{Q,i}=(\mathbf0_{p\times1},\mathbf I_p)
\widehat{\bm c}_i$,
\begin{equation}\label{eq:pqr-pilot-coefficient-rate}
 \max_i\|\widehat{\bm\vartheta}_{Q,i}\|=O_p(\eta_{NT}).
\end{equation}
Since $\widetilde{\mathbf E}=\mathbf M_X\mathbf E$, pairwise differencing
eliminates the fitted intercept and yields
\begin{equation}\label{eq:pqr-residual-difference}
 \frac{\widetilde E_{ti}-\widetilde E_{si}}{d_i}
 =e_{ti}-e_{si}-(\bm f_t-\bm f_s)^\top\widehat{\bm\vartheta}_{Q,i}.
\end{equation}
For fixed $\bm\vartheta_Q\in\R^p$ and $x>0$, define
\[
 \begin{aligned}
 \mathcal U_{x,\bm\vartheta_Q}(\bm z)&=\binom T2^{-1}
  \sum_{s<t}\ind\{|z_t-z_s-(\bm f_t-\bm f_s)^\top\bm\vartheta_Q|\leq x\},
   \qquad\bm z\in\R^T,\\
 U_{i,T}(x,\bm\vartheta_Q)&=\mathcal U_{x,\bm\vartheta_Q}(e_{1i},\ldots,e_{Ti}).
 \end{aligned}
\]
Let
\[
 \mathcal G_t=\sigma(e_{1i},\ldots,e_{ti}),\qquad
 M_t=\E\{U_{i,T}(x,\bm\vartheta_Q)\mid\mathcal G_t,\mathbf F\},\qquad
 \Delta_{ti}=M_t-M_{t-1}.
\]
For $\bm z,\bm z'\in\R^T$,
\[
 z_s=z_s'\ (s\ne t)
 \quad\Longrightarrow\quad
 |\mathcal U_{x,\bm\vartheta_Q}(\bm z)-\mathcal U_{x,\bm\vartheta_Q}(\bm z')|
 \leq\frac{T-1}{\binom T2}=\frac2T.
\]
Conditionally on $(\mathcal G_{t-1},\mathbf F)$, $\Delta_{ti}$ is centered
and has range length at most $2/T$. Exponential tilting and iterated
conditioning give
\[
 \begin{aligned}
 \E(e^{\lambda\Delta_{ti}}\mid\mathcal G_{t-1},\mathbf F)
 &\leq e^{\lambda^2/(2T^2)},\\
 \E\!\left[
 e^{\lambda\{U_{i,T}(x,\bm\vartheta_Q)-\E(U_{i,T}(x,\bm\vartheta_Q)\mid\mathbf F)\}}
 \mid\mathbf F\right]
 &\leq e^{\lambda^2/(2T)}.
 \end{aligned}
\]
Exponential Markov's inequality yields, for every $y>0$,
\begin{equation}\label{eq:pqr-bounded-differences}
 \Pp\!\left\{|U_{i,T}(x,\bm\vartheta_Q)
  -\E(U_{i,T}(x,\bm\vartheta_Q)\mid\mathbf F)|>y\mid\mathbf F\right\}
 \leq2\inf_{\lambda>0}e^{-\lambda y+\lambda^2/(2T)}
 =2e^{-Ty^2/2}.
\end{equation}
For $H_N(x,a)=\Pp(|D_{0,N}-a|\leq x)$, with
$H_N(x,0)=H_N(x)$, symmetry gives
\[
 \begin{aligned}
 \partial_aH_N(x,0)&=f_{\Delta,N}(x)-f_{\Delta,N}(-x)=0,\\
 \partial_a^2H_N(x,a)&=f_{\Delta,N}'(a+x)-f_{\Delta,N}'(a-x),\\
 |H_N(x,a)-H_N(x,0)|
 &=\left|\int_0^a(a-u)\partial_a^2H_N(x,u)\,du\right|\leq Ca^2.
 \end{aligned}
\]
Write $\widehat{\mathbf\Sigma}_f=T^{-1}\sum_t
(\bm f_t-\bar{\bm f})(\bm f_t-\bar{\bm f})^\top$. Then
\[
 \binom T2^{-1}\sum_{s<t}
       \{(\bm f_t-\bm f_s)^\top\bm\vartheta_Q\}^2
 =\frac{2T}{T-1}\bm\vartheta_Q^\top\widehat{\mathbf\Sigma}_f\bm\vartheta_Q
 \leq C\|\bm\vartheta_Q\|^2.
\]
Consequently,
\begin{equation}\label{eq:pqr-expectation-perturbation}
 \sup_x\left|\E\{U_{i,T}(x,\bm\vartheta_Q)\mid\mathbf F\}-H_N(x)\right|
 \leq C\|\bm\vartheta_Q\|^2.
\end{equation}
Let $\mathcal B_K$ be a deterministic $\delta_{NT}$-net of
$\{\|\bm\vartheta_Q\|\leq K\eta_{NT}\}$, where
\[
 \begin{gathered}
 \delta_{NT}=\frac{\eta_{NT}}{8C_f\sqrt L},\qquad
 \max_t\|\bm f_t\|\leq C_f\sqrt L,\qquad
 |\mathcal B_K|\leq(CK\sqrt L)^p,\\
 \max_{s,t}|(\bm f_t-\bm f_s)^\top(\bm\vartheta_Q-\bm\vartheta_Q^\circ)|
 \leq2C_f\sqrt L\,\delta_{NT}=\eta_{NT}/4,
 \end{gathered}
\]
where $\|\bm\vartheta_Q-\bm\vartheta_Q^\circ\|\leq\delta_{NT}$.
For fixed $A>0$ and $M>1/4$, set
$q_\pm=q_{D,N}\pm(M-\tfrac14)\eta_{NT}$ and
$C_1^2/2>A+p+3$. By \eqref{eq:pqr-bounded-differences},
\[
 \begin{aligned}
 &\Pp\left\{\max_{\substack{i,\bm\vartheta_Q\in\mathcal B_K\\x\in\{q_-,q_+\}}}
 |U_{i,T}(x,\bm\vartheta_Q)-\E(U_{i,T}(x,\bm\vartheta_Q)\mid\mathbf F)|
      >C_1\eta_{NT}\,\middle|\,\mathbf F\right\}\\
 &\hspace{25mm}\leq4N(CK\sqrt L)^p e^{-C_1^2L/2}
 \leq C_K(NT)^{-A}.
 \end{aligned}
\]
Define
\[
 \mathcal E_{K,NT}=\mathcal F_{NT}\cap
 \left\{\begin{gathered}
 \max_i\|\widehat{\bm\vartheta}_{Q,i}\|\leq K\eta_{NT},\\
 \max_{\substack{i,\bm\vartheta_Q\in\mathcal B_K\\x\in\{q_-,q_+\}}}
 |U_{i,T}(x,\bm\vartheta_Q)-\E\{U_{i,T}(x,\bm\vartheta_Q)\mid\mathbf F\}|
       \leq C_1\eta_{NT}
 \end{gathered}\right\}.
\]
On $\mathcal E_{K,NT}$,
\[
 U_{i,T}(x-\eta_{NT}/4,\bm\vartheta_Q^\circ)
 \leq U_{i,T}(x,\widehat{\bm\vartheta}_{Q,i})
 \leq U_{i,T}(x+\eta_{NT}/4,\bm\vartheta_Q^\circ).
\]
Take
\[
 M=\tfrac14+(C_1+2)/c_D,\qquad
 M\eta_{NT}\leq h_D,\qquad CK^2\eta_{NT}\leq1.
\]
By \eqref{eq:pqr-expectation-perturbation}, on $\mathcal E_{K,NT}$,
\[
 \begin{aligned}
 U_{i,T}(q_{D,N}-M\eta_{NT},\widehat{\bm\vartheta}_{Q,i})
 &\leq\tau_D-\{c_D(M-\tfrac14)-C_1\}\eta_{NT}
                 +CK^2\eta_{NT}^2\leq\tau_D-\eta_{NT},\\
 U_{i,T}(q_{D,N}+M\eta_{NT},\widehat{\bm\vartheta}_{Q,i})
 &\geq\tau_D+\{c_D(M-\tfrac14)-C_1\}\eta_{NT}
                 -CK^2\eta_{NT}^2\geq\tau_D+\eta_{NT}.
 \end{aligned}
\]
By \eqref{eq:pqr-quantile} and \eqref{eq:pqr-residual-difference},
\[
 \mathcal E_{K,NT}\subseteq
 \{\max_i|\widehat q_i/d_i-q_{D,N}|\leq M\eta_{NT}\}.
\]
By \eqref{eq:pqr-pilot-coefficient-rate},
\[
 \begin{aligned}
 \Pp\{\max_i|\widehat q_i/d_i-q_{D,N}|>M\eta_{NT}\}
 &\leq\Pp(\mathcal E_{K,NT}^c)\\
 &\leq
 \Pp(\mathcal F_{NT}^c)+\frac CB
 +2N(p+1)e^{-cK^2L/B}+C_K(NT)^{-A}.
 \end{aligned}
\]
Taking $K^2>2B/c$ and then $B\to\infty$ proves
\eqref{eq:pqr-quantile-rate}; moreover,
\[
 \Pp\{\min_i\widehat q_i\leq0\}
 \leq\Pp\{\max_i|\widehat q_i/d_i-q_{D,N}|>q_\ell/2\}
 \longrightarrow0.
\]
For \eqref{eq:pqr-scale-rate}, put
\[
 e_i=\frac{\widehat q_i}{q_{D,N}d_i}-1,\qquad
 \pi_{D,j}=\frac{d_j^2}{Nm_{D,N}},\qquad e_*=\max_i|e_i|.
\]
On $\{e_*\leq1/2\}$,
\[
 \begin{aligned}
 \log\frac{\widehat d_{Q,i}^{\,g}}{d_i}
 &=\log(1+e_i)-\frac1N\sum_j\log(1+e_j),\\
 \log\frac{\widehat d_{Q,i}}{d_i/\sqrt{m_{D,N}}}
 &=\log(1+e_i)-\frac12
          \log\left\{\sum_j\pi_{D,j}(1+e_j)^2\right\}.
 \end{aligned}
\]
Since $\sum_j\pi_{D,j}=1$,
\[
 (1-e_*)^2\leq\sum_j\pi_{D,j}(1+e_j)^2\leq(1+e_*)^2,\qquad
 \left|\log\sum_j\pi_{D,j}(1+e_j)^2\right|\leq4e_*.
\]
Using $|\log(1+x)|\leq2|x|$ for $|x|\leq1/2$,
\[
 \max_i\left\{
 \left|\log\frac{\widehat d_{Q,i}^{\,g}}{d_i}\right|
 +\left|\log\frac{\widehat d_{Q,i}}{d_i/\sqrt{m_{D,N}}}\right|
 \right\}
 \leq C\max_i|e_i|
 \leq C\max_i|\widehat q_i/d_i-q_{D,N}|=O_p(\eta_{NT}).
\qedhere
\]
\end{proof}

\subsection{Uniform expansion of the weighted estimating equations}
\label{app:linearization}

Condition on a design in $\mathcal F_{NT}$; all constants below are
uniform over this event. Define
\[
 \bm b=\mathbf D^{-1/2}\bm\alpha,\qquad
 \ell_t=\bm f_t^\top\mathbf Q_T^{-1}\bar{\bm f},\qquad
 h_t=1-\ell_t,\qquad
 \mathbf\Delta_T=\mathbf Q_T^{-1}
          \frac1T\sum_t\bm f_t\tau_t\bm v_t^\top.
\]
Here $\mathbf\Delta_T\in\R^{p\times N}$; its transposed rows and columns are
\[
 \bm\Delta_{T,j\cdot}=\mathbf\Delta_T^\top\bm e_j\in\R^N,\qquad
 \bm\Delta_{T,\cdot i}=\mathbf\Delta_T\bm e_i\in\R^p.
\]
The standardized projected observations satisfy
\begin{align}
 \bm x_t=\mathbf D^{-1/2}\bm Z_t
 &=\tau_t\bm v_t+h_t\bm b-\mathbf\Delta_T^\top\bm f_t,
        \label{eq:exact-projected-data}\\
 \bar{\bm x}&=T^{-1}\sum_t\bm x_t,\nonumber\\
 T^{-1}\sum_t h_t=T^{-1}\sum_t h_t^2&=\omega_T,\qquad
 T^{-1}\sum_t\ell_t=T^{-1}\sum_t\ell_t^2=\kappa_T,
        \label{eq:projection-identities}\\
 T^{-1}\sum_th_t\bm f_t&=\bm0,\qquad
 T^{-1}\sum_th_t\ell_t=0.\nonumber
\end{align}
By \eqref{eq:projection-identities},
\[
 \begin{gathered}
 T^{-1}\sum_t(h_t^2+\ell_t^2+\|\bm f_t\|^2)\leq C,\qquad
 \max_t(|h_t|+|\ell_t|+\|\bm f_t\|)\leq C\sqrt L,\\
 T^{-1}\sum_t\{h_t^4+h_t^2\|\bm f_t\|^2\}\leq CL.
 \end{gathered}
\]
For $a_t=g(\tau_t)$ and $a=\E g(\tau)$, set
\begin{equation}\label{eq:oracle-projection-quantities}
 \begin{gathered}
 \bar a_{h,T}=T^{-1}\sum_th_t^2a_t,\qquad
 \bm m_{h,T}=T^{-1}\sum_th_ta_t\bm f_t,\\
 c_t=h_tk(\tau_t)-\tau_t\bm f_t^\top\mathbf Q_T^{-1}\bm m_{h,T},
 \qquad
 \sigma_{w,T}^2=\frac{T^{-1}\sum_tc_t^2}{\bar a_{h,T}^2}.
 \end{gathered}
\end{equation}
Set $\sigma_{w,T}^2=1$ when $\bar a_{h,T}=0$ and use
\[
 \mathcal O_{w,NT}=\mathcal F_{NT}\cap
       \{\bar a_{h,T}>0,\ \textstyle\sum_tc_t^2>0\}
\]
whenever $\sigma_{w,T}$ is a divisor. The profile convention in \eqref{eq:normalized-profile} gives
\begin{equation}\label{eq:profile-scale-covariance}
 \widehat A_{h,w}^{\,Q}=\rho_Q\widehat A_{h,w}^{\,g},\qquad
 \widehat B_{c,w}^{\,Q}=\widehat B_{c,w}^{\,g},\qquad
 \widehat d_{Q,i}=\rho_Q\widehat d_{Q,i}^{\,g}.
\end{equation}
Thus $S_{cw,i}=S_{w,i}$ for $c>0$, and the trace and geometric
representatives give the same statistic and alpha estimate.

For the spherical linear forms, let the coefficients satisfy
\[
 \sigma(c_t^\circ:t\leq T)\subseteq\sigma(\bm\tau,\mathbf F),
 \qquad V_T^\circ:=T^{-1}\sum_t(c_t^\circ)^2=O_p(1).
\]
Conditional on $(\bm\tau,\mathbf F)$, the spherical directions remain independent.

\begin{lemma}[Spherical linear forms and factor projection]
\label{lem:radial-moments}
Under Assumptions~\ref{ass:radial-primitive}--\ref{ass:weight-primitive},
\begin{align}
 \max_i\left|T^{-1}\sum_tc_t^\circ v_{ti}\right|
       &=O_p(\eta_{NT}),
       \label{eq:conditional-linear-bound}\\
 \max_{j,i}|(\mathbf\Delta_T)_{ji}|+
       \|\bar{\bm x}-\omega_T\bm b\|_\infty
       &=O_p(\eta_{NT}).
       \label{eq:projection-coefficient-bound}
\end{align}
\end{lemma}

\begin{proof}[Proof of Lemma~\ref{lem:radial-moments}]
The spherical moment-generating-function bound gives, conditionally on
the radial variables and factor design,
\begin{align*}
 \E\!\left[\exp\!\left\{\frac{\lambda}{T}
                \sum_tc_t^\circ v_{ti}\right\}
                \,\middle|\,\bm\tau,\mathbf F\right]
 &=\prod_t\E\!\left(e^{\lambda c_t^\circ v_{ti}/T}
                \,\middle|\,\bm\tau,\mathbf F\right)
 \leq\exp\!\left(\frac{\lambda^2V_T^\circ}{2T}\right),\\
 \Pp\!\left(\max_i\left|T^{-1}\sum_tc_t^\circ v_{ti}\right|>x
                \,\middle|\,\bm\tau,\mathbf F\right)
 &\leq2N\inf_{\lambda>0}
       \exp\!\left(-\lambda x+\frac{\lambda^2V_T^\circ}{2T}\right)
 =2N\exp\!\left(-\frac{Tx^2}{2V_T^\circ}\right).
\end{align*}
Interpret the probability as zero when $V_T^\circ=0$. For $K,A>0$ and
$x_K=\{2K(A+1)L/T\}^{1/2}$,
\[
 \Pp\!\left(\max_i\left|T^{-1}\sum_tc_t^\circ v_{ti}\right|>x_K\right)
 \leq\Pp(V_T^\circ>K)+2N(NT)^{-A-1}.
\]
Tightness of $V_T^\circ$, followed by $K\to\infty$, proves
\eqref{eq:conditional-linear-bound}.

For the coordinate vector $\bm e_j$, put
$b_{tj}^{\Delta}=\bm e_j^\top\mathbf Q_T^{-1}\bm f_t$.
The normalized design and $\E\tau^2\leq C$ give
\begin{gather*}
 T^{-1}\sum_t(b_{tj}^{\Delta})^2=(\mathbf Q_T^{-1})_{jj}\leq C,
 \qquad
 \E\!\left\{T^{-1}\sum_t(b_{tj}^{\Delta}\tau_t)^2
                  \,\middle|\,\mathbf F\right\}\leq C,\\
 \bar{\bm x}-\omega_T\bm b=T^{-1}\sum_th_t\tau_t\bm v_t,
 \qquad
 \E\!\left\{T^{-1}\sum_t(h_t\tau_t)^2
                  \,\middle|\,\mathbf F\right\}
       =\omega_T\E\tau^2\leq C.
\end{gather*}
Conditional Markov inequalities and a union bound over the fixed $p$ give
\[
 \Pp\!\left(\max_{j,i}|(\mathbf\Delta_T)_{ji}|
       +\|\bar{\bm x}-\omega_T\bm b\|_\infty
           >C_A\sqrt K\,\eta_{NT}\,\middle|\,\mathbf F\right)
 \leq\frac{C}{K}+2N(p+1)(NT)^{-A-1}.
\]
Letting $K\to\infty$ establishes
\eqref{eq:projection-coefficient-bound}.
\end{proof}

For the radial variance, define
\begin{equation}\label{eq:radial-tail-envelope}
 \begin{gathered}
 X_N=\tau_N^2+\tau_N^{-2},\qquad
 \Psi(K)=\sup_N\E\{X_N\ind(X_N>K)\},\quad K>0,\\
 \beta_{NT}(K)=\sqrt{\frac{KL}{T}}+\Psi(K).
 \end{gathered}
\end{equation}
The radial assumptions give $\sup_N\E X_N<\infty$ and $\Psi(K)\to0$, since
$\E\{\tau_N^{-2}\ind(\tau_N^{-2}>K)\}\leq K^{-1}\E\tau_N^{-4}$.
These bounds are uniform over admissible designs and profiles.

\begin{lemma}[Oracle sensitivity and variance]\label{lem:oracle-variance}
Under Assumptions~\ref{ass:radial-primitive}--\ref{ass:weight-primitive},
$c\leq\mathcal V_{w,T}\leq C$ and
\begin{equation}\label{eq:oracle-sensitivity-rate}
 |\bar a_{h,T}-a\omega_T|+\|\bm m_{h,T}\|=O_p(\eta_{NT}).
\end{equation}
On $\{\bar a_{h,T}>0\}$, the conditional variance satisfies
\begin{equation}\label{eq:oracle-conditional-variance-lower}
 \sigma_{w,T}^2
 \geq\left(\frac1T\sum_t\frac{h_t^2}{\tau_t^2}\right)^{-1}.
\end{equation}
For any deterministic $K=K_{NT}\to\infty$ with $KL/T\to0$,
\begin{equation}\label{eq:oracle-population-variance-rate}
 |\sigma_{w,T}^2-\mathcal V_{w,T}|
   =O_p\{\eta_{NT}+\beta_{NT}(K)\}=o_p(1).
\end{equation}
Moreover, $\Pp(\mathcal O_{w,NT})\to1$, and $\sigma_{w,T}^{-1}=O_p(1)$.
\end{lemma}

\begin{proof}[Proof of Lemma~\ref{lem:oracle-variance}]
The profile bound gives
\[
 \Var\{g(\tau)\}\leq C(1+\E\tau^{-4})\leq C.
\]
Independence over time and $\sum_th_t\bm f_t=\bm0$ give
\begin{gather*}
 \E(\bar a_{h,T}-a\omega_T)^2
   =\frac{\Var\{g(\tau)\}}{T^2}\sum_th_t^4\leq CL/T,\\
 \bm m_{h,T}=\frac1T\sum_th_t\{g(\tau_t)-a\}\bm f_t,\qquad
 \E\|\bm m_{h,T}\|^2
   =\frac{\Var\{g(\tau)\}}{T^2}
            \sum_th_t^2\|\bm f_t\|^2\leq CL/T.
\end{gather*}
Thus \eqref{eq:oracle-sensitivity-rate} holds and
$\Pp(\bar a_{h,T}>0)\to1$.

By \eqref{eq:oracle-projection-quantities} and
$\sum_th_t\bm f_t=\bm0$,
\[
 \frac1T\sum_t\frac{h_tc_t}{\tau_t}
 =\bar a_{h,T}
   -\left(\frac1T\sum_th_t\bm f_t\right)^\top
                   \mathbf Q_T^{-1}\bm m_{h,T}
 =\bar a_{h,T}.
\]
Cauchy--Schwarz therefore yields
\[
 \bar a_{h,T}^{\,2}
 \leq
 \left(\frac1T\sum_t\frac{h_t^2}{\tau_t^2}\right)
 \left(\frac1T\sum_tc_t^2\right),
\]
The first factor has mean $\omega_T\E\tau^{-2}\leq C$ and variance
$O(L/T)$. Hence \eqref{eq:oracle-conditional-variance-lower},
$\sigma_{w,T}^{-1}=O_p(1)$ and $\Pp(\mathcal O_{w,NT})\to1$ follow.

Put
\[
 c_t^0=h_tk(\tau_t),\qquad
 d_t^c=\tau_t\bm f_t^\top\mathbf Q_T^{-1}\bm m_{h,T},\qquad
 c_t=c_t^0-d_t^c.
\]
Conditional Markov inequalities and
$k(\tau)^2\leq C(\tau^2+\tau^{-2})$ give
\begin{gather*}
 \frac1T\sum_t(c_t^0)^2=O_p(1),\qquad
 \frac1T\sum_t\tau_t^2\|\bm f_t\|^2=O_p(1),\\
 \frac1T\sum_t(d_t^c)^2
 \leq C\|\bm m_{h,T}\|^2\frac1T\sum_t\tau_t^2\|\bm f_t\|^2
 =O_p(L/T),\\
 \left|\frac1T\sum_t\{c_t^2-(c_t^0)^2\}\right|
 \leq2\left\{\frac1T\sum_t(c_t^0)^2\right\}^{1/2}
          \left\{\frac1T\sum_t(d_t^c)^2\right\}^{1/2}
                +\frac1T\sum_t(d_t^c)^2
 =O_p(\eta_{NT}).
\end{gather*}
On $\{|\bar a_{h,T}-a\omega_T|\leq a\omega_T/2\}$,
\[
 |\bar a_{h,T}^{-2}-(a\omega_T)^{-2}|
 \leq C|\bar a_{h,T}-a\omega_T|=O_p(\eta_{NT}),
\]
and consequently
\begin{equation}\label{eq:oracle-variance-reduction}
 \sigma_{w,T}^2
 =\frac{T^{-1}\sum_th_t^2k(\tau_t)^2}{a^2\omega_T^2}
                                    +O_p(\eta_{NT}).
\end{equation}

Conditionally on $\mathbf F$, let $Y_t=k(\tau_t)^2$,
$X_t=\tau_t^2+\tau_t^{-2}$ and $Y_t^{(K)}=Y_t\ind(X_t\leq K)$.
Uniformly on $\mathcal F_{NT}$,
\[
 \E[(Y_t^{(K)})^2]
 \leq C\E\{X_t^2\ind(X_t\leq K)\}\leq CK,
 \qquad
 \E\{Y_t\ind(X_t>K)\}\leq C\Psi(K).
\]
Independence and
$T^{-2}\sum_th_t^4\leq CL/T$ give
\[
 \Var\!\left\{\frac1T\sum_th_t^2
              (Y_t^{(K)}-\E Y_t^{(K)})\,\middle|\,\mathbf F\right\}
 \leq C\frac{KL}{T}.
\]
The complementary centered average has expected absolute value at most
$C\Psi(K)$. Chebyshev's and Markov's inequalities yield
\begin{equation}\label{eq:weighted-radial-ui-rate}
 \frac1T\sum_th_t^2\{k(\tau_t)^2-\E k(\tau)^2\}
   =O_p\!\left\{\sqrt{\frac{KL}{T}}+\Psi(K)\right\}.
\end{equation}
Combining \eqref{eq:oracle-variance-reduction} and
\eqref{eq:weighted-radial-ui-rate} proves
\eqref{eq:oracle-population-variance-rate}, uniformly over profiles fixed
conditional on $\mathbf F$.

Finally, Cauchy--Schwarz and $a\geq c$ yield
\begin{equation}\label{eq:variance-lower-bound}
 a^2\leq\E k(\tau)^2\,\E\tau^{-2},\qquad
 0<c\leq\frac1{\omega_T\E\tau^{-2}}
       \leq\mathcal V_{w,T}
       =\frac{\E k(\tau)^2}{a^2\omega_T}\leq C.
\end{equation}
The upper bound follows from $\E k(\tau)^2\leq C\E X_N$ and
$a,\omega_T\geq c$.
\end{proof}

For $\bm\delta\in\R^N$, define
\begin{align}
 \mathbf\Lambda_{\bm\delta}
    &=\diag(e^{-\delta_1},\ldots,e^{-\delta_N}),&
 s(\bm x,\bm\delta)
    &=N^{-1/2}\|\mathbf\Lambda_{\bm\delta}\bm x\|,\nonumber\\
 \bm\phi(\bm x,\bm\delta)
    &=g\{s(\bm x,\bm\delta)\}\bm x.
 \label{eq:joint-score-definitions}
\end{align}
By Lemma~\ref{lem:pqr-scale}, the PQR perturbation satisfies
\[
 \widehat\delta_{Q,i}
     =\log(\widehat d_{Q,i}^{\,g}/d_i),\qquad
 \|\widehat{\bm\delta}_Q\|_\infty=O_p(\eta_{NT}).
\]
Define
\[
 \mathbf J=\partial_{\bm x}\bm\phi,\qquad
 \mathbf H=\partial_{\bm\delta}\bm\phi.
\]
For $\mathbf M\in\{\mathbf J,\mathbf H\}$, let
$\mathbf M_t=\mathbf M(\bm e_t,\bm0)$, where
$\bm e_t=\tau_t\bm v_t$ and $s_t=\|\bm e_t\|/\sqrt N$. Thus
\begin{align}
 \mathbf J_t=\partial_{\bm x}\bm\phi
    &=g(s_t)\mathbf I_N+
       \frac{g'(s_t)}{Ns_t}\bm e_t\bm e_t^\top,\nonumber\\
 H_{t,ij}=(\partial_{\bm\delta}\bm\phi)_{ij}
    &=-\frac{g'(s_t)}{Ns_t}e_{ti}e_{tj}^{2}.
 \label{eq:four-jacobians}
\end{align}
Define the finite collection of design weights and the derivative rate by
\[
 \begin{gathered}
 \mathcal B_T=\{(h_t),(h_t^2),(h_tf_{tj}):1\leq j\leq p\},\\
 j_{NT}=\frac{L^3}{\sqrt T}+L^3\sqrt{\frac{b_N}N}.
 \end{gathered}
\]
These weights satisfy, uniformly on $\mathcal F_{NT}$,
\[
 \max_{b\in\mathcal B_T}T^{-1}\sum_t|b_t|\leq C,\qquad
 \max_{t,b\in\mathcal B_T}|b_t|\leq CL,\qquad
 \max_{b\in\mathcal B_T}T^{-1}\sum_tb_t^2\leq CL.
\]
Collect the empirical derivative errors as
\[
 \mathfrak J_{NT}
 =\max_{b\in\mathcal B_T}
   \left\{\left\|T^{-1}\sum_tb_t(\mathbf J_t-a_t\mathbf I_N)
                                              \right\|_\infty
       +\left\|T^{-1}\sum_tb_t\mathbf H_t\right\|_\infty\right\}.
\]
The corresponding unperturbed score error is
\[
 \mathfrak P_{NT}
   =\max_{b\in\mathcal B_T}\left\|T^{-1}\sum_tb_t
         \{\bm\phi(\bm e_t,\bm0)-k(\tau_t)\bm v_t\}\right\|_\infty.
\]
The radial curvature envelope is retained inside the time average:
\begin{equation}\label{eq:average-curvature-envelope}
 \mathfrak M_{NT}
 =\max_{b\in\mathcal B_T}T^{-1}\sum_t|b_t|(1+\tau_t^{-3}),
 \qquad
 \Pp(\mathfrak M_{NT}>K\mid\mathbf F)\leq C/K.
\end{equation}
Since $\E\tau^{-3}<\infty$ and
$\max_bT^{-1}\sum_t|b_t|\leq C$, $\mathfrak M_{NT}=O_p(1)$.

For $C,c>0$, let $\mathcal D_{NT}(C,c)$ be the event
\[
 \sup_{\substack{t\leq T,\ 0<\|\bm u\|_\infty+\|\bm\delta\|_\infty
                                       \leq cT^{-1/4}\\
                 \mathbf M\in\{\mathbf J,\mathbf H\}}}
 \frac{\|\mathbf M(\bm e_t+\bm u,\bm\delta)-\mathbf M_t\|_\infty}
 {(1+\tau_t^{-3})(\|\bm u\|_\infty+\|\bm\delta\|_\infty)}
 \leq CL.
\]

\begin{lemma}[Empirical Jacobians and local derivatives]
\label{lem:jacobians}
Under Assumptions~\ref{ass:radial-primitive}--\ref{ass:weight-primitive},
\begin{align}
 \mathfrak J_{NT}&=O_p(j_{NT}),
          \label{eq:empirical-jacobian-bounds}\\
 \mathfrak P_{NT}&=O_p\left(L^3\sqrt{\frac{b_N}{NT}}\right).
          \label{eq:true-radial-replacement}
\end{align}
For each fixed $A>0$, some $C_A,c_A>0$ satisfy
\[
 \Pp\{\mathcal D_{NT}(C_A,c_A)^c\mid\mathbf F\}
 \leq C_A(NT)^{-A}+\zeta_{\tau,T}=o(1)
 \quad\text{on }\mathcal F_{NT}.
\]
\end{lemma}

\Needspace{3\baselineskip}
\begin{proof}[Proof of Lemma~\ref{lem:jacobians}]
Replace $\varpi_t$ by one in \eqref{eq:four-jacobians} and denote
the resulting matrices by a superscript zero:
\begin{align*}
 J_{t,ij}^0&=a_t\ind\{i=j\}
                +N^{-1}\tau_tg'(\tau_t)v_{ti}v_{tj},\\
 H_{t,ij}^0&=-N^{-1}\tau_t^2g'(\tau_t)v_{ti}v_{tj}^2.
\end{align*}
For $\varpi\in[1/2,2]$, the profile conditions imply
\begin{align*}
 |g(\tau\varpi)-g(\tau)|
   &\leq C(\tau^{-1}+\tau^{-2})|\varpi-1|,\\
 |\tau g'(\tau\varpi)/\varpi-\tau g'(\tau)|
   &\leq C(\tau^{-1}+\tau^{-2})|\varpi-1|,\\
 |\tau^2g'(\tau\varpi)/\varpi-\tau^2g'(\tau)|
   &\leq C(1+\tau^{-1})|\varpi-1|.
\end{align*}
Moreover,
$N^{-1}\sum_j|v_{tj}|\leq\varpi_t$ and
$N^{-1}\sum_jv_{tj}^2=\varpi_t^2$. Hence, on
$\mathcal A_{NT}^{\mathrm{sph}}(K_A)$,
\[
 \|\mathbf M_t-\mathbf M_t^0\|_\infty
 \leq C_AL(1+\tau_t^{-2})|\varpi_t-1|,
 \qquad \mathbf M\in\{\mathbf J,\mathbf H\}.
\]
Averaging before bounding the radial multiplier gives
\[
 \max_{b\in\mathcal B_T}
 \left\|T^{-1}\sum_tb_t(\mathbf M_t-\mathbf M_t^0)\right\|_\infty
 \leq C_AL\mathfrak M_{NT}\max_t|\varpi_t-1|
 =O_p\left(L^{3/2}\sqrt{\frac{b_N}N}\right).
\]

It suffices to control the entries of $\mathbf J_t^0-a_t\mathbf I_N$ and
$\mathbf H_t^0$ through
\[
 \tau_tg'(\tau_t)v_{ti}v_{tj},\quad
 \tau_t^2g'(\tau_t)v_{ti}v_{tj}^2.
\]
Write either component as $Y_{t,ij}$. Its second moment is uniformly
bounded by $\E\tau^{-4}\leq C$ and the spherical moment bounds.
For fixed large $K_A$, define
\[
 \mathcal E_t=
 \left\{\begin{gathered}
   \tau_t^{-1}\leq T^{1/4},\qquad \max_i|v_{ti}|\leq K_A\sqrt L,\\
   |\varpi_t-1|\leq K_A\sqrt{b_NL/N}
 \end{gathered}\right\}.
\]
Since $L/\sqrt T\to0$, the profile bounds imply
\[
 \sup_{t,i,j}\E Y_{t,ij}^2\leq C,\qquad
 |Y_{t,ij}|\ind_{\mathcal E_t}\leq C_A\sqrt T\,L.
\]
Uniform integrability gives
\[
 \begin{aligned}
 \E\{(1+\tau^{-2})\ind(\tau^{-1}>T^{1/4})\}
 &\leq2T^{-1/2}
        \E\{\tau^{-4}\ind(\tau^{-1}>T^{1/4})\}\\
 &\leq2T^{-1/2}\zeta_{\tau,T}=o(T^{-1/2}).
 \end{aligned}
\]
Cauchy--Schwarz and sufficiently large $K_A$ yield
\[
 \left|\E(Y_{t,ij}\ind_{\mathcal E_t})-\E Y_{t,ij}\right|
 \leq C_AT^{-1/2}\zeta_{\tau,T}+C_A(NT)^{-A-3}.
\]
For $\widetilde Y_{t,ij}
=Y_{t,ij}\ind_{\mathcal E_t}-\E(Y_{t,ij}\ind_{\mathcal E_t})$
and $X_t=b_t\widetilde Y_{t,ij}$, conditional independence gives
\[
 \E X_t=0,\qquad
 \sum_t\E X_t^2\leq CTL,\qquad
 |X_t|\leq C_A\sqrt T\,L^2.
\]
For centered summands with variance sum $V$ and bound $M$, the
exponential series gives
\[
 \log\E e^{\lambda\sum_tX_t}
 \leq\frac{\lambda^2V}{2(1-|\lambda|M/3)},\qquad |\lambda|M<3,
\]
Thus $\Pp(|\sum_tX_t|>\sqrt{2Vx}+Mx/3)\leq2e^{-x}$.
With $x=C_AL$, a union bound over at most $CN^2$ entries and fixed weights gives
\[
 \max_{i,j,b\in\mathcal B_T}
 \left|T^{-1}\sum_tb_t(Y_{t,ij}-\E Y_{t,ij})\right|
 =O_p(L^3/\sqrt T).
\]
The truncation event fails with probability at most
$\zeta_{\tau,T}+C_A(NT)^{-A}=o(1)$. Restoring the $N^{-1}$ factors
gives the same rate in maximum absolute row-sum norm.

By sign symmetry, $\E\mathbf H_t^0=\mathbf0$, whereas
\[
 \E(\mathbf J_t^0-a_t\mathbf I_N)
       =N^{-1}\E\{\tau g'(\tau)\}\mathbf R_N.
\]
Since $(\mathbf R_N^2)_{ii}\leq b_N$,
\[
 \max_i\sum_j|R_{N,ij}|
 \leq\sqrt{N\max_i(\mathbf R_N^2)_{ii}}\leq\sqrt{Nb_N},
 \qquad
 \|\E(\mathbf J_t^0-a_t\mathbf I_N)\|_\infty
 \leq C\sqrt{b_N/N}.
\]
Combining these estimates proves
\eqref{eq:empirical-jacobian-bounds}.

For the radial replacement, write
\[
 D_{ti}=\tau_t\{g(\tau_t\varpi_t)-g(\tau_t)\}v_{ti}.
\]
Since $\mathcal E_t$ is sign invariant and $D_{ti}$ is odd,
\[
 \E(D_{ti}\ind_{\mathcal E_t}\mid\tau_t,\mathbf F)=0.
\]
Let $\rho_{N,A}=K_A\sqrt{b_NL/N}$. On $\mathcal E_t$,
\[
 |D_{ti}|\leq C\rho_{N,A}(1+\tau_t^{-1})|v_{ti}|,\qquad
 |D_{ti}|\ind_{\mathcal E_t}
     \leq C_AT^{1/4}L\sqrt{b_N/N}.
\]
Independence of $\tau_t$ and $\bm v_t$ and
$\E\tau^{-2}<\infty$ imply
\[
 \E(D_{ti}^2\ind_{\mathcal E_t}\mid\mathbf F)
 \leq C\rho_{N,A}^2\leq C_A b_NL/N.
\]
Apply Bernstein's inequality and a union bound to
$b_tD_{ti}\ind_{\mathcal E_t}$, using variance sum $C_ATb_NL^2/N$
and summand bound $C_AT^{1/4}L^2\sqrt{b_N/N}$:
\[
 \mathfrak P_{NT}
 =O_p\left\{
 L^{3/2}\sqrt{\frac{b_N}{NT}}
 +\frac{L^3}{T^{3/4}}\sqrt{\frac{b_N}N}\right\}
 =O_p\left(L^3\sqrt{\frac{b_N}{NT}}\right).
\]
Exact centering eliminates truncation bias, proving
\eqref{eq:true-radial-replacement}.

On $\mathcal A_{NT}^{\mathrm{sph}}(K_A)\cap\mathcal R_{\tau,T}$,
take $\|\bm u\|_\infty+\|\bm\delta\|_\infty\leq r\leq c_AT^{-1/4}$.
For sufficiently small $c_A$,
\[
 \begin{gathered}
 e^{-r}(\tau_t\varpi_t-\|\bm u\|_\infty)
 \leq s(\bm e_t+\bm u,\bm\delta)
 \leq e^r(\tau_t\varpi_t+\|\bm u\|_\infty),\\
 r/\tau_t\leq c_A,\qquad
 c\tau_t\leq s(\bm e_t+\bm u,\bm\delta)\leq C\tau_t,\\
 \|\bm e_t+\bm u\|_\infty\leq C_A\tau_t\sqrt L,\qquad
 N^{-1/2}\|\bm e_t+\bm u\|\leq C\tau_t.
 \end{gathered}
\]
Within this neighborhood, differentiate along
$(\bm x+\lambda\bm a,\bm\delta+\lambda\bm d)$ for
$\bm\xi=(\bm a,\bm d)$, $\|\bm\xi\|_\infty\leq1$. With
$\bm y=\mathbf\Lambda_{\bm\delta}\bm x$ and $s=\|\bm y\|/\sqrt N$,
\begin{align*}
 \bm y'&=\mathbf\Lambda_{\bm\delta}(\bm a-\bm d\odot\bm x),&
 \bm y''&=\mathbf\Lambda_{\bm\delta}
                  (\bm d^{\odot2}\odot\bm x-2\bm d\odot\bm a),\\
 s'&=\frac{\bm y^\top\bm y'}{Ns},&
 s''&=\frac{\|\bm y'\|^2+\bm y^\top\bm y''}{Ns}
                         -\frac{(s')^2}{s}.
\end{align*}
The preceding radial bounds yield
\[
 \begin{gathered}
 \max_i|y_i|\leq C_A\tau_t\sqrt L,\qquad
 \max_i(|y_i'|+|y_i''|)\leq C_A(1+\tau_t\sqrt L),\\
 N^{-1/2}(\|\bm y'\|+\|\bm y''\|)\leq C(1+\tau_t),\qquad
 |s'|\leq C(1+\tau_t),\qquad
 |s''|\leq C(1+\tau_t+\tau_t^{-1}).
 \end{gathered}
\]
Differentiating the location score gives
\[
 \phi_i''
 =g''(s)(s')^2x_i+g'(s)s''x_i+2g'(s)s'a_i.
\]
The profile derivative bounds imply
\[
 \max_i|\phi_i''|\leq C_A\sqrt L(1+\tau_t^{-3}).
\]
Polarization and homogeneity convert these bounds for equal
directions into the bilinear bound
\[
 \|D^2\bm\phi[\bm\xi,\bm\zeta]\|_\infty
 \leq C_AL(1+\tau_t^{-3})
             \|\bm\xi\|_\infty\|\bm\zeta\|_\infty.
\]
Integration along the segment from $(\bm e_t,\bm0)$ to
$(\bm e_t+\bm u,\bm\delta)$ proves
\[
 \mathcal A_{NT}^{\mathrm{sph}}(K_A)\cap\mathcal R_{\tau,T}
       \subseteq\mathcal D_{NT}(C_A,c_A),
\]
and hence the stated probability bound.

Taylor's integral remainder $\bm{\mathcal R}_t$ satisfies
\begin{equation}\label{eq:joint-taylor-remainder}
 \max_{b\in\mathcal B_T}
 \left\|T^{-1}\sum_tb_t\bm{\mathcal R}_t\right\|_\infty
 \leq C_AL\mathfrak M_{NT}
       \left(\max_t\|\bm u_t\|_\infty+\|\bm\delta\|_\infty\right)^2,
\end{equation}
for perturbations in this neighborhood, with
$\mathfrak M_{NT}=O_p(1)$ by $\E\tau^{-3}<\infty$.
\end{proof}

To identify the statistical solution, put
\[
 r_0=\frac{L^2}{\sqrt T},\qquad
 \mathcal B_{NT}^0=\{\bm u:\|\bm u\|_\infty\leq r_0\},\qquad
 \mathcal B_{NT}^{Q}=\bm b+\mathcal B_{NT}^0.
\]
The growth conditions imply
\[
 \frac{\eta_{NT}}{r_0}\to0,\qquad
 T^{1/4}\sqrt L\,r_0\to0,\qquad
 \frac{e_{NT}}{\eta_{NT}}
 =\frac{L^{9/2}}{\sqrt T}
        +L^{7/2}\sqrt{\frac{b_N}{N}}\to0.
\]
Center the projected observations as
\[
 \bm y_t=\bm x_t-h_t\bm b
        =\bm e_t-\mathbf\Delta_T^\top\bm f_t,
\]
and define
\begin{align*}
 \bm L_{h,T}(\bm z,\bm\delta)
 &=T^{-1}\sum_t h_t\bm\phi(\bm x_t-h_t\bm z,\bm\delta),\\
 \bm L_{h,T}^c(\bm u,\bm\delta)
 &=\bm L_{h,T}(\bm b+\bm u,\bm\delta)
   =T^{-1}\sum_t h_t\bm\phi(\bm y_t-h_t\bm u,\bm\delta).
\end{align*}
The centered observations and the PQR scale are invariant to shifts in
the alpha vector.

\begin{lemma}[Local root with fixed PQR scale]\label{lem:local-root}
Under Assumptions~\ref{ass:radial-primitive}--\ref{ass:weight-primitive},
with probability tending to one, the estimating equation
$\bm L_{h,T}(\bm z,\widehat{\bm\delta}_Q)=\bm0$ has a unique root
$\widehat{\bm z}$ in $\mathcal B_{NT}^{Q}$. Its fitted radii are positive,
and $\|\widehat{\bm z}-\bm b\|_\infty=O_p(\eta_{NT})$ uniformly over
deterministic $\bm\alpha$.
\end{lemma}

\begin{proof}[Proof of Lemma~\ref{lem:local-root}]
At the centered origin, the combined residual and scale perturbation is
\[
 U_{0,NT}
 =\max_t\|\mathbf\Delta_T^\top\bm f_t\|_\infty
       +\|\widehat{\bm\delta}_Q\|_\infty
 =O_p(L/\sqrt T),
\]
by \eqref{eq:projection-coefficient-bound} and
\eqref{eq:pqr-scale-rate}. Taylor's formula gives
\begin{align*}
 \bm L_{h,T}^c(\bm0,\widehat{\bm\delta}_Q)
 &=T^{-1}\sum_t h_tk(\tau_t)\bm v_t
       -\mathbf\Delta_T^\top\bm m_{h,T}+\bm R_{0,T},\\
 \|\bm R_{0,T}\|_\infty
 &=O_p\left\{L^3\sqrt{\frac{b_N}{NT}}
                   +j_{NT}\eta_{NT}+\frac{L^3}{T}\right\}
 =O_p(e_{NT}).
\end{align*}
Using \eqref{eq:conditional-linear-bound},
$\|\mathbf\Delta_T^\top\bm m_{h,T}\|_\infty=O_p(\eta_{NT}^2)$, and
$e_{NT}=o(\eta_{NT})$,
\begin{equation}\label{eq:center-score-order}
 \|\bm L_{h,T}^c(\bm0,\widehat{\bm\delta}_Q)\|_\infty
 =O_p(\eta_{NT}).
\end{equation}

Choose $C_F$ with $\max_t|h_t|\leq C_F\sqrt L$.
Since $C_F\sqrt L\,r_0+U_{0,NT}=o_p(T^{-1/4})$, the local bounds hold
uniformly on $\mathcal B_{NT}^0$ with probability $1-o(1)$:
\[
 \inf_{\bm u\in\mathcal B_{NT}^0}
 s(\bm y_t-h_t\bm u,\widehat{\bm\delta}_Q)\geq c\tau_t>0
 \qquad(t\leq T)
\]
and
\begin{align*}
 J_{L,T}
 &:=\sup_{\bm u\in\mathcal B_{NT}^0}
 \|\partial_{\bm u}\bm L_{h,T}^c
       (\bm u,\widehat{\bm\delta}_Q)+a\omega_T\mathbf I_N\|_\infty\\
 &\leq\mathfrak J_{NT}+|\bar a_{h,T}-a\omega_T|
       +CL\mathfrak M_{NT}(C_F\sqrt L\,r_0+U_{0,NT})\\
 &=O_p\{j_{NT}+L^{3/2}r_0+L^2/\sqrt T\}=o_p(1).
\end{align*}
Let $\mathcal E_{NT}^{\mathrm{loc}}$ be the intersection of
$\mathcal F_{NT}\cap\{\min_i\widehat q_i>0\}$ and this radial event with
\[
 J_{L,T}\leq a\omega_T/4,\qquad
 \|\bm L_{h,T}^c(\bm0,\widehat{\bm\delta}_Q)\|_\infty
       \leq a\omega_T r_0/4.
\]
On $\mathcal E_{NT}^{\mathrm{loc}}$, which has probability tending to one,
Jacobian integration gives
\[
 \bm L_{h,T}^c(\bm u,\widehat{\bm\delta}_Q)
 =\bm L_{h,T}^c(\bm0,\widehat{\bm\delta}_Q)
       -a\omega_T\bm u+\bm R_T^L(\bm u),\qquad
 \|\bm R_T^L(\bm u)\|_\infty\leq J_{L,T}\|\bm u\|_\infty.
\]
Each coordinate score has opposite signs on its corresponding faces.
Brouwer's theorem gives a zero $\widehat{\bm u}$. Any two zeros satisfy
\[
 a\omega_T\|\bm u-\bm u'\|_\infty
 \leq J_{L,T}\|\bm u-\bm u'\|_\infty,
\]
so they coincide. At this unique zero,
\[
 (a\omega_T-J_{L,T})\|\widehat{\bm u}\|_\infty
 \leq\|\bm L_{h,T}^c(\bm0,\widehat{\bm\delta}_Q)\|_\infty,
\]
Hence $\widehat{\bm z}=\bm b+\widehat{\bm u}$ and
$\|\widehat{\bm z}-\bm b\|_\infty=O_p(\eta_{NT})$,
uniformly over deterministic alpha vectors.

The local zero admits an observable selection. Define
\[
 \widehat{\mathcal C}_{NT}
 =\left\{\bm a\in\R^N:
    \left\|(\widehat{\mathbf D}_Q^{\,g})^{-1/2}
       (\bm a-\widehat{\bm\alpha}_{\LS})\right\|_\infty
          \leq r_0/2\right\},
\]
Since $\mathbf D^{-1/2}\widehat{\bm\alpha}_{\LS}
=\bar{\bm x}/\omega_T$, \eqref{eq:projection-coefficient-bound} gives
\[
 \|\mathbf D^{-1/2}\widehat{\bm\alpha}_{\LS}-\bm b\|_\infty
       =O_p(\eta_{NT}),\qquad
 \|(\widehat{\mathbf D}_Q^{\,g})^{-1/2}
       (\mathbf D^{1/2}\widehat{\bm z}
                    -\widehat{\bm\alpha}_{\LS})\|_\infty
       =O_p(\eta_{NT})=o_p(r_0).
\]
Moreover,
\[
 \sup_{\bm a\in\widehat{\mathcal C}_{NT}}
       \|\mathbf D^{-1/2}\bm a-\bm b\|_\infty
 \leq
 \|\mathbf D^{-1/2}\widehat{\bm\alpha}_{\LS}-\bm b\|_\infty
       +\tfrac12r_0e^{\|\widehat{\bm\delta}_Q\|_\infty}
 =\{1/2+o_p(1)\}r_0.
\]
Consequently,
\[
 \begin{gathered}
 \mathcal E_{NT}^{Q}
 =\mathcal E_{NT}^{\mathrm{loc}}\cap
       \left\{\mathbf D^{1/2}\widehat{\bm z}
       \in\widehat{\mathcal C}_{NT}
       \subseteq\mathbf D^{1/2}\mathcal B_{NT}^{Q}\right\},\\
 \Pp(\mathcal E_{NT}^{Q})\longrightarrow1.
 \end{gathered}
\]
Select the unique zero if it exists and all radii are positive on the cube.
Compactness and continuity give a measurable selection with
\[
 \begin{gathered}
 \widehat{\bm\alpha}_w=\mathbf D^{1/2}\widehat{\bm z}
       \quad\text{on }\mathcal E_{NT}^{Q},\\
 \mathbf Y^c=\mathbf Y+\one_T\bm c^\top
 \quad\Longrightarrow\quad
 \widehat{\mathcal C}_{NT}(\mathbf Y^c)
   =\widehat{\mathcal C}_{NT}(\mathbf Y)+\bm c,\\
 \widehat{\bm\alpha}_w(\mathbf Y^c)
   =\widehat{\bm\alpha}_w(\mathbf Y)+\bm c
       \quad\text{on a defined fit}.
 \end{gathered}
\]
The OLS center translates, while PQR scales and residual scores are
invariant; hence fitted residuals are unchanged.
\end{proof}

\begin{lemma}[Uniform centered alpha expansion]\label{lem:weighted-linearization}
Under Assumptions~\ref{ass:radial-primitive}--\ref{ass:weight-primitive},
the local root in Lemma~\ref{lem:local-root} satisfies the following bounds uniformly
over deterministic alpha vectors:
\begin{equation}\label{eq:root-rates}
 \|\widehat{\bm z}-\bm b\|_\infty
       +\|\widehat{\bm\delta}_Q\|_\infty=O_p(\eta_{NT}),\qquad
 \widehat{\bm z}=\mathbf D^{-1/2}\widehat{\bm\alpha}_w.
\end{equation}
On an extension, some $\bm G_w\mid(\bm\tau,\mathbf F)
\sim N(\bm0,\mathbf R_N)$ satisfies
\begin{equation}\label{eq:weighted-bahadur}
 \widehat{\bm z}-\bm b
 =\frac{\sigma_{w,T}}{\sqrt T}\bm G_w+\bm\epsilon_{w,NT},
 \qquad\|\bm\epsilon_{w,NT}\|_\infty=O_p(e_{NT}).
\end{equation}
\end{lemma}

\begin{proof}[Proof of Lemma~\ref{lem:weighted-linearization}]
Lemma~\ref{lem:local-root} and the PQR bound give
\eqref{eq:root-rates}. Write
$\widehat{\bm u}=\widehat{\bm z}-\bm b$ and
\[
 \bm u_t=-h_t\widehat{\bm u}-\mathbf\Delta_T^\top\bm f_t,\qquad
 \max_t\|\bm u_t\|_\infty=O_p(L/\sqrt T).
\]
The centered score equation and Taylor's formula yield
\begin{align}
 \bm0={}&T^{-1}\sum_th_tk(\tau_t)\bm v_t
       -\bar a_{h,T}\widehat{\bm u}
       -\mathbf\Delta_T^\top\bm m_{h,T}+\bm R_T,\nonumber\\
 \|\bm R_T\|_\infty
 ={}&O_p\left\{L^3\sqrt{\frac{b_N}{NT}}
                     +j_{NT}\eta_{NT}+\frac{L^3}T\right\}
 =O_p(e_{NT}).
 \label{eq:expanded-projected-score}
\end{align}
Define
$\mathbf J_T[b]=T^{-1}\sum_tb_t(\mathbf J_t-a_t\mathbf I_N)$.
The remainder is
\[
 \begin{aligned}
 \bm R_T
 &=T^{-1}\sum_th_t\{\bm\phi(\bm e_t,\bm0)-k(\tau_t)\bm v_t\}
       -\mathbf J_T[h^2]\widehat{\bm u}\\
 &\quad-\sum_{j=1}^p\mathbf J_T[h f_j]\bm\Delta_{T,j\cdot}
       +\left(T^{-1}\sum_th_t\mathbf H_t\right)
                               \widehat{\bm\delta}_Q
       +T^{-1}\sum_th_t\bm{\mathcal R}_t.
 \end{aligned}
\]
The pathwise matrix norm inequality and
\eqref{eq:joint-taylor-remainder} give
\[
 \begin{aligned}
 \|\bm R_T\|_\infty
 &\leq\mathfrak P_{NT}
       +\mathfrak J_{NT}
        \left(\|\widehat{\bm u}\|_\infty
          +\sum_j\|\bm\Delta_{T,j\cdot}\|_\infty
          +\|\widehat{\bm\delta}_Q\|_\infty\right)\\
 &\quad+C_AL\mathfrak M_{NT}
       \left(\max_t\|\bm u_t\|_\infty
                              +\|\widehat{\bm\delta}_Q\|_\infty\right)^2.
 \end{aligned}
\]

\Needspace{3\baselineskip}
Define the projection-adjusted oracle location
\begin{equation}\label{eq:oracle-location-exact}
 \bm u_w^o=\frac1{T\bar a_{h,T}}\sum_tc_t\bm v_t.
\end{equation}
Indeed,
\[
 \mathbf\Delta_T^\top\bm m_{h,T}
 =T^{-1}\sum_t\tau_t\bm v_t\bm f_t^\top
                           \mathbf Q_T^{-1}\bm m_{h,T},\qquad
 \widehat{\bm u}-\bm u_w^o
 =\bar a_{h,T}^{-1}\bm R_T=O_p(e_{NT})
       \quad\text{in }\|\cdot\|_\infty.
\]
Adjoin independent $\chi_t\sim\chi_N$, independent of the original
variables, and set
\[
 s_t^G=\chi_t/\sqrt N,\qquad
 \bm g_t=s_t^G\bm v_t
       =\mathbf R_N^{1/2}(\chi_t\bm U_t^\circ).
\]
By Gaussian polar decomposition, the $\bm g_t$ are independent
$N(\bm0,\mathbf R_N)$ and independent of $(\bm\tau,\mathbf F)$.
For $W_T=\sum_tc_t^2$, define
\begin{equation}\label{eq:conditional-gaussian-construction}
 \bm G_w=
 \begin{cases}
  W_T^{-1/2}\sum_tc_t\bm g_t,&W_T>0,\\
  \bm g_1,&W_T=0.
 \end{cases}
\end{equation}
Since $c_t$ is $\sigma(\bm\tau,\mathbf F)$-measurable,
$\bm G_w\mid(\bm\tau,\mathbf F)\sim N(\bm0,\mathbf R_N)$.

On $\{W_T>0\}$, put
\[
 \bm D_{G,T}=W_T^{-1/2}\sum_tc_t(1-s_t^G)\bm v_t,\qquad
 H_{G,T}=\frac{\sum_tc_t^2(1-s_t^G)^2}{W_T},
\]
and set both quantities to zero when $W_T=0$. Conditional on
$(\bm\chi,\bm\tau,\mathbf F)$, \eqref{eq:spherical-moments} yields
\[
 \E\{e^{\lambda D_{G,T,i}}\mid\bm\chi,\bm\tau,\mathbf F\}
 \leq e^{\lambda^2H_{G,T}/2},\qquad
 \Pp\{\|\bm D_{G,T}\|_\infty>x\mid\bm\chi,\bm\tau,\mathbf F\}
 \leq2N e^{-x^2/(2H_{G,T})}.
\]
The second bound is understood as zero when $H_{G,T}=0$.
Since $|s_t^G-1|\leq|(s_t^G)^2-1|$ and
$\E(\chi_t^2/N-1)^2=2/N$,
\[
 \E(H_{G,T}\mid\bm\tau,\mathbf F)\leq2/N.
\]
Conditioning on the auxiliary radii and applying Markov's inequality
gives, for $K>0$,
\[
 \Pp\left\{\|\bm D_{G,T}\|_\infty
           >\sqrt{\frac{4K\log(2N)}N}\right\}
 \leq\frac2K+\frac1{2N}.
\]
It follows that
$\|\bm D_{G,T}\|_\infty=O_p(\sqrt{\log N/N})$. On
$\mathcal O_{w,NT}$, Lemma~\ref{lem:oracle-variance} now implies
\begin{equation}\label{eq:spherical-gaussian-coupling}
 \left\|\frac1{T\bar a_{h,T}}
             \sum_tc_t(\bm v_t-\bm g_t)\right\|_\infty
 =\frac{\sigma_{w,T}}{\sqrt T}\|\bm D_{G,T}\|_\infty
 =O_p\left(\sqrt{\frac{\log N}{NT}}\right).
\end{equation}

Finally,
\[
 \frac1{T\bar a_{h,T}}\sum_tc_t\bm g_t
      =\frac{\sigma_{w,T}}{\sqrt T}\bm G_w,\qquad
 \sqrt{\frac{\log N}{NT}}\leq e_{NT},
\]
Since $b_N\geq1$, \eqref{eq:oracle-location-exact} and
\eqref{eq:expanded-projected-score} yield \eqref{eq:weighted-bahadur}.
The event
\[
 \mathcal H_{w,NT}
 =\mathcal O_{w,NT}\cap\mathcal E_{NT}^{Q}
\]
satisfies $\Pp(\mathcal H_{w,NT})\to1$ uniformly in $\bm\alpha$.
\end{proof}
\subsection{Fitted radial moments and studentization}
\label{app:studentization}

The geometric fitted radius and its normalized version are
\[
 \widehat r_{w,t}^{\,g}=\rho_Q\widehat r_{w,t},\qquad
 \widehat s_t=\widehat r_{w,t}^{\,g}/\sqrt N
 =s(\bm x_t-h_t\widehat{\bm z},\widehat{\bm\delta}_Q).
\]
Define
\[
 \begin{gathered}
 \mathcal H_w=\{g,\ k^2,\ s\mapsto sk(s),\ s\mapsto s^2,
                                      \ s\mapsto s^{-2}\},\\
 \mathcal W_T=\{(1),(h_t^2),(h_tf_{tj}),(f_{tj}f_{t\ell}):
                              1\leq j,\ell\leq p\}.
 \end{gathered}
\]
For $H\in\mathcal H_w$ and $(b_t^\circ)\in\mathcal W_T$, put
\[
 F_{H,b,T}=T^{-1}\sum_tb_t^\circ\{H(\widehat s_t)-H(\tau_t)\}.
\]

\begin{lemma}[Fitted radial moments]\label{lem:fitted-radial-moments}
Under Assumptions~\ref{ass:radial-primitive}--\ref{ass:weight-primitive},
consider residuals $\bm e_t+\bm u_t$ and a geometric scale perturbation
$\widehat{\bm\delta}$. Put
$\widehat s_t=s(\bm e_t+\bm u_t,\widehat{\bm\delta})$. If
\[
 \|\widehat{\bm\delta}\|_\infty=O_p(\eta_{NT}),\qquad
 \|\bm u_t\|_\infty\leq K_T\eta_{NT}(1+\|\bm f_t\|),\quad K_T=O_p(1),
\]
the fitted moment errors satisfy
\begin{gather*}
 \max_{H,b}|F_{H,b,T}|
     =O_p\{L/\sqrt T+\sqrt{b_NL/N}\}=O_p(d_{NT}),\\
 \max_H|F_{H,1,T}|
     =O_p\{\eta_{NT}+\sqrt{b_NL/N}\}.
\end{gather*}
\end{lemma}

\begin{proof}[Proof of Lemma~\ref{lem:fitted-radial-moments}]
On $\{\|\widehat{\bm\delta}\|_\infty\leq1\}$, the triangle inequality gives
\begin{align*}
 |\widehat s_t-\tau_t\varpi_t|
 &\leq N^{-1/2}\|(\mathbf\Lambda_{\widehat{\bm\delta}}-\mathbf I_N)
                               \bm e_t\|
       +N^{-1/2}\|\mathbf\Lambda_{\widehat{\bm\delta}}\bm u_t\|\\
 &\leq C\tau_t\varpi_t\|\widehat{\bm\delta}\|_\infty
                     +C\|\bm u_t\|_\infty,\\
 |\widehat s_t-\tau_t|
 &\leq C\{\tau_t|\varpi_t-1|+\tau_t\|\widehat{\bm\delta}\|_\infty
                  +K_T\eta_{NT}(1+\|\bm f_t\|)\}.
\end{align*}
By \eqref{eq:inverse-fourth-tail},
\[
 \Pp(\mathcal R_{\tau,T}^c)
 =\Pp\{\max_t\tau_t^{-1}>T^{1/4}\}
 \leq\zeta_{\tau,T}\longrightarrow0.
\]
On $\mathcal R_{\tau,T}$, the relative effect of the largest residual perturbation
is bounded by $CK_TLT^{-1/4}$. With
$\rho_{1,NT}=\sqrt{b_NL/N}+\eta_{NT}+LT^{-1/4}$, the growth condition gives
\begin{gather*}
 Q_{NT}:=\max_t|\widehat s_t/\tau_t-1|=O_p(\rho_{1,NT}),\qquad
 \rho_{1,NT}\longrightarrow0,\\
 \mathcal M_{NT}=
 \{Q_{NT}\leq1/2,\ \|\widehat{\bm\delta}\|_\infty\leq1\},\\
 \Pp(\mathcal M_{NT}^c)
 \leq\Pp\!\left(\frac{Q_{NT}}{\rho_{1,NT}}>\frac1{2\rho_{1,NT}}\right)
       +\Pp\!\left(\frac{\|\widehat{\bm\delta}\|_\infty}{\eta_{NT}}
                                      >\eta_{NT}^{-1}\right)\to0.
\end{gather*}
The profile bounds give
\[
 k'(s)=g(s)+sg'(s),\quad |k'(s)|\leq C(1+s^{-1}+s^{-2}),\qquad
 \max_{H\in\mathcal H_w}|H'(s)|\leq C(s+1+s^{-2}+s^{-3}).
\]
On $\mathcal M_{NT}$, put $s_{t,u}=\tau_t+u(\widehat s_t-\tau_t)$.
For $0\leq u\leq1$,
\[
 \tfrac12\tau_t\leq s_{t,u}\leq\tfrac32\tau_t,\qquad
 H(\widehat s_t)-H(\tau_t)
 = (\widehat s_t-\tau_t)\int_0^1H'(s_{t,u})\,du.
\]
Define
\begin{align*}
 A_{b,T}&=T^{-1}\sum_t|b_t^\circ|
                    (\tau_t^2+\tau_t+\tau_t^{-1}+\tau_t^{-2}),\\
 B_{b,T}&=T^{-1}\sum_t|b_t^\circ|(1+\|\bm f_t\|)
                    (\tau_t+1+\tau_t^{-2}+\tau_t^{-3}).
\end{align*}
Integration yields
\[
 \max_H|F_{H,b,T}|
 \leq C(\max_t|\varpi_t-1|+\|\widehat{\bm\delta}\|_\infty)A_{b,T}
                       +CK_T\eta_{NT}B_{b,T}.
\]
The design identities give
\begin{gather*}
 \max_{b\in\mathcal W_T}\frac1T\sum_t|b_t^\circ|\leq C,\qquad
 \max_{b\in\mathcal W_T,t}|b_t^\circ|\leq CL,\\
 \max_{b\in\mathcal W_T}\frac1T\sum_t|b_t^\circ|(1+\|\bm f_t\|)
 \leq C(1+\max_t\|\bm f_t\|)\leq C\sqrt L,\\
 \frac1T\sum_t(1+\|\bm f_t\|)\leq C.
\end{gather*}
Radial--design independence and conditional Markov inequalities yield
\begin{gather*}
 \max_b\E(A_{b,T}\mid\mathbf F)+\E(B_{1,T}\mid\mathbf F)\leq C,\qquad
 \max_b\E(B_{b,T}/\sqrt L\mid\mathbf F)\leq C,\\
 \max_b A_{b,T}+B_{1,T}=O_p(1),\qquad
 \max_b B_{b,T}=O_p(\sqrt L),\\
 \max_{H,b}|F_{H,b,T}|=O_p\{\sqrt{b_NL/N}+\sqrt L\eta_{NT}\},\qquad
 \max_H|F_{H,1,T}|=O_p\{\sqrt{b_NL/N}+\eta_{NT}\}.
\end{gather*}
\end{proof}

Write
\begin{gather*}
 \widehat a_h=T^{-1}\sum_th_t^2g(\widehat s_t),\qquad
 \widehat b_h=T^{-1}\sum_th_t^2k(\widehat s_t)^2,\\
 \widehat{\bm m}_h=T^{-1}\sum_th_tg(\widehat s_t)\bm f_t,\qquad
 \widehat c_t=h_tk(\widehat s_t)
                 -\widehat s_t\bm f_t^\top\mathbf Q_T^{-1}\widehat{\bm m}_h,\\
 \widehat b_c=T^{-1}\sum_t\widehat c_t^2,\qquad
 \widehat{\mathcal V}_w^{\,g}
 =\rho_Q^2\frac{\widehat B_{c,w}}{N\widehat A_{h,w}^2}
 =\frac{\widehat b_c}{\widehat a_h^2}.
\end{gather*}
Since $\widehat A_{h,w}^{\,g}=\widehat a_h/\sqrt N$ and
$\widehat B_{c,w}^{\,g}=\widehat b_c$, \eqref{eq:df-variance} becomes
\[
 \widehat{\operatorname{se}}_{w,i}
 =\widehat d_{Q,i}^{\,g}
       \sqrt{\widehat{\mathcal V}_w^{\,g}/T}.
\]
On the event of positive denominators, define
\begin{align*}
 r_{w,i}&=\frac{\sqrt T\epsilon_{w,NT,i}}{\sigma_{w,T}},&
 s_{w,i}&=\frac{\widehat d_{Q,i}^{\,g}}{d_i}
       \left(\frac{\widehat{\mathcal V}_w^{\,g}}
                        {\sigma_{w,T}^2}\right)^{1/2}.
\end{align*}

\begin{lemma}[Studentization]\label{lem:weighted-studentization}
Under the conditions of Lemma~\ref{lem:weighted-linearization},
$\widehat{\mathcal V}_w^{\,g}>0$ with probability tending to one.
For any deterministic $K=K_{NT}\to\infty$ with $KL/T\to0$,
\begin{equation}\label{eq:variance-rates}
 |\sigma_{w,T}^2-\mathcal V_{w,T}|
       =O_p\{\eta_{NT}+\beta_{NT}(K)\},\qquad
 |\widehat{\mathcal V}_w^{\,g}-\sigma_{w,T}^2|=O_p(d_{NT}).
\end{equation}
On the local-root event,
\begin{equation}\label{eq:gaussian-bridge-constructed}
 Z_{w,i}^{\circ}
 :=\frac{\widehat\alpha_{w,i}-\alpha_i}
          {\widehat{\operatorname{se}}_{w,i}}
 =\frac{G_{w,i}+r_{w,i}}{s_{w,i}},
\end{equation}
where
\begin{equation}\label{eq:bridge-error-rates}
 \max_i|r_{w,i}|=O_p(\sqrt T e_{NT}),\qquad
 \max_i|s_{w,i}-1|=O_p(d_{NT}).
\end{equation}
Moreover,
\begin{equation}\label{eq:studentized-se-consistency}
 \max_i\left|
 \frac{\sqrt T\,\widehat{\operatorname{se}}_{w,i}}
      {d_i\sqrt{\mathcal V_{w,T}}}-1\right|
   =O_p\{d_{NT}+\beta_{NT}(K)\}=o_p(1).
\end{equation}
These conclusions hold for arbitrary deterministic alpha vectors.
\end{lemma}

\begin{proof}[Proof of Lemma~\ref{lem:weighted-studentization}]
At the alpha root,
\[
 \bm u_t=h_t(\bm b-\widehat{\bm z})-\mathbf\Delta_T^\top\bm f_t,
 \qquad \|\bm u_t\|_\infty\leq K_T\eta_{NT}(1+\|\bm f_t\|),
 \quad K_T=O_p(1).
\]
Set
\begin{gather*}
 b_h^o=T^{-1}\sum_th_t^2k(\tau_t)^2,\qquad
 \bm n_h^o=T^{-1}\sum_th_t\tau_tk(\tau_t)\bm f_t,\qquad
 \mathbf M_h^o=T^{-1}\sum_t\tau_t^2\bm f_t\bm f_t^\top,\\
 \widehat{\bm n}_h=T^{-1}\sum_th_t\widehat s_tk(\widehat s_t)\bm f_t,
 \qquad
 \widehat{\mathbf M}_h=T^{-1}\sum_t\widehat s_t^2\bm f_t\bm f_t^\top,\\
 b_c^o=T^{-1}\sum_tc_t^2
 =b_h^o-2\bm m_{h,T}^\top\mathbf Q_T^{-1}\bm n_h^o
       +\bm m_{h,T}^\top\mathbf Q_T^{-1}\mathbf M_h^o
                                    \mathbf Q_T^{-1}\bm m_{h,T},\\
 \widehat b_c=\widehat b_h
       -2\widehat{\bm m}_h^\top\mathbf Q_T^{-1}\widehat{\bm n}_h
       +\widehat{\bm m}_h^\top\mathbf Q_T^{-1}\widehat{\mathbf M}_h
                                    \mathbf Q_T^{-1}\widehat{\bm m}_h.
\end{gather*}
Lemma~\ref{lem:fitted-radial-moments} gives the rate $d_{NT}$
for each fitted-minus-latent moment. Since $p$ is fixed,
\begin{align*}
 &|\widehat a_h-\bar a_{h,T}|+|\widehat b_h-b_h^o|
    +\|\widehat{\bm m}_h-\bm m_{h,T}\|
    +\|\widehat{\bm n}_h-\bm n_h^o\|
    +\|\widehat{\mathbf M}_h-\mathbf M_h^o\|_{\op}
       =O_p(d_{NT}),\\
 &\E\|\mathbf M_h^o\|_{\op}
    \leq\frac{\E\tau^2}{T}\sum_t\|\bm f_t\|^2\leq C,
 \qquad \mathbf M_h^o=O_p(1),\qquad b_h^o=O_p(1),\\
 &\bm n_h^o
 =\frac1T\sum_th_t\bm f_t\{\tau_tk(\tau_t)-\E[\tau k(\tau)]\}
       =O_p(1),\qquad \bm m_{h,T}=O_p(\eta_{NT}).
\end{align*}
For $\bm n_h^o$, use
\[
 \E|\tau k(\tau)|\leq(\E\tau^2\,\E k(\tau)^2)^{1/2}\leq C,\qquad
 T^{-1}\sum_t|h_t|\|\bm f_t\|\leq C.
\]
Since $\|\mathbf Q_T^{-1}\|_{\op}\leq C$, expansion gives
\begin{align*}
 |\widehat b_c-b_c^o|
 &\leq|\widehat b_h-b_h^o|
     +C\|\widehat{\bm m}_h-\bm m_{h,T}\|
                     (\|\widehat{\bm n}_h\|+\|\bm n_h^o\|)\\
 &\quad+C(\|\widehat{\bm m}_h\|+\|\bm m_{h,T}\|)
                           \|\widehat{\bm n}_h-\bm n_h^o\|\\
 &\quad+C\|\widehat{\mathbf M}_h\|_{\op}
        (\|\widehat{\bm m}_h\|+\|\bm m_{h,T}\|)
                           \|\widehat{\bm m}_h-\bm m_{h,T}\|\\
 &\quad+C\|\bm m_{h,T}\|^2
                         \|\widehat{\mathbf M}_h-\mathbf M_h^o\|_{\op}
 =O_p(d_{NT}).
\end{align*}
On $\{\bar a_{h,T}\geq a\omega_T/2,
 |\widehat a_h-\bar a_{h,T}|\leq a\omega_T/4\}$,
\begin{align*}
 |\widehat{\mathcal V}_w^{\,g}-\sigma_{w,T}^2|
 &=\left|\frac{\widehat b_c}{\widehat a_h^2}
                       -\frac{b_c^o}{\bar a_{h,T}^2}\right|\\
 &\leq C\{|\widehat b_c-b_c^o|
                   +b_c^o|\widehat a_h-\bar a_{h,T}|\}
 =O_p(d_{NT}),\\
 |\widehat{\mathcal V}_w^{\,g}-\mathcal V_{w,T}|
 &\leq|\widehat{\mathcal V}_w^{\,g}-\sigma_{w,T}^2|
       +|\sigma_{w,T}^2-\mathcal V_{w,T}|
 =O_p\{d_{NT}+\beta_{NT}(K)\}.
\end{align*}
The event has probability tending to one by Lemma~\ref{lem:oracle-variance}.
Projection orthogonality and Cauchy--Schwarz give
\begin{gather*}
 \frac1T\sum_t\frac{h_t\widehat c_t}{\widehat s_t}
 =\widehat a_h-\left(\frac1T\sum_th_t\bm f_t\right)^\top
                          \mathbf Q_T^{-1}\widehat{\bm m}_h=\widehat a_h,\\
 \widehat a_h^2
 \leq\left(\frac1T\sum_t\frac{h_t^2}{\widehat s_t^2}\right)
                   \widehat b_c,\qquad
 \widehat a_h>0, \min_t\widehat s_t>0
     \ \Longrightarrow\ \widehat b_c>0.
\end{gather*}
Choose $v_0>0$ such that $\mathcal V_{w,T}\geq v_0$ uniformly.
Lemma~\ref{lem:oracle-variance} gives
$\Pp(\sigma_{w,T}^2\geq v_0/2)\to1$.
On $\{\sigma_{w,T}^2\geq v_0/2,
 |\widehat{\mathcal V}_w^{\,g}-\sigma_{w,T}^2|\leq v_0/4\}$,
\begin{gather*}
 \left|\sqrt{\widehat{\mathcal V}_w^{\,g}/\sigma_{w,T}^2}-1\right|
 \leq\frac{|\widehat{\mathcal V}_w^{\,g}-\sigma_{w,T}^2|}
                 {\sigma_{w,T}^2}=O_p(d_{NT}),\\
 D_{Q,T}:=\max_i|\widehat d_{Q,i}^{\,g}/d_i-1|
 \leq e^{\|\widehat{\bm\delta}_Q\|_\infty}
                   \|\widehat{\bm\delta}_Q\|_\infty=O_p(\eta_{NT}),\\
 \max_i|s_{w,i}-1|
 \leq D_{Q,T}+(1+D_{Q,T})
       \left|\sqrt{\widehat{\mathcal V}_w^{\,g}/\sigma_{w,T}^2}-1\right|
 =O_p(d_{NT}),\\
 \max_i|r_{w,i}|\leq\sqrt{2T/v_0}\,
              \|\bm\epsilon_{w,NT}\|_\infty=O_p(\sqrt T e_{NT}).
\end{gather*}
The same argument with $\mathcal V_{w,T}\geq v_0$ proves
\eqref{eq:studentized-se-consistency}.
By \eqref{eq:weighted-bahadur} and \eqref{eq:profile-scale-covariance},
\begin{gather*}
 \widehat d_{Q,i}^{\,g}\sqrt{\widehat{\mathcal V}_w^{\,g}}
 =\frac{\widehat d_{Q,i}}{\rho_Q}\,
       \rho_Q\sqrt{\frac{\widehat B_{c,w}}{N\widehat A_{h,w}^2}},\\
 Z_{w,i}^{\circ}
 =\frac{\sqrt T\,d_i(\widehat z_i-b_i)}
              {\widehat d_{Q,i}^{\,g}\sqrt{\widehat{\mathcal V}_w^{\,g}}}
 =\frac{G_{w,i}+r_{w,i}}{s_{w,i}}.
\end{gather*}
\end{proof}

For the true geometric radius,
\begin{equation}\label{eq:true-radial-variance}
 \frac{J(w)}N:=\frac{\E w(r)^2}{N[\E\{w(r)/r\}]^2}
 =J_{\tau,w}+O\{(b_N/N)^{1/2}\}.
\end{equation}
Write $r/\sqrt N=\tau\varpi_N$. For $H\in\{g,k^2\}$, the profile
bounds and Lemma~\ref{lem:spherical-geometry} give
\begin{align*}
 |H(\tau \varpi_N)-H(\tau)|
 &\leq C(\tau^2+\tau+\tau^{-1}+\tau^{-2})
          |\varpi_N-1|(1+\varpi_N+\varpi_N^{-3}),\\
 \E\{|\varpi_N-1|(1+\varpi_N+\varpi_N^{-3})\}
 &\leq\{\E(\varpi_N-1)^2\}^{1/2}
       \{\E(1+\varpi_N+\varpi_N^{-3})^2\}^{1/2}
 \leq C\sqrt{b_N/N}.
\end{align*}
Independence of $\tau$ and $\varpi_N$ implies
\begin{gather*}
 \max_{H\in\{g,k^2\}}
       |\E H(\tau \varpi_N)-\E H(\tau)|\leq C\sqrt{b_N/N},\\
 a_r=\E g(\tau \varpi_N),\qquad |a_r-a|\leq C\sqrt{b_N/N},\qquad
 A_w=a_r/\sqrt N,\quad a_r\geq a/2,\\
 \left|\frac{J(w)}N-J_{\tau,w}\right|
 =\left|\frac{\E k(\tau \varpi_N)^2}{a_r^2}
          -\frac{\E k(\tau)^2}{a^2}\right|
 \leq C\{\sqrt{b_N/N}+|a_r-a|\}\leq C\sqrt{b_N/N}.
\end{gather*}

\subsection{Residual degrees-of-freedom calibration}
\label{app:df-calibration}

For $\bm\varepsilon_i=(\varepsilon_{1i},\ldots,\varepsilon_{Ti})^\top$,
\eqref{eq:radial-model} and $\mathbf M_X^2=\mathbf M_X$ give
\begin{align*}
 \E\{\|\mathbf M_X\bm\varepsilon_i\|^2
                         \mid\mathbf F,\bm\tau\}
 &=d_i^2\tr\{\mathbf M_X\diag(\tau_1^2,\ldots,\tau_T^2)\}
   =d_i^2\sum_{t=1}^T(\mathbf M_X)_{tt}\tau_t^2,\\
 \E\{\|\mathbf M_X\bm\varepsilon_i\|^2\mid\mathbf F\}
 &=d_i^2\E\tau^2\,\tr(\mathbf M_X)=n_T\sigma_i^2,
\end{align*}
where $\sigma_i^2=d_i^2\E\tau^2$. For fixed $p$ and $T\geq2(p+1)$,
\begin{gather}
 \tfrac12\leq c_{\mathrm{df},T}\leq1,\qquad
 0\leq1-\sqrt{c_{\mathrm{df},T}}
   =\frac{\rank(\mathbf X)}
          {T(1+\sqrt{c_{\mathrm{df},T}})}
   \leq\frac{p+1}{T},\notag\\
 0\leq c_{\mathrm{df},T}^{-1/2}-1
   =\frac{\rank(\mathbf X)}
     {T\sqrt{c_{\mathrm{df},T}}(1+\sqrt{c_{\mathrm{df},T}})}
   \leq\frac{2(p+1)}T,\notag\\*
 \frac{(p+1)L}{T}\longrightarrow0.\label{eq:df-multiplier-bounds}
\end{gather}

\Needspace{3\baselineskip}
\subsection{Proofs of the main distributional results}
\label{app:main-statistic}

\begin{proof}[Proof of Theorem~\ref{thm:main-statistic}]
By \eqref{eq:alpha-equivariance}, it suffices to consider $\bm\alpha=\bm0$.
Let $\mathcal C_{w,NT}$ combine the design, local-root and positivity
events in Lemmas~\ref{lem:local-root}
and~\ref{lem:weighted-studentization}. Then
\[
 \Pp(\mathcal C_{w,NT}^c)\longrightarrow0.
\]
On $\mathcal C_{w,NT}$,
\eqref{eq:gaussian-bridge-constructed}--\eqref{eq:bridge-error-rates} yield
\[
 Z_{w,i}^{\circ}=\frac{G_{w,i}+r_{w,i}}{s_{w,i}},\qquad
 \bm G_w\mid(\bm\tau,\mathbf F)\sim N(\bm0,\mathbf R_N),
\]
where $\max_i|r_{w,i}|=O_p(\sqrt T e_{NT})$ and
$\max_i|s_{w,i}-1|=O_p(d_{NT})$.
Conditional Gaussian tails and a union bound imply
\[
 \Pp\{\max_i|G_{w,i}|>x\mid\bm\tau,\mathbf F\}
       \leq2N e^{-x^2/2},\qquad
 \max_i|G_{w,i}|=O_p(\sqrt L).
\]
Since $d_{NT}\to0$,
\[
 \mathcal C_{w,NT}^{\circ}
 =\mathcal C_{w,NT}\cap\{\max_i|s_{w,i}-1|\leq1/2\},\qquad
 \Pp\{(\mathcal C_{w,NT}^{\circ})^c\}\longrightarrow0.
\]
On $\mathcal C_{w,NT}^{\circ}$,
\begin{align*}
 \max_i|Z_{w,i}^{\circ}-G_{w,i}|
 &\leq2\max_i|r_{w,i}|
       +2\max_i|s_{w,i}-1|\max_i|G_{w,i}|\\
 &=O_p(\sqrt T e_{NT}+\sqrt L\,d_{NT})=O_p(r_{NT}).
\end{align*}
Since $\Pp\{(\mathcal C_{w,NT}^{\circ})^c\}\to0$, the same stochastic
order holds unconditionally.

For deterministic $K=K_{NT}\to\infty$ with $KL/T\to0$,
\eqref{eq:studentized-se-consistency} gives, uniformly in $i$,
\[
 \frac{\sqrt T\,\widehat{\operatorname{se}}_{w,i}}
      {d_i\sqrt{\mathcal V_{w,T}}}
 =\frac{\widehat d_{Q,i}^{\,g}}{d_i}
       \left(\frac{\widehat{\mathcal V}_{w}^{\,g}}
                         {\mathcal V_{w,T}}\right)^{1/2}
 =1+O_p\{d_{NT}+\beta_{NT}(K)\}=1+o_p(1).
\]
Equation~\eqref{eq:df-multiplier-bounds} adds an $O(T^{-1})$ calibration
error, proving \eqref{eq:standard-error-consistency}.

For the centered calibrated pivots,
$Z_{w,i}^{\circ,\mathrm{df}}
 =\sqrt{c_{\mathrm{df},T}}Z_{w,i}^{\circ}$. Therefore
\begin{align}
 \max_i|Z_{w,i}^{\circ,\mathrm{df}}-G_{w,i}|
 &\leq\max_i|Z_{w,i}^{\circ}-G_{w,i}|
       +\frac{p+1}{T}\max_i|G_{w,i}|\nonumber\\
 &=O_p\{r_{NT}+(p+1)\sqrt L/T\}=O_p(r_{NT}),
 \label{eq:df-gaussian-transfer}
\end{align}
where $p$ is fixed and $r_{NT}\geq L^5/\sqrt T$.
The explicit rates satisfy
\begin{align*}
 r_{NT}\sqrt L
 &=\frac{L^{11/2}}{\sqrt T}
       +L^{9/2}\sqrt{\frac{b_N}{N}}
       +\frac{L^2}{\sqrt T}
       +L^{3/2}\sqrt{\frac{b_N}{N}}\\
 &\leq C\left\{\sqrt{\frac{L^{11}}T}
                  +\sqrt{\frac{b_NL^9}N}\right\}\longrightarrow0.
\end{align*}
On the local-root event,
\[
 S_{w,i}^{\mathrm{df}}
  =\frac{\alpha_i}{\widehat{\operatorname{se}}_{w,i}^{\,\mathrm{df}}}
          +Z_{w,i}^{\circ,\mathrm{df}}.
\]
This proves \eqref{eq:random-drift-representation}. All bounds involve
centered standardized errors and relative scales, and are uniform over
$\bm\alpha$ and $d_i>0$.
\end{proof}

\begin{proof}[Proof of Corollary~\ref{cor:radial-local-power}]
For either fitted standard error $\widehat{\operatorname{se}}_{w,i}^{\dagger}$
and statistic $S_{w,i}^{\dagger}$, the local alternative and
\eqref{eq:standard-error-consistency} give
\[
 \frac{\alpha_i}{\widehat{\operatorname{se}}_{w,i}^{\dagger}}
 =\frac{\sqrt T\alpha_i/d_i}
         {\sqrt T\widehat{\operatorname{se}}_{w,i}^{\dagger}/d_i}
       \xrightarrow{p}
       m_i^\circ:=\frac{b_i^\circ}{\sqrt{\mathcal V_w}}.
\]
Theorem~\ref{thm:main-statistic} gives
\[
 S_{w,i}^{\dagger}=m_i^\circ+G_{w,i}+o_p(1),
 \qquad \Pp(G_{w,i}\leq x)=\Phi(x).
\]
Slutsky's theorem and continuity of the normal distribution imply
\[
 S_{w,i}^{\dagger}\Rightarrow N(m_i^\circ,1),\qquad
 \Pp(S_{w,i}^{\dagger}>c)\to\barPhi(c-m_i^\circ).
\]
For $b_i^\circ>0$, differentiation gives
\[
 \frac{\partial}{\partial\mathcal V_w}
       \barPhi(c-b_i^\circ/\sqrt{\mathcal V_w})
 =-\phi(c-m_i^\circ)\frac{b_i^\circ}{2\mathcal V_w^{3/2}}<0.
\]
Only the tested coordinate is restricted to a local alternative.
\end{proof}

\subsection{Verification of the radial conditions}
\label{app:elliptical-examples}

Let $H_{G,N}=\chi_N/\sqrt N$, where $\chi_a$ has the chi distribution
with $a>0$ degrees of freedom. Let $B$ be uniform on $(0,1)$ and assume
\[
 0<a_-\leq A\leq a_+<\infty,\qquad \E A^2=1.
\]
The radial variables for the multivariate normal distribution,
standardized multivariate $t$ distribution, common-scale mixture of
multivariate normal distributions, constant-radius distribution, and
uniform distribution on an ellipsoid are, respectively,
\begin{align*}
 \tau_{\mathrm G}&=H_{G,N},&
 \tau_{\mathrm S,\nu}&=\frac{\sqrt{\nu-2}\,H_{G,N}}{\chi_\nu},&
 \tau_{\mathrm M}&=AH_{G,N},\\
 \tau_{\mathrm C}&=1,&
 \tau_{\mathrm B}&=\sqrt{\frac{N+2}{N}}\,B^{1/N}.&&
\end{align*}
Each product has independent factors. Fix $\nu>2$ for the multivariate
$t$ law and choose
\[
 m>4,\qquad N_0=4\lceil m/2\rceil,\qquad
 2<s\leq3,\quad s<\nu\ \text{for the $t$ law}.
\]

\begin{lemma}[Radial examples]\label{lem:radial-examples}
The five radial laws satisfy the uniform-integrability conditions of
Assumption~\ref{ass:radial-primitive} for $N\geq N_0$, and
$\E\tau^2=1$. When $\mathbf\Sigma_N$ is positive definite, the
uniform distribution on its ellipsoid is not a nondegenerate
independent normal scale mixture.
\end{lemma}

\begin{proof}[Proof of Lemma~\ref{lem:radial-examples}]
The identity $\E\chi_a^s=2^{s/2}\Gamma\{(a+s)/2\}/\Gamma(a/2)$,
$s>-a$, gives, for $0<m<N$,
\[
 \E H_{G,N}^{-m}
 =\left(\frac N2\right)^{m/2}
       \frac{\Gamma\{(N-m)/2\}}{\Gamma(N/2)}.
\]
For the fixed order $m>4$, let $j=\lceil m/2\rceil$.
When $N\geq N_0=4j$, comparison with the even moment gives
\begin{gather*}
 N-2\ell\geq N/2\quad(1\leq\ell\leq j),\\
 \begin{aligned}
 (\E H_{G,N}^{-m})^{1/m}
 &\leq(\E H_{G,N}^{-2j})^{1/(2j)}
   =\left\{\frac{N^j\Gamma(N/2-j)}
                    {2^j\Gamma(N/2)}\right\}^{1/(2j)}\\
 &=\left\{\prod_{\ell=1}^{j}\frac{N}{N-2\ell}\right\}^{1/(2j)}
   \leq(2^j)^{1/(2j)}=\sqrt2.
 \end{aligned}
\end{gather*}
For the chosen order $2<s\leq3$,
\[
 \E H_{G,N}^2=1,\qquad
 \E H_{G,N}^s\leq(\E H_{G,N}^4)^{s/4}
       =(1+2/N)^{s/4}\leq(3/2)^{3/4}.
\]
Independence gives the multivariate $t$ radial moments:
\begin{align*}
 \E\tau_{\mathrm S,\nu}^{\,s}
 &=\left(\frac{\nu-2}{N}\right)^{s/2}
    \frac{\Gamma\{(N+s)/2\}\Gamma\{(\nu-s)/2\}}
         {\Gamma(N/2)\Gamma(\nu/2)},\qquad 0<s<\nu,\\
 \E\tau_{\mathrm S,\nu}^{-m}
 &=\left(\frac{N}{\nu-2}\right)^{m/2}
    \frac{\Gamma\{(\nu+m)/2\}\Gamma\{(N-m)/2\}}
         {\Gamma(\nu/2)\Gamma(N/2)},\qquad 0<m<N.
\end{align*}
For the chosen $s<\nu$ and $N\geq N_0$, the moment bounds for
$H_{G,N}$ therefore imply
\begin{gather*}
 \E\tau_{\mathrm S,\nu}^{\,s}
 =(\nu-2)^{s/2}\E H_{G,N}^s\,
      2^{-s/2}\frac{\Gamma\{(\nu-s)/2\}}{\Gamma(\nu/2)}
 \leq C_{\nu,s},\\
 \E\tau_{\mathrm S,\nu}^{-m}
 =(\nu-2)^{-m/2}\E H_{G,N}^{-m}\,
      2^{m/2}\frac{\Gamma\{(\nu+m)/2\}}{\Gamma(\nu/2)}
 \leq C_{\nu,m},\\
 \E\tau_{\mathrm S,\nu}^{\,2}
 =(\nu-2)\E H_{G,N}^2\,
       \frac{\Gamma(\nu/2-1)}{2\Gamma(\nu/2)}=1.
\end{gather*}
The inverse-moment order requires $m<N$ and imposes no additional
restriction on the fixed degrees of freedom $\nu>2$.

For the bounded scale mixture and constant-radius laws,
\begin{gather*}
 \E\tau_{\mathrm M}^{\,s}
       =\E A^s\E H_{G,N}^s\leq a_+^s(1+2/N)^{s/4},\qquad
 \E\tau_{\mathrm M}^{-m}
       =\E A^{-m}\E H_{G,N}^{-m}\leq a_-^{-m}2^{m/2},\\
 \E\tau_{\mathrm M}^{\,2}=\E A^2\,\E H_{G,N}^2=1,\qquad
 \E\tau_{\mathrm C}^{\,s}
       =\E\tau_{\mathrm C}^{-m}=\E\tau_{\mathrm C}^{\,2}=1.
\end{gather*}
For $B\sim\operatorname{Unif}(0,1)$, direct integration gives
\begin{align*}
 \E\tau_{\mathrm B}^{\,s}
 &=\left(\frac{N+2}{N}\right)^{s/2}
          \int_0^1x^{s/N}\,dx
   =\left(\frac{N+2}{N}\right)^{s/2}\frac{N}{N+s}
   \leq(3/2)^{3/2},\\
 \E\tau_{\mathrm B}^{-m}
 &=\left(\frac{N}{N+2}\right)^{m/2}
          \int_0^1x^{-m/N}\,dx
   =\left(\frac{N}{N+2}\right)^{m/2}\frac{N}{N-m}
   \leq2,\\
 \E\tau_{\mathrm B}^{\,2}
       &=\frac{N+2}{N}\frac{N}{N+2}=1.
\end{align*}
Thus each law satisfies, for its fixed parameters,
\[
 \E\tau^2=1,\qquad \sup_{N\geq N_0}\E\tau_N^s<\infty,\qquad
 \sup_{N\geq N_0}\E\tau_N^{-m}<\infty,\qquad s>2,\quad m>4.
\]
These bounds imply
\begin{gather*}
 \sup_{N\geq N_0}
 \E\{\tau_N^2\ind(\tau_N^2>K)\}
 \leq K^{-(s-2)/2}\sup_{N\geq N_0}\E\tau_N^s
 \longrightarrow0,\\
 \sup_{N\geq N_0}
 \E\{\tau_N^{-4}\ind(\tau_N^{-4}>K)\}
 \leq K^{-(m-4)/4}\sup_{N\geq N_0}\E\tau_N^{-m}
 \longrightarrow0.
\end{gather*}
Consequently, every fixed $\nu>2$ in the standardized multivariate
$t$ family satisfies both uniform-integrability conditions.

When $\mathbf\Sigma_N\succ0$,
$\mathbf\Sigma_N^{-1/2}\mathbf D^{1/2}\mathbf R_N^{1/2}$ is orthogonal,
so
\[
 \mathbf\Sigma_N^{-1/2}\bm\varepsilon
 \overset{d}{=}\sqrt{N+2}\,B^{1/N}\bm U^\circ,\qquad
 \Pp(B^{1/N}\leq u)=u^N,\quad 0\leq u\leq1.
\]
This law is uniform on
$\{\bm x:\bm x^\top\mathbf\Sigma_N^{-1}\bm x\leq N+2\}$.
Its bounded support excludes a nondegenerate independent normal scale mixture.
\end{proof}

\section{Proofs for false discovery control}
\label{app:weighted-fdr-proofs}

Let $\bm G\sim N(\bm0,\mathbf R)$ with $\diag(\mathbf R)=\mathbf I_N$,
and let $\mathbf R_{\mathcal I}$ be the principal submatrix for a fixed
set $\mathcal I$. For the first three lemmas, define
\[
 1\leq m=|\mathcal I|\leq N,\qquad
 V_G(u)=\sum_{i\in\mathcal I}\ind\{G_i\geq z(u)\},\quad 0<u<1,
 \qquad V_G(1)=m.
\]

\begin{lemma}[Gaussian tail-count variance]
\label{lem:fdr-gaussian-count}
If $\|\mathbf R_{\mathcal I}\|_{\op}\leq\Lambda$, then, for $0<u<1$,
\begin{equation}\label{eq:fdr-gaussian-count-var}
 \E V_G(u)=mu,\qquad
 \Var\{V_G(u)\}\leq\Lambda mu(1-u)\leq\Lambda mu.
\end{equation}
\end{lemma}

\begin{proof}[Proof of Lemma~\ref{lem:fdr-gaussian-count}]
Let $\{H_\ell:\ell\geq0\}$ be the orthonormal Hermite basis of
$L^2(\phi)$, normalized by
\[
 e^{tx-t^2/2}
   =\sum_{\ell\geq0}\frac{t^\ell}{\sqrt{\ell!}}H_\ell(x).
\]
The centered indicator $f_u(x)=\ind\{x\geq z(u)\}-u$ has the
mean-square expansion
\[
 f_u=\sum_{\ell\geq1}a_{\ell,u}H_\ell,
 \qquad
 \sum_{\ell\geq1}a_{\ell,u}^2=u(1-u).
\]
For standard bivariate normal $(X,Y)$ with correlation $\rho$,
the identity
$\E e^{tX+sY-(t^2+s^2)/2}=e^{\rho ts}$ gives
\begin{equation}\label{eq:fdr-hermite-covariance}
 \E\{H_\ell(X)H_j(Y)\}
       =\ind(\ell=j)\rho^\ell.
\end{equation}
Gaussian exponential moments justify coefficient comparison, including
$\rho=\pm1$.

By the Schur product theorem,
$\mathbf H_\ell=\mathbf R_{\mathcal I}^{\circ\ell}\succeq0$.
Since $\mathbf H_1\preceq\Lambda\mathbf I_m$ and
$\diag(\mathbf R_{\mathcal I})=\mathbf I_m$,
\[
 (\Lambda\mathbf I_m-\mathbf H_\ell)\circ\mathbf R_{\mathcal I}
       =\Lambda\mathbf I_m-\mathbf H_{\ell+1}\succeq0
       \quad\text{whenever}\quad
       \mathbf H_\ell\preceq\Lambda\mathbf I_m.
\]
Thus $0\preceq\mathbf H_\ell\preceq\Lambda\mathbf I_m$ for all
$\ell\geq1$. For $f_{u,M}=\sum_{\ell=1}^M a_{\ell,u}H_\ell$,
\eqref{eq:fdr-hermite-covariance} gives
\[
 \E\left\{\sum_{i\in\mathcal I}f_{u,M}(G_i)\right\}^2
 =\sum_{\ell=1}^M a_{\ell,u}^2
           \one_m^\top\mathbf H_\ell\one_m
 \leq\Lambda m\sum_{\ell=1}^M a_{\ell,u}^2.
\]
For each fixed $m$, the triangle inequality in $L^2$ gives
\[
 \left\|\sum_{i\in\mathcal I}
          \{f_{u,M}(G_i)-f_u(G_i)\}\right\|_{L^2}
 \leq m\|f_{u,M}-f_u\|_{L^2(\phi)}\longrightarrow0.
\]
Passing to the limit in the preceding variance bound proves
\[
 \Var\{V_G(u)\}
 \leq\Lambda m\sum_{\ell\geq1}a_{\ell,u}^2
 =\Lambda mu(1-u).
\]
Finally, the unit marginal variances imply
$\E V_G(u)=\sum_{i\in\mathcal I}\barPhi\{z(u)\}=mu$.
\end{proof}

For the sparse-correlation bound, suppose
$\max_{i\ne j}|R_{ij}|\leq\varrho$, where $0<\varrho<1$, and write
\[
 s=\max_i|\{j\ne i:R_{ij}\ne0\}|,\qquad
 \beta_\varrho=\frac{2\varrho}{1+\varrho}\in(0,1).
\]

\begin{lemma}[Gaussian tail counts under sparse correlation]
\label{lem:fdr-sparse-count}
Under these correlation restrictions, there is a constant $C_\varrho$
depending only on $\varrho$ such that
\begin{equation}\label{eq:fdr-sparse-count-var}
 \Var\{V_G(u)\}\leq mu+C_\varrho msu^{2-\beta_\varrho},
                         \qquad 0<u<1.
\end{equation}
\end{lemma}

\begin{proof}[Proof of Lemma~\ref{lem:fdr-sparse-count}]
For standard bivariate normal $(X,Y)$ with $|\rho|\leq\varrho$,
\[
 \begin{aligned}
 f_\rho(z+v,z+w)
 ={}&\frac1{2\pi\sqrt{1-\rho^2}}
 \exp\left\{-\frac{z^2}{1+\rho}-\frac{z(v+w)}{1+\rho}\right.\\
 &\left.\hspace{24mm}
 -\frac{(v-\rho w)^2+(1-\rho^2)w^2}{2(1-\rho^2)}\right\},
 \qquad z\geq1,\quad v,w\geq0.
 \end{aligned}
\]
Dropping the nonpositive quadratic term in the exponent and integrating
over the positive quadrant gives
\[
 \begin{aligned}
 \Pp(X\geq z,Y\geq z)
 &\leq\frac{e^{-z^2/(1+\rho)}}{2\pi\sqrt{1-\rho^2}}
        \left\{\int_0^\infty e^{-zv/(1+\rho)}\,dv\right\}^{2}\\
 &=\frac{(1+\rho)^2}{2\pi z^2\sqrt{1-\rho^2}}
                e^{-z^2/(1+\rho)}
 \leq C_\varrho z^{-2}e^{-z^2/(1+\varrho)}.
 \end{aligned}
\]
For $u=\barPhi(z)$ and $z\geq1$,
\[
 \begin{aligned}
 u&\geq\int_z^{z+1/z}\phi(x)\,dx
 \geq z^{-1}\phi(z+z^{-1})
 =z^{-1}\phi(z)e^{-1-1/(2z^2)}
 \geq c z^{-1}e^{-z^2/2},\\
 z^{-2}e^{-z^2/(1+\varrho)}
 &\leq C_\varrho z^{2/(1+\varrho)-2}
                          u^{2/(1+\varrho)}
 \leq C_\varrho u^{2/(1+\varrho)}.
 \end{aligned}
\]
For $z<1$,
\[
 u\geq\barPhi(1),\qquad
 \Pp(X\geq z,Y\geq z)
 \leq1\leq\{\barPhi(1)\}^{-2/(1+\varrho)}u^{2/(1+\varrho)}.
\]
Therefore
\begin{equation}\label{eq:fdr-bivariate-tail}
 \Pp\{X\geq z(u),Y\geq z(u)\}
       \leq C_\varrho u^{2/(1+\varrho)},\qquad 0<u<1.
\end{equation}
Gaussian independence and the definition of $s$ give
\[
 \begin{gathered}
 R_{ij}=0\ \Longrightarrow\ G_i\perp G_j
 \ \Longrightarrow\
 \Cov\{\ind(G_i\geq z(u)),\ind(G_j\geq z(u))\}=0,\\
 \bigl|\{(i,j)\in\mathcal I^2:i\ne j,\ R_{ij}\ne0\}\bigr|\leq ms.
 \end{gathered}
\]
By \eqref{eq:fdr-bivariate-tail},
\[
 \begin{aligned}
 \Var\{V_G(u)\}
 &=mu(1-u)+\sum_{\substack{i,j\in\mathcal I\\i\ne j,\ R_{ij}\ne0}}
       [\Pp\{G_i\geq z(u),G_j\geq z(u)\}-u^2]\\
 &\leq mu+C_\varrho msu^{2/(1+\varrho)}
   =mu+C_\varrho msu^{2-\beta_\varrho}.
 \end{aligned}
 \qedhere
\]
\end{proof}

For $0<u_0<1$ and $0<\epsilon\leq1/2$, define the uniform-count event
\[
 \mathcal B(u_0,\epsilon)
 =\left\{\sup_{u_0\leq u\leq1}
          \frac{V_G(u)}{Nu}
       \leq(1+\epsilon)\left(\frac mN+\epsilon\right)\right\}.
\]
The normalization by $N$ gives an additive approximation that remains
applicable when $m/N$ tends to zero.

\begin{lemma}[Uniform Gaussian tail counts]
\label{lem:fdr-grid}
Under Lemma~\ref{lem:fdr-gaussian-count},
\begin{equation}\label{eq:fdr-grid-bound}
 \Pp\{\mathcal B(u_0,\epsilon)^c\}
 \leq\frac{(1+\epsilon)\Lambda m}{N^2u_0\epsilon^3}.
\end{equation}
Under Lemma~\ref{lem:fdr-sparse-count},
\begin{equation}\label{eq:fdr-sparse-grid-bound}
 \Pp\{\mathcal B(u_0,\epsilon)^c\}
 \leq\frac{C_\varrho m}{N^2\epsilon^3}
           \left(\frac1{u_0}+\frac{s}{u_0^{\beta_\varrho}}\right).
\end{equation}
\end{lemma}

\begin{proof}[Proof of Lemma~\ref{lem:fdr-grid}]
Define the geometric grid
\[
 \begin{gathered}
 J=\left\lceil\frac{\log(1/u_0)}{\log(1+\epsilon)}\right\rceil,
 \qquad u_j=u_0(1+\epsilon)^j\ (0\leq j<J),\qquad u_J=1,\\
 \mathcal D=\bigcap_{j=0}^J
       \{V_G(u_j)\leq(m+\epsilon N)u_j\},\qquad
 u_{j+1}\leq(1+\epsilon)u_j\quad(0\leq j<J).
 \end{gathered}
\]
On $\mathcal D$, for $u\in[u_j,u_{j+1}]$,
\[
 \frac{V_G(u)}{Nu}
 \leq\frac{V_G(u_{j+1})}{Nu}
 \leq\left(\frac mN+\epsilon\right)\frac{u_{j+1}}u
 \leq(1+\epsilon)\left(\frac mN+\epsilon\right),
 \qquad \mathcal D\subseteq\mathcal B(u_0,\epsilon).
\]
Since $V_G(1)=m$, Chebyshev's inequality and
\eqref{eq:fdr-gaussian-count-var} give
\[
 \begin{aligned}
 \Pp\{\mathcal B(u_0,\epsilon)^c\}
 &\leq\Pp(\mathcal D^c)
 \leq\sum_{j=0}^{J-1}
     \Pp\{V_G(u_j)-mu_j>\epsilon Nu_j\}\\
 &\leq\frac{\Lambda m}{N^2\epsilon^2}
                \sum_{j=0}^{J-1}u_j^{-1}
 \leq\frac{\Lambda m}{N^2\epsilon^2u_0}
                \sum_{j=0}^{\infty}(1+\epsilon)^{-j}
 =\frac{(1+\epsilon)\Lambda m}{N^2u_0\epsilon^3}.
 \end{aligned}
\]
Under sparse correlation, \eqref{eq:fdr-sparse-count-var} gives
\[
 \Pp\{V_G(u_j)-mu_j>\epsilon Nu_j\}
 \leq\frac{m}{N^2\epsilon^2}
       \left\{\frac1{u_j}
             +\frac{C_\varrho s}{u_j^{\beta_\varrho}}\right\}.
\]
For $\beta=\beta_\varrho\in(0,1)$,
\[
 \begin{aligned}
 1-(1+\epsilon)^{-\beta}
 &=\beta\int_0^\epsilon(1+x)^{-\beta-1}\,dx
 \geq\beta(3/2)^{-\beta-1}\epsilon,\\
 \sum_{j=0}^{J-1}u_j^{-\beta}
 &\leq\frac{u_0^{-\beta}}{1-(1+\epsilon)^{-\beta}}
 \leq\frac{(3/2)^{\beta+1}}{\beta\epsilon u_0^\beta}.
 \end{aligned}
\]
Consequently,
\[
 \begin{aligned}
 \Pp\{\mathcal B(u_0,\epsilon)^c\}
 &\leq\frac{m}{N^2\epsilon^2}\sum_{j=0}^{J-1}u_j^{-1}
       +\frac{C_\varrho ms}{N^2\epsilon^2}
                     \sum_{j=0}^{J-1}u_j^{-\beta_\varrho}\\
 &\leq\frac{C_\varrho m}{N^2\epsilon^3}
           \left(\frac1{u_0}+\frac{s}{u_0^{\beta_\varrho}}\right).
 \end{aligned}
\]
These are \eqref{eq:fdr-grid-bound} and
\eqref{eq:fdr-sparse-grid-bound}.
\end{proof}

Write $\lambda(x)=\phi(x)/\barPhi(x)$ for the inverse Mills ratio.

\begin{lemma}[Normal-tail inflation]\label{lem:normal-tail-inflation}

For $z\in\R$ and $d\geq0$,
\begin{equation}\label{eq:fdr-tail-shift}
 \frac{\barPhi(z-d)}{\barPhi(z)}
       \leq e^{d\{1+\max(z,0)\}}.
\end{equation}
\end{lemma}

\begin{proof}[Proof of Lemma~\ref{lem:normal-tail-inflation}]
For $x\geq0$, let $E_x\sim\operatorname{Exp}(x+1)$. Then
\[
 \E E_x=\frac1{x+1},\qquad
 \E E_x^2=\frac2{(x+1)^2},
\]
and Jensen's inequality gives
\[
 \begin{aligned}
 \frac{\barPhi(x)}{\phi(x)}
 &=\int_0^\infty e^{-xt-t^2/2}\,dt
 =\frac1{x+1}\E e^{E_x-E_x^2/2}\\
 &\geq\frac1{x+1}
        \exp\{\E E_x-\E(E_x^2)/2\}
 =\frac1{x+1}\exp\left\{\frac{x}{(x+1)^2}\right\}
 \geq\frac1{x+1}.
 \end{aligned}
\]
For $x<0$,
\[
 \lambda(x)=\frac{\phi(x)}{\barPhi(x)}
 \leq\frac{\phi(0)}{\barPhi(0)}=\sqrt{2/\pi}<1,
 \qquad \lambda(x)\leq1+\max(x,0)\quad(x\in\R).
\]
Hence, for $z\in\R$ and $d\geq0$,
\[
 \log\frac{\barPhi(z-d)}{\barPhi(z)}
 =\int_{z-d}^z\lambda(x)\,dx
 \leq\int_{z-d}^z\{1+\max(z,0)\}\,dx
 =d\{1+\max(z,0)\}.\qedhere
\]
\end{proof}

\begin{proof}[Proof of Theorem~\ref{thm:fdr-main}]
For $N_0=0$, set $V_{G,0}\equiv0$ and $\mathcal B_{NT}=\Omega$.
Write $S_i=S_{w,i}^{\mathrm{df}}$, let $\widehat{\operatorname{se}}_i$
be its fitted standard error, and put
\[
 \nu_i=\frac{\sqrt T\alpha_i}{d_i\sqrt{\mathcal V_{w,T}}},\qquad
 \widehat\nu_i=\frac{\alpha_i}{\widehat{\operatorname{se}}_i},\qquad
 G_i=G_{w,i},\qquad
 \widehat k=\widehat k_w,\qquad
 \mathcal C_{NT}=\sigma(\mathbf F,\bm\tau).
\]
For the event $\mathcal C_{w,NT}$ in the proof of
Theorem~\ref{thm:main-statistic}, set
$\eta_{\mathrm{root},NT}:=\Pp(\mathcal C_{w,NT}^c)\to0$.
On $\mathcal C_{w,NT}$, $p_{w,i}=\barPhi(S_i)$ and
\begin{equation}\label{eq:fdr-derived-representation}
 S_i=\widehat\nu_i+G_i+\xi_i,\qquad
 \bm G\mid\mathcal C_{NT}\sim N(\bm0,\mathbf R_N),\qquad
 \Xi_{NT}:=\frac{\max_i|\xi_i|}{r_{NT}}=O_p(1).
\end{equation}
Also, writing
\[
 D_{NT}:=\max_i\left|
   \frac{\sqrt T\,\widehat{\operatorname{se}}_i}
        {d_i\sqrt{\mathcal V_{w,T}}}-1\right|,
\]
Theorem~\ref{thm:main-statistic} gives $D_{NT}=o_p(1)$.
Set $\Xi_{NT}=D_{NT}=0$ on $\mathcal C_{w,NT}^c$, retaining
$\eta_{\mathrm{root},NT}$ in the exceptional-event bounds below.

Let
\[
 \mathcal A_{NT}
 =\mathcal F_{NT}\cap
       \{\min_{i\in\mathcal S_N}\nu_i\geq2\sqrt{\log N}\}
       \in\mathcal C_{NT},\qquad
 \eta_{C,NT}=\Pp(\mathcal A_{NT}^c).
\]
By Assumption~\ref{ass:fdr-signals},
\[
 \eta_{C,NT}
 \leq\Pp(\mathcal F_{NT}^c)
       +\Pp\{\min_{i\in\mathcal S_N}\nu_i<2\sqrt{\log N}\}
       \longrightarrow0.
\]
On $\mathcal C_{w,NT}$, $\widehat\nu_i\leq0$ for $i\in\mathcal H_0$.
For $(u_0,z_0)=(u_{0,N},z_{0,N})$ in \eqref{eq:fdr-floor},
the normal Chernoff bound gives
\[
 \begin{gathered}
 0<u_0=\frac{qk_N}{N}\leq\frac q2<\frac12,\qquad z_0>0,\\
 z_0^2\leq2\log\frac{N}{qk_N}
          \leq2\log N+2\log(1/q),\qquad
 r_{NT}(1+z_0)\longrightarrow0.
 \end{gathered}
\]
Choose the deterministic error envelope
\begin{align}
 a_{NT}&=\left(\frac{r_{NT}}{1+z_0}\right)^{1/2},\qquad
 \frac{a_{NT}}{r_{NT}}
     =\{r_{NT}(1+z_0)\}^{-1/2}\longrightarrow\infty,\nonumber\\
 \delta_{NT}^{\mathrm{tail}}&=a_{NT}(1+z_0)
     =\{r_{NT}(1+z_0)\}^{1/2}\longrightarrow0.
 \label{eq:fdr-derived-envelope}
\end{align}
Set
\[
 \mathcal E_{NT}
  =\mathcal C_{w,NT}
        \cap\{\Xi_{NT}\leq a_{NT}/r_{NT}\}
        \cap\{D_{NT}\leq1/4\},\qquad
 \eta_{E,NT}=\Pp(\mathcal E_{NT}^c).
\]
Tightness of $\Xi_{NT}$ and consistency of $D_{NT}$ imply
\[
 \eta_{E,NT}
 \leq\eta_{\mathrm{root},NT}
       +\Pp\{\Xi_{NT}>a_{NT}/r_{NT}\}
       +\Pp(D_{NT}>1/4)\longrightarrow0.
\]
Conditional Gaussian bounds use only $\mathcal C_{NT}$. Define
\begin{equation}\label{eq:fdr-count-def}
 \begin{gathered}
 P_i=p_{w,i},\qquad R_N=|\widehat{\mathcal H}|,\\
 V_N=|\widehat{\mathcal H}\cap\mathcal H_0|,\qquad
 V_N(u)=\sum_{i\in\mathcal H_0}\ind(P_i\leq u).
 \end{gathered}
\end{equation}
On $\mathcal A_{NT}\cap\mathcal E_{NT}$, for
$i\in\mathcal H_0$ and $u\in[u_0,q]$,
\begin{equation}\label{eq:fdr-null-inclusion}
 P_i\leq u\ \Longrightarrow\ S_i\geq z(u)
 \ \Longrightarrow\ G_i\geq z(u)-\widehat\nu_i-\xi_i
                         \geq z(u)-a_{NT}.
\end{equation}
Set $v_{NT}(u)=\barPhi\{z(u)-a_{NT}\}$. Since
$z(u)\leq z_0$ and $z_0>0$, \eqref{eq:fdr-tail-shift} gives
\begin{equation}\label{eq:fdr-v-inflation}
 u_0\leq u\leq v_{NT}(u)<1,\qquad
 \frac{v_{NT}(u)}u
 \leq e^{a_{NT}\{1+\max(z(u),0)\}}
 \leq e^{\delta_{NT}^{\mathrm{tail}}}.
\end{equation}
Let $V_{G,0}=V_G$ for $\mathcal I=\mathcal H_0$, and define
\[
 \vartheta_N=
 \begin{cases}
 b_N/k_N,& b_N/k_N\to0,\\[1mm]
 k_N^{-1}+
 \displaystyle\frac{s_N}{N^{1-\beta_\varrho}k_N^{\beta_\varrho}},
       &\text{under the sparse alternative}.
 \end{cases}
\]
Use the first branch when both conditions hold.
Assumption~\ref{ass:fdr-signals} gives $\vartheta_N\to0$, and
\[
 \begin{gathered}
 \frac{b_NN_0}{N^2u_0}
       \leq\frac{b_N}{qk_N},\qquad
 \frac{N_0}{N^2u_0}\leq\frac1{qk_N},\\
 \frac{N_0s_N}{N^2u_0^{\beta_\varrho}}
       \leq
       \frac{s_N}
       {q^{\beta_\varrho}N^{1-\beta_\varrho}k_N^{\beta_\varrho}}.
 \end{gathered}
\]
For $0<\epsilon_N\leq1/2$, let
$\mathcal B_{NT}=\mathcal B(u_0,\epsilon_N)$ for $V_{G,0}$, so that
\begin{equation}\label{eq:fdr-G-uniform}
 \sup_{u_0\leq v\leq1}\frac{V_{G,0}(v)}{Nv}
       \leq(1+\epsilon_N)\left(\frac{N_0}{N}+\epsilon_N\right)
           \quad\hbox{on }\mathcal B_{NT}.
\end{equation}
Under the spectral alternative, Lemma~\ref{lem:fdr-grid} gives
\[
 \begin{gathered}
 \|\mathbf R_{N,\mathcal H_0}\|_{\op}\leq b_N,\\
 \Pp(\mathcal B_{NT}^c\mid\mathcal C_{NT})
 \leq\frac{Cb_NN_0}{N^2u_0\epsilon_N^3}
 \leq C\vartheta_N\epsilon_N^{-3}.
 \end{gathered}
\]
Under the sparse alternative, \eqref{eq:fdr-sparse-grid-bound} gives
\[
 \Pp(\mathcal B_{NT}^c\mid\mathcal C_{NT})
 \leq\frac{CN_0}{N^2\epsilon_N^3}
       \left\{\frac1{u_0}
                +\frac{s_N}{u_0^{\beta_\varrho}}\right\}
 \leq C\vartheta_N\epsilon_N^{-3}.
\]
Thus
\begin{equation}\label{eq:fdr-B-event}
 \Pp(\mathcal A_{NT}\cap\mathcal B_{NT}^c)
 =\E\{\ind(\mathcal A_{NT})
                \Pp(\mathcal B_{NT}^c\mid\mathcal C_{NT})\}
 \leq C\vartheta_N\epsilon_N^{-3}.
\end{equation}
By \eqref{eq:fdr-null-inclusion}--\eqref{eq:fdr-v-inflation} and
\eqref{eq:fdr-G-uniform}, on
$\mathcal A_{NT}\cap\mathcal E_{NT}\cap\mathcal B_{NT}$,
\begin{equation}\label{eq:fdr-observed-uniform}
 \begin{aligned}
 V_N(u)&\leq V_{G,0}\{v_{NT}(u)\}\\
 &\leq Nv_{NT}(u)(1+\epsilon_N)
          \left(\frac{N_0}{N}+\epsilon_N\right)\\
 &\leq Nu(1+\epsilon_N)
          \left(\frac{N_0}{N}+\epsilon_N\right)
          e^{\delta_{NT}^{\mathrm{tail}}},
          \qquad u_0\leq u\leq q.
 \end{aligned}
\end{equation}
On $\mathcal A_{NT}\cap\mathcal E_{NT}$,
\[
 \widehat\nu_i
  \geq\frac{\nu_i}{1+D_{NT}}
  \geq\frac85\sqrt{\log N},\qquad i\in\mathcal S_N.
\]
Set $\ell_{NT}=\frac85\sqrt{\log N}-z_0-a_{NT}$ and choose
$0<c_0<\frac85-\sqrt2$. Eventually,
\[
 \frac{\ell_{NT}}{\sqrt{\log N}}
 \geq\frac85-\left\{2+\frac{2\log(1/q)}{\log N}\right\}^{1/2}
           -\frac{a_{NT}}{\sqrt{\log N}}\geq c_0.
\]
On $\mathcal A_{NT}\cap\mathcal E_{NT}$, for $i\in\mathcal S_N$,
\[
 \begin{aligned}
 G_i\geq-\ell_{NT}
 &\ \Longrightarrow\
 S_i\geq\widehat\nu_i+G_i-a_{NT}
      \geq\frac85\sqrt{\log N}-\ell_{NT}-a_{NT}=z_0\\
 &\ \Longrightarrow\
 P_i\leq u_0.
 \end{aligned}
\]
Define $Q_N=\sum_{i\in\mathcal S_N}\ind(G_i\geq-\ell_{NT})$.
By the conditional Gaussian law,
\[
 \begin{aligned}
 \E(2k_N-Q_N\mid\mathcal C_{NT})
 &=\sum_{i\in\mathcal S_N}
       \Pp(G_i<-\ell_{NT}\mid\mathcal C_{NT})\\
 &=2k_N\barPhi(\ell_{NT}),\\
 \Pp(Q_N<k_N\mid\mathcal C_{NT})
 &\leq\frac{\E(2k_N-Q_N\mid\mathcal C_{NT})}{k_N}
 =2\barPhi(\ell_{NT}).
 \end{aligned}
\]
Consequently,
\begin{equation}\label{eq:fdr-signal-markov}
 \Pp\{\mathcal A_{NT}\cap(Q_N<k_N)\}
 \leq h_N:=2\barPhi(\ell_{NT})
 \leq2e^{-\ell_{NT}^2/2}\leq2N^{-c_0^2/2}.
\end{equation}
On $\mathcal A_{NT}\cap\mathcal E_{NT}\cap\{Q_N\geq k_N\}$,
\[
 \sum_{i=1}^N\ind(P_i\leq u_0)\geq Q_N\geq k_N
 \ \Longrightarrow\
 P_{(k_N)}\leq u_0=\frac{qk_N}{N}
 \ \Longrightarrow\
 \widehat k\geq k_N.
\]
By \eqref{eq:bh},
\[
 \widehat k=0
 \ \Longrightarrow\
 P_{(1)}>q/N>0
 \ \Longrightarrow\
 R_N=V_N=V_N(0)=0.
\]
For $\widehat k>0$,
\[
 R_N\geq\widehat k,\qquad
 R_N>\widehat k
 \ \Longrightarrow\
 P_{(R_N)}\leq\frac{q\widehat k}{N}\leq\frac{qR_N}{N}
 \ \Longrightarrow\
 R_N\leq\widehat k.
\]
Hence
\begin{equation}\label{eq:fdr-self-consistency}
 R_N=\widehat k,\qquad V_N=V_N(qR_N/N).
\end{equation}
On $\mathcal A_{NT}\cap\mathcal E_{NT}\cap\mathcal B_{NT}
\cap\{Q_N\geq k_N\}$,
\[
 \begin{gathered}
 R_N\geq k_N,\qquad \frac{qR_N}{N}\in[u_0,q],\\
 \FDP=\frac{V_N(qR_N/N)}{R_N}
 \leq q(1+\epsilon_N)
       \left(\frac{N_0}{N}+\epsilon_N\right)
       e^{\delta_{NT}^{\mathrm{tail}}},
 \end{gathered}
\]
by \eqref{eq:fdr-count-def}, \eqref{eq:fdr-observed-uniform}, and
\eqref{eq:fdr-self-consistency}. Set
\[
 \mathcal K_{NT}=\mathcal A_{NT}\cap\mathcal E_{NT}
                  \cap\mathcal B_{NT}\cap\{Q_N\geq k_N\}.
\]
By \eqref{eq:fdr-B-event} and \eqref{eq:fdr-signal-markov},
\[
 \begin{aligned}
 \Pp(\mathcal K_{NT}^c)
 &\leq\Pp(\mathcal A_{NT}^c)+\Pp(\mathcal E_{NT}^c)
       +\Pp(\mathcal A_{NT}\cap\mathcal B_{NT}^c)
       +\Pp\{\mathcal A_{NT}\cap(Q_N<k_N)\}\\
 &\leq\Delta_{NT}:=\eta_{C,NT}+\eta_{E,NT}+h_N
                         +C\vartheta_N\epsilon_N^{-3}.
 \end{aligned}
\]
Since $0\leq\FDP\leq1$,
\begin{align}
 \Pp\!\left\{\FDP>
       q(1+\epsilon_N)\left(\frac{N_0}{N}+\epsilon_N\right)
         e^{\delta_{NT}^{\mathrm{tail}}}\right\}
       &\leq\Delta_{NT},\label{eq:fdr-prob-bound}\\
 \FDR
 &=\E\{\FDP\ind(\mathcal K_{NT})\}
        +\E\{\FDP\ind(\mathcal K_{NT}^c)\}\nonumber\\*
 &\leq q(1+\epsilon_N)\left(\frac{N_0}{N}+\epsilon_N\right)
         e^{\delta_{NT}^{\mathrm{tail}}}+\Delta_{NT}.
 \label{eq:fdr-mean-bound}
\end{align}
Choose $\epsilon_N=\min(1/2,\vartheta_N^{1/4})$.
By \eqref{eq:fdr-derived-envelope}, eventually
\[
 \begin{gathered}
 \epsilon_N=\vartheta_N^{1/4},\\
 (1+\epsilon_N)\left(\frac{N_0}{N}+\epsilon_N\right)
       e^{\delta_{NT}^{\mathrm{tail}}}-\frac{N_0}{N}
       \leq C(\vartheta_N^{1/4}+\delta_{NT}^{\mathrm{tail}})
       \longrightarrow0,\\
 \Delta_{NT}
 \leq\eta_{C,NT}+\eta_{E,NT}
          +2N^{-c_0^2/2}+C\vartheta_N^{1/4}\longrightarrow0.
 \end{gathered}
\]
Equations~\eqref{eq:fdr-prob-bound}--\eqref{eq:fdr-mean-bound} imply
\[
 \begin{aligned}
 \FDR-q\frac{N_0}{N}
 &\leq C(\vartheta_N^{1/4}+\delta_{NT}^{\mathrm{tail}})
       +\eta_{C,NT}+\eta_{E,NT}+2N^{-c_0^2/2}\longrightarrow0,\\
 \Pp\!\left\{\FDP>q\frac{N_0}{N}+\epsilon\right\}
 &\leq\Delta_{NT}
       +\ind\{C(\vartheta_N^{1/4}+\delta_{NT}^{\mathrm{tail}})>\epsilon\}
       \longrightarrow0,\qquad\epsilon>0.
 \end{aligned}
\]
This proves \eqref{eq:fdr-asymptotic}.
\end{proof}

\begin{proof}[Proof of Corollary~\ref{cor:strong-fund-recovery}]
Use the notation from Theorem~\ref{thm:fdr-main} and put
\[
 m_{NT}=\min_{i\in\mathcal D_N}\nu_i,\qquad
 M_{NT}=m_{NT}-z_0,\qquad
 \mathcal U_{NT}=\{D_{NT}z_0\leq1\}.
\]
By \eqref{eq:studentized-se-consistency}, with $K_{NT}$ from
\eqref{eq:recovery-tail-rate},
\[
 D_{NT}=O_p\{d_{NT}+\beta_{NT}(K_{NT})\}.
\]
Since $z_0=O(\sqrt{\log N})$, the growth condition and
\eqref{eq:recovery-tail-rate} imply
\[
 \begin{gathered}
 D_{NT}z_0
 =O_p\!\left\{d_{NT}\sqrt L+
       \sqrt{\frac{K_{NT}L^2}{T}}
       +\sqrt L\,\Psi(K_{NT})\right\}=o_p(1),\\
 \eta_{U,NT}:=\Pp(\mathcal U_{NT}^c)\longrightarrow0.
 \end{gathered}
\]
On $\mathcal E_{NT}$ all target alphas are positive and
\begin{equation}\label{eq:fdr-recovery-random-margin}
 \min_{i\in\mathcal D_N}\widehat\nu_i-z_0
 \geq\frac{m_{NT}}{1+D_{NT}}-z_0
 =\frac{M_{NT}-D_{NT}z_0}{1+D_{NT}}.
\end{equation}
Let $n_{NT}=\min(N,T)$ and choose
\[
 c_{NT}=\max\{j\in\mathbb N:j\leq n_{NT},\quad
                         \Pp(M_{NT}<4j)\leq j^{-1}\}.
\]
Since $M_{NT}\xrightarrow{p}\infty$, the nonempty set eventually
contains each fixed $j$, so $c_{NT}\to\infty$. Let
\[
 \mathcal T_{NT}=\{M_{NT}\geq4c_{NT}\},\qquad
 \eta_{D,NT}:=\Pp(\mathcal T_{NT}^c)
        \leq c_{NT}^{-1}\longrightarrow0,
\]
\eqref{eq:fdr-recovery-random-margin} gives, on
$\mathcal E_{NT}\cap\mathcal U_{NT}\cap\mathcal T_{NT}$,
\[
 \min_{i\in\mathcal D_N}\widehat\nu_i-z_0
   \geq\frac{4c_{NT}-1}{5/4}\geq2c_{NT}.
\]
Also $a_{NT}\leq c_{NT}$ eventually. On the intersection with
$\mathcal A_{NT}\cap\{Q_N\geq k_N\}$,
\eqref{eq:fdr-self-consistency} gives $R_N\geq k_N$, and therefore
\[
 \begin{gathered}
 G_i\geq-c_{NT}
 \ \Longrightarrow\
 S_i\geq\widehat\nu_i+G_i-a_{NT}
      \geq z_0+2c_{NT}-c_{NT}-a_{NT}\geq z_0,\\
 P_i\leq u_0\leq qR_N/N
 \ \Longrightarrow\ i\in\widehat{\mathcal H},
 \qquad i\in\mathcal D_N.
 \end{gathered}
\]
The conditional Gaussian law gives
\begin{align*}
 \E\left\{\frac1{|\mathcal D_N|}
       \sum_{i\in\mathcal D_N}\ind(G_i<-c_{NT})
             \,\middle|\,\mathcal C_{NT}\right\}
 &=\barPhi(c_{NT})\leq e^{-c_{NT}^2/2}.
\end{align*}
Combining this bound with \eqref{eq:fdr-signal-markov} yields
\begin{equation}\label{eq:fdr-power-bound}
 \E\{1-\TDP(\mathcal D_N)\}
 \leq\eta_{C,NT}+\eta_{E,NT}+\eta_{U,NT}+c_{NT}^{-1}
              +2N^{-c_0^2/2}+e^{-c_{NT}^2/2}\longrightarrow0.
\end{equation}
Markov's inequality now gives
$\Pp\{\TDP(\mathcal D_N)<1-x\}\leq
x^{-1}\E\{1-\TDP(\mathcal D_N)\}\to0$ for each $x>0$.
\end{proof}

\section{Proofs of the efficiency results}\label{app:efficiency}

This appendix derives the optimal radial weight and the benchmark
efficiency gaps.

\begin{proof}[Proof of Theorem~\ref{thm:optimal-radial-weight}]
Put $u(r)=w(r)/A_w$ and $u_\star(r)=(m_{-2}r)^{-1}$. Then
\begin{gather*}
 \E\{u(r)/r\}=1,\qquad
 \E u(r)^2=A_w^{-2}\E w(r)^2<\infty,\\
 \E\{u_\star(r)/r\}=m_{-2}^{-1}\E r^{-2}=1,\qquad
 \E u_\star(r)^2=m_{-2}^{-1}<\infty,\\
 \E|u(r)u_\star(r)|
 \leq\{\E u(r)^2\,\E u_\star(r)^2\}^{1/2}<\infty.
\end{gather*}
The common normalization makes the cross term vanish:
\begin{equation*}
 \E\{[u(r)-u_\star(r)]u_\star(r)\}
 =\frac1{m_{-2}}
    \left[\E\{u(r)/r\}-\E\{u_\star(r)/r\}\right]=0.
\end{equation*}
Consequently,
\[
 \begin{aligned}
 J(w)
 &=\E u(r)^2\\
 &=\E u_\star(r)^2
   +2\E\{[u(r)-u_\star(r)]u_\star(r)\}
   +\E\{[u(r)-u_\star(r)]^2\}\\
 &=\frac1{m_{-2}}
   +\E\left\{\left[\frac{w(r)}{A_w}
                         -\frac1{m_{-2}r}\right]^2\right\}.
 \end{aligned}
\]
The minimum and equality condition follow:
\[
 \begin{gathered}
 J(w_\star)=m_{-2}^{-1}=\inf_wJ(w),\\
 J(w)=J(w_\star)
 \ \Longleftrightarrow\
 \E\{[u(r)-u_\star(r)]^2\}=0
 \ \Longleftrightarrow\
 \Pp\!\left\{w(r)=\frac{A_w}{m_{-2}}r^{-1}\right\}=1.
 \end{gathered}
\]
Applying the same calculation to $r=\tau$, $w=k$, and
$A_w=\E\{k(\tau)/\tau\}$ gives
\eqref{eq:profile-efficiency-gap}.\qedhere
\end{proof}

\Needspace{3\baselineskip}
\begin{proof}[Proof of Corollary~\ref{cor:benchmark-efficiency-gaps}]
Let $u_0=a_{\SSm}^{-1}$ and $u_\star(\tau)=(\zeta\tau)^{-1}$.
Cauchy--Schwarz gives
\[
 \begin{gathered}
 0<a_{\SSm}\leq\zeta^{1/2}<\infty,\qquad
 \E|\tau|\leq\mu_2^{1/2}<\infty,\\
 \E\{u_0/\tau\}=u_0a_{\SSm}=1,
 \qquad
 \E\{u_\star(\tau)/\tau\}=1,\\
 \E\{u_\star(\tau)^2\}=\zeta^{-1},\qquad
 \E\{u_0u_\star(\tau)\}=\zeta^{-1}.
 \end{gathered}
\]
Hence
\begin{align*}
 \E\{\tau-u_\star(\tau)\}^2
   &=\mu_2-\frac2\zeta+\frac1\zeta
     =\mu_2-\zeta^{-1},\\
 \E\{u_0-u_\star(\tau)\}^2
   &=u_0^2-\frac2\zeta+\frac1\zeta
     =a_{\SSm}^{-2}-\zeta^{-1}.
\end{align*}
By \eqref{eq:benchmark-variances},
\[
 \mathcal V_{\LS,T}-\mathcal V_{\IN,T}
 =\omega_T^{-1}(\mu_2-\zeta^{-1})
 =\omega_T^{-1}\E\{\tau-u_\star(\tau)\}^2\geq0.
\]
Since $\omega_T+\kappa_T=1$,
\[
 \begin{aligned}
 \omega_T^2\mathcal V_{\SSm,T}
 &=u_0^2-2\kappa_Tu_0\E\tau+\kappa_T\mu_2\\
 &=\omega_Tu_0^2+\kappa_T
          (u_0^2-2u_0\E\tau+\mu_2)\\
 &=\omega_Tu_0^2+\kappa_T\E(u_0-\tau)^2.
 \end{aligned}
\]
Subtracting $\mathcal V_{\IN,T}=(\omega_T\zeta)^{-1}$ yields
\[
 \begin{aligned}
 \mathcal V_{\SSm,T}-\mathcal V_{\IN,T}
 &=\frac{u_0^2-\zeta^{-1}}{\omega_T}
       +\frac{\kappa_T}{\omega_T^2}\E(u_0-\tau)^2\\
 &=\frac1{\omega_T}\E\{u_0-u_\star(\tau)\}^2
       +\frac{\kappa_T}{\omega_T^2}\E(u_0-\tau)^2\geq0.
 \end{aligned}
\]
The equality conditions are
\[
 \begin{gathered}
 \mu_2-\zeta^{-1}=0
 \ \Longleftrightarrow\
 \Pp\{\tau=(\zeta\tau)^{-1}\}=1
 \ \Longleftrightarrow\
 \Pp\{\tau=\zeta^{-1/2}\}=1,\\
 a_{\SSm}^{-2}-\zeta^{-1}=0
 \ \Longleftrightarrow\
 \zeta=a_{\SSm}^2
 \ \Longleftrightarrow\
 \Var(\tau^{-1})=0,\\
 \Pp(\tau=t_0)=1
 \ \Longrightarrow\
 u_0=t_0=u_\star(\tau),\qquad
 \E(u_0-\tau)^2=0.
 \end{gathered}
\]
This proves both the nonnegative gaps and their equality conditions.
\end{proof}

For positive local alphas, the common-cutoff power comparisons follow from
Corollary~\ref{cor:radial-local-power}, the variance gaps, and
$c_{\mathrm{df},T}\to1$.

\end{document}